\documentclass[manuscript]{acmart}
\DeclareUnicodeCharacter{2009}{\,}
\usepackage[qm]{qcircuit}
\usepackage{caption}
\usepackage{graphicx}
\usepackage{soul}
\usepackage{mathtools}
\usepackage{subcaption}
\usepackage{float}
\usepackage[algo2e, linesnumbered, ruled, vlined]{algorithm2e}
\usepackage{tikz}
\usepackage{tikz-3dplot} 
\usetikzlibrary{trees, shadows, arrows.meta, shapes.multipart, backgrounds, calc, fit, positioning}
\usetikzlibrary{decorations.pathreplacing}

\definecolor{OIgray}{RGB}{153,153,153}

\colorlet{neutral-left}{OIgray!15!white} 

\definecolor{PastelIBMBlue}{RGB}{173,200,255}     
\definecolor{PastelIBMPurple}{RGB}{188,180,255}   
\definecolor{PastelIBMPink}{RGB}{244,182,204}     
\definecolor{PastelIBMOrange}{RGB}{255,194,150}   
\definecolor{PastelIBMYellow}{RGB}{255,225,150}   

\newcommand{\ket}[1]{\lvert #1 \rangle} 
\newcommand{\bra}[1]{\langle #1 \rvert}
\newcommand{\xdownarrow}[1]{%
  {\left\downarrow\vbox to #1{}\right.\kern-\nulldelimiterspace}
}

\newtheorem{theorem}{Theorem}
\newtheorem{lemma}{Lemma}

\AtBeginDocument{%
  }

\setcopyright{acmlicensed}
\copyrightyear{2025}
\acmYear{2025}
\acmDOI{XXXXXXX.XXXXXXX}

\acmJournal{POMACS}
\acmVolume{37}
\acmNumber{4}
\acmArticle{111}
\acmMonth{8}

\usepackage{xcolor}
\usepackage{booktabs}
\usepackage{makecell}

\newcommand{\RightCellDownArrow}[2][\Large]{%
  \par\noindent
  \hfill
  \makebox[0.465\linewidth]{\rotatebox[origin=c]{270}{$\longrightarrow$}}
   \par\vspace{#2}%
}

\newcommand{\LeftCellDownArrow}[2][\Large]{%
  \par\noindent
  \makebox[0.56\linewidth][l]{\rotatebox[origin=c]{270}{$\longrightarrow$}}%
  \hfill
  \par\vspace{#2}%
}

\begin{document}

\title{Efficient Quantum State Preparation with Bucket Brigade QRAM}

\author{Alessandro Berti}
\email{alessandro.berti@df.unipi.it}
\orcid{0000-0001-9144-9572}
\affiliation{%
  \institution{Department of Computer Science and Department of Physics, University of Pisa}
  \city{Pisa}
  \country{Italy}
}

\author{Francesco Ghisoni}
\email{francesco.ghisoni01@universitadipavia.it}
\orcid{0000-0002-0123-2933}
\affiliation{%
  \institution{Department of Physics, University of Pavia}
  \city{Pavia}
  \country{Italy}
}

\renewcommand{\shortauthors}{\footnotesize \textsc{\textbf{Preprint}}}

\begin{abstract}
The preparation of data in quantum states is a critical component in the design of quantum algorithms. The cost of this step can significantly limit the realization of quantum advantage in domains such as machine learning, finance, and chemistry. One of the main approaches to achieve efficient state preparation is through the use of Quantum Random Access Memory (QRAM), a theoretical device for coherent data access with several proposed hardware implementations. In this work, we present a framework that integrates the hardware model of the Bucket Brigade QRAM (BBQRAM) with the classical data structure of the Segment Tree to achieve efficient state preparation. We introduce a memory layout that embeds a segment tree within BBQRAM memory cells by preserving the segment tree's hierarchy and supporting data retrieval in logarithmic time via specialized access primitives. We demonstrate that our method encodes a matrix $A \in \mathbb{R}^{M \times N}$ in a quantum register of $\Theta(\log_2(MN))$ qubits in $\mathcal{O}(\log_2^2(MN))$ time, {requiring a constant number of working qubits (under fixed precision) and $\mathcal{O}(MN)$ memory cells within the BBQRAM architecture.} We further illustrate the method through a numerical example. This framework provides theoretical support for quantum algorithms that assume negligible data loading overhead and establishes a foundation for designing classical-to-quantum encoding algorithms that are aware of the underlying hardware QRAM architecture.
\end{abstract}


\keywords{Bucket Brigade, QRAM, Efficient State Preparation, Segment Tree}

\maketitle

\section{Introduction}\label{sec:introduction}
Efficient quantum state preparation is fundamental to obtaining advantages from quantum algorithms that process large data~\cite{hann2021practicality, luongo2021quantumalgorithms, prakash2014quantum}. Without proper state preparation, the time required for encoding data in a quantum state can dominate the overall computational cost, resulting in a \emph{data bottleneck} that negates theoretical speedups. 
Many applications across diverse domains depend on the efficient encoding of large datasets into quantum states. Examples include quantum linear algebra~\cite{Lloyd_2014_QPCA, bernasconiMatMul}, quantum artificial intelligence~\cite{Wiebe_2012_data_fitting, Rebentrost_2014_QSVM, biamonte2017quantum, schuld2018supervised, berti2024role, bernasconi2024quantum}, quantum finance~\cite{rebentrost2022_finance, https://doi.org/10.4230/lipics.tqc.2022.2}, and quantum simulations of complex materials and molecules~\cite{Babbush_2018, Low_2024, Berry_2019, fomichev2024}. Quantum algorithms typically assume that input data is already available in a quantum state at negligible cost. However, in practice, implementations must account for the overhead associated with state preparation to realize actual speedups.
To prevent this step from becoming a computational bottleneck, its complexity must scale at most polylogarithmically with the data size~\cite{kerenidis_et_al:LIPIcs.ITCS.2017.49}.

{Quantum Random Access Memory (QRAM)}~\cite{Jaques2025qramsurveycritique} {enables data retrieval in superposition, a key primitive for quantum state preparation and many quantum algorithms. The literature distinguishes two fundamentally different approaches to realize QRAM: \emph{circuit-based QRAM} and \emph{hardware QRAM}. Circuit-based QRAM implements the memory access within a quantum circuit using a standard gate set. This approach faces an inherent gate-count lower bound that scales linearly in the number of memory entries, even when the circuit depth remains logarithmic; such a cost becomes prohibitive for dense data at large scale. Hardware QRAM, on the other hand, decouples memory storage from quantum processor and supports repeated coherent access to the same classical data without rebuilding a circuit for each query. However, QRAM alone does not guarantee efficient state preparation. Without a structured organization of data in memory, retrieval costs can grow significantly and reduce or negate the computational gains of a quantum algorithm.}

In~\cite{kerenidis_et_al:LIPIcs.ITCS.2017.49}, the authors address this challenge using KP-trees, a tree-like structure of precomputed amplitude norms. It is worth mentioning that the KP-Tree data structure is equivalent to the standard Segment Tree, a well-known data structure in computer science; the difference between the two is primarily notational and does not affect performance. To maintain consistency with standard terminology and to facilitate communication between the quantum computing and computer science communities, we refer to this data structure as \textit{Segment Tree} throughout this manuscript. The authors use a segment tree to organize and precompute the amplitudes required for state preparation and assume QRAM queries that efficiently retrieve the stored values from the segment tree. Such work establishes an important theoretical foundation for achieving efficient state preparation, but it does not examine the overhead introduced by a hardware QRAM model, leaving a loose complexity bound. 
The core challenge is therefore providing an efficient state preparation procedure that is aware of the underlying QRAM architecture, allowing a clear performance analysis to evaluate possible quantum advantages across various application domains.

In this work, we address this open problem under fault-tolerant quantum computing assumptions by embedding the Segment Tree data structure within the \emph{Bucket Brigade} QRAM architecture (BBQRAM)~\cite{giovannetti2008architectures}, a well-studied hardware QRAM model that exhibits logarithmic access time in the number of stored values. While the Segment Tree is a fundamental support data structure, embedding it directly into BBQRAM can be non-trivial. A naïve mapping of tree nodes to memory cells can inflate retrieval costs, undermining the efficiency gains necessary for efficient state preparation. To overcome this, we propose a memory layout that guarantees efficient data retrieval in superposition from BBQRAM memory cells. Based on this layout, we define quantum primitives for retrieving precomputed amplitudes stored in the segment tree. These primitives enable the construction of an efficient amplitude encoding algorithm for matrices that can represent the input data of a given quantum algorithm. In particular, given a matrix $A \in \mathbb{R}^{M \times N}$ with elements $a_{i,j}$, we preprocess $A$ to construct a segment tree of squared norms and embed it into BBQRAM according to our memory layout. Using the defined retrieval primitives, we implement a reversible procedure that prepares the quantum state $
\frac{1}{\lVert A \rVert_F} \sum_{i=0}^{M-1}\sum_{j=0}^{N-1} a_{i,j} \, \ket{i}\ket{j}$ in $\mathcal{O}(\log_2^2(MN))$ time in $\Theta(\log_2(MN))$ qubits, {requiring a constant number of working qubits under a fixed precision assumption and $\mathcal{O}(MN)$ memory cells within the BBQRAM architecture.}

This work makes the following contributions: \begin{enumerate}
\item \textbf{BBQRAM-aware state preparation algorithm.} We present a framework that integrates the Bucket Brigade QRAM architecture with the Segment Tree data structure, demonstrating how theoretical state preparation algorithms for matrices can be realized within a hardware QRAM model. This bridges the gap between quantum algorithms and how they interact with quantum memory architectures.

\item \textbf{Memory layout design.} We design and formally analyze a mapping from segment tree nodes to BBQRAM memory cells that preserves the tree's hierarchical structure and ensures logarithmic access time.

\item \textbf{Quantum retrieval primitives.} We define quantum primitives for retrieving precomputed amplitudes and sign bits in superposition from the BBQRAM, which serve as fundamental building blocks for efficient amplitude encoding algorithms.

\item \textbf{Explicit polylogarithmic time complexity bound.} While prior work~\cite{kerenidis_et_al:LIPIcs.ITCS.2017.49} established the existence of efficient state preparation with unspecified polylogarithmic complexity {(i.e., $\mathcal{O}(\mathrm{polylog}(MN))$), our work sharpens this bound by proving that the state preparation algorithm runs in $\mathcal{O}(\log_2^2(MN))$ time.}

\item \textbf{Comprehensive numerical example.} We provide a detailed step-by-step numerical example that demonstrates the algorithm execution, illustrating the mapping and retrieval processes for practical implementation guidance.
\end{enumerate}

We organize the manuscript as follows. In Section~\ref{sec:related_works}, we provide an overview of QRAM-based techniques for the preparation of a quantum state. In Section~\ref{sec:preliminaries}, we introduce the preliminaries, including notation and essential quantum operations central to our work. In Section~\ref{sec:background}, we review the BBQRAM architecture with a focus on routing and retrieval complexity, and we outline the classical Segment Tree data structure for amplitude precomputation. Section~\ref{sec:efficient_state_preparation} details our mapping of the segment tree into BBQRAM memory cells, describes the retrieval primitives, and presents the efficient state preparation algorithm. In Section~\ref{sec:numerical_example}, we provide a numerical example that reviews the algorithm step by step. Finally, Section~\ref{sec:conclusion} summarizes the main contributions and discusses directions for future research.

\section{Related works}\label{sec:related_works}
\begin{table*}[t]
\centering
\resizebox{\textwidth}{!}{%
\begin{tabular}{@{} l c c c c c c c c c c @{}}
\toprule
\thead{Work} &
\thead{Data\\type} &
\thead{Encoding} &
\thead{\# QRAMs} &
\thead{QRAM\\Model} &
\thead{Retrieval\\Cost} &
\thead{Qubits} &
\thead{\# Ancillas} &
\thead{Memory\\Cells} &
\thead{Quantum Time} &
\thead{Classical Time} \\
\midrule
\cite{kerenidis_et_al:LIPIcs.ITCS.2017.49} & Matrix & Amplitude & $1$ & Oracle & $\mathcal{O}(\log_2(MN))$ & $\mathcal{O}(\log_2(MN))$ & --- & --- & $\mathcal{O}(\operatorname{polylog}(MN))$ & $\mathcal{O}(MN)$ \\[3pt]
\cite{casares2020} & Matrix & Amplitude & $1$ & Circuit (BB) & --- & $\mathcal{O}(\log_2(MN))$ & --- & --- & $\mathcal{O}(\operatorname{polylog}(MN))$ & $\mathcal{O}(MN)$ \\[3pt]
\cite{Zhang_2025} & Vector & Amplitude \& Block & $1$ & Circuit (BB) & $\mathcal{O}(\log_2(N))$ & $\mathcal{O}(\log_2(N))$ & $\mathcal{O}(N)$ & --- & $\mathcal{O}(\log_2(N))$ & $\mathcal{O}(N)$ \\[3pt]
\cite{prakash2014quantum} & Sparse vector & Amplitude & $2$ & Oracle & --- & $\mathcal{O}(\log_2(N))$ & --- & --- & $\tilde{\mathcal{O}}\!\left(k \sqrt{\mathrm{nnz}(x)}\, \lVert x \rVert_{\infty}\right)$ & $\mathcal{O}(\mathrm{nnz}(x))$ \\[3pt]
\cite{PhysRevA.110.032439} & Sparse vector & Amplitude & --- & Circuit & --- & $\mathcal{O}(\log_2(N))$ & $\mathcal{O}(1)$ & --- & $\mathcal{O}(s \log_2(N))$ & --- \\[3pt]
\cite{Clader_2022} & Matrix & Block & $1$ & Circuit (BB) & --- & $\mathcal{O}(\log_2(N))$ & $\mathcal{O}(N)$ & --- & $\mathcal{O}(\log_2(N))$ & $\mathcal{O}(N)$ \\
\midrule
\textbf{This work} & Matrix & Amplitude & $1$ & Hardware (BB) & $\mathcal{O}(\log_2(MN))$ & $\Theta(\log_2(MN))$ & --- & $\mathcal{O}(MN)$ & $\mathcal{O}(\log_2^2(MN))$ & $\mathcal{O}(MN)$ \\
\bottomrule
\end{tabular}%
}
\caption{{Comparison of QRAM-based state preparation approaches. In the \textbf{Data type} column, matrix refers to $A\in\mathbb{R}^{M \times N}$, vector refers to $x\in\mathbb{R}^{N}$, and sparse vector refers to $x\in\mathbb{R}^{N}$ with sparsity $s$ or $\mathrm{nnz}(x)$ non-zero elements. The \textbf{QRAM model} column distinguishes three settings: \emph{Oracle} treats the QRAM as an abstract black-box memory primitive, \emph{Circuit (BB)} implements the Bucket Brigade architecture as part of the quantum circuit using ancillary qubits, and \emph{Hardware (BB)} models the Bucket Brigade QRAM as a separate hardware device. \textbf{Retrieval cost} denotes the cost of a single coherent memory access (query or routing). \textbf{Qubits} refers to the number of qubits in the output register. \textbf{\# Ancillas} refers to additional qubits required by the QRAM architecture or the preparation circuit. \textbf{Memory cells} refers to the number of memory cells in the hardware QRAM. \textbf{Quantum time} denotes the overall asymptotic time for the full state preparation or block-encoding procedure. \textbf{Classical time} refers to the offline preprocessing time. A dash (---) indicates the metric is not applicable or not reported.}}
\label{tab:qram_comparison}
\end{table*}

{This section reviews related works on quantum state preparation. We organize the discussion by the QRAM model each work assumes, progressing from abstract oracle treatments, through circuit-level implementations, to concrete architectural models. Table}~\ref{tab:qram_comparison} {summarizes the comparison.}

{In}~\cite{kerenidis_et_al:LIPIcs.ITCS.2017.49, Wossnig_2018}{, the authors propose a QRAM-based state preparation scheme for amplitude encoding of a matrix $A \in \mathbb{R}^{M \times N}$. For each row, a segment tree stores the squared magnitudes $a_{i,j}^2$ together with the corresponding signs at its leaves, while internal nodes accumulate partial sums. An additional segment tree stores the squared row norms $\lVert A_i \rVert^2$ at its leaves, providing coherent access to the row-level normalization factors. A single QRAM instance gives coherent access to every level of these trees, yielding a query complexity of $\mathcal{O}(\log_2 MN)$ and an overall time of $\mathcal{O}(\operatorname{polylog}(MN))$. Because the work treats the QRAM as an abstract oracle, it does not analyze routing costs; the classical preprocessing to construct all segment trees scales as $\mathcal{O}(MN)$. In}~\cite{prakash2014quantum}{, the author introduces the Augmented QRAM, which combines two oracle QRAM instances with a classical key--value map implemented via hash functions. This model prepares the amplitude encoding of a sparse vector $x \in \mathbb{R}^{N}$ in time $\tilde{\mathcal{O}}(k\sqrt{\mathrm{nnz}(x)}\,\lVert x \rVert_{\infty})$, where $\mathrm{nnz}(x)$ denotes the number of non-zero elements and $k$ the number of copies, with a classical preprocessing cost of $\mathcal{O}(\mathrm{nnz}(x))$. As in}~\cite{kerenidis_et_al:LIPIcs.ITCS.2017.49}{, the QRAM is treated as a black-box primitive and no hardware architecture model is specified.}

{Several works move beyond the oracle abstraction and analyze the preparation procedure at the circuit level, focusing on reducing the cost of implementing the multiplexer circuit by employing techniques such as space--time trade-offs and exploiting structure in the data to reduce the cost of performing $U_{\mathsf{QRAM}}$}~\cite{PhysRevA.110.032439, PhysRevLett.129.230504}{. In general, these works use the so called \textit{circuit based QRAM}}~\cite{Jaques2025qramsurveycritique}{, where they treat the QRAM as a set of ancillary qubits that interact with the main register. While this is a valid approach, loading dense matrices inherently requires resources that scale exponentially with the number of qubits --- a cost that, in the circuit-based model, manifests as a trade-off between the number of ancillas and the circuit depth --- and these techniques tend to be most practical in the case of sparse matrices}~\cite{PhysRevLett.129.230504}.
{In}~\cite{casares2020}{, the author adopt the scheme of}~\cite{kerenidis_et_al:LIPIcs.ITCS.2017.49} {and describe it in terms of a Bucket Brigade QRAM circuit. The authors do not explicitly analyze the routing overhead, the overall time complexity remains $\mathcal{O}(\operatorname{polylog}(MN))$, and the work does not formally specify how to embed the matrix $A$ and its associated segment trees within the Bucket Brigade memory layout. In}~\cite{Zhang_2025}{, the authors study amplitude encoding of a vector $x \in \mathbb{R}^{N}$ using an explicit circuit-level model inspired by the Bucket Brigade architecture. The approach computes rotation angles classically and then feeds them into a circuit that treats the Bucket Brigade structure as a collection of ancillary qubits implementing the routing logic; the QRAM is therefore not accessed as a memory oracle but rather used as part of the preparation circuit itself. The state preparation time scales as $\mathcal{O}(\log_2 N)$, the work explicitly derives the routing procedure, and precision-dependent gate costs are analyzed. In}~\cite{PhysRevA.110.032439}{, the authors consider sparse quantum state preparation using a circuit QRAM, achieving a time complexity of $\mathcal{O}(s \log_2 N)$ for a vector with sparsity $s$ and using only two ancillary qubits. In}~\cite{Clader_2022}{, the authors analyze resource requirements for block-encoding of matrices using a circuit-level Bucket Brigade QRAM with $\mathcal{O}(\log_2 N)$ time complexity. In these last two works, the analyses focus on gate counts and circuit depth but do not account for routing costs or architectural constraints of a hardware QRAM model}.
{We observe that, since the BBQRAM architecture employs unitary operation, it can be formalized within the standard circuit model using the Clifford+T gate set, resulting in a circuit-based BBQRAM with logarithmic depth and a size linear in the number of ancillary qubits}~\cite{hann2021resilience, arunachalam2015robustness, doriguello2025practicalityquantumsievingalgorithms}.{ However, recent critiques suggest that such circuit-based implementations may be impractical for fault-tolerant applications. Specifically, a fault-tolerant QRAM requires $\Omega(2^n)$ total logical gates, of which $\Omega(\sqrt{2^n})$ must be non-Clifford, and necessitates active quantum error correction operating in parallel across $\Omega(2^n)$ logical qubits}~\cite{Jaques2025qramsurveycritique}.

{In contrast to the above approaches, our work treats the QRAM as a separate hardware device that stores data in the computational basis, with the quantum processing unit (QPU) interacting with it through a set of working qubits. We formulate the state preparation algorithm in terms of data stored within a hardware BBQRAM model, provide a formal description of how to embed an arbitrary matrix $A \in \mathbb{R}^{M \times N}$ into the memory, express the entire preparation routine through coherent retrieval operations, and derive a resource analysis that accounts for the number of QRAM accesses, the routing overhead, and the gate-level operations on the working qubits. This model enables polylogarithmic-time data loading and, in principle, the reuse of stored data across different quantum routines. While the physical realization of such a device and its interface with a generic quantum processor pose engineering challenges}~\cite{Jaques2025qramsurveycritique}{, several recent works demonstrate promising progress toward practical implementations}~\cite{Hann_2019, cesa2025, dalzell2025, Wang_2025, PRXQuantum.5.020312, PRXQuantum.2.030319, deriso2025}.

\section{Preliminaries}\label{sec:preliminaries}
We assume basic knowledge of quantum computing; for an introduction to the subject, we refer the reader to~\cite{nielsen2010quantum}. Throughout this manuscript, we adopt the notation
\begin{equation*}
    \ket{\psi}_{\mathrm{name}}^{\mathrm{size}},
\end{equation*}
where the subscript identifies the name of the quantum register $\ket{\psi}$, and the superscript indicates its size in qubits. The size and name are omitted when they are clear from context.

In Section~\ref{sec:basis_and_amplitude}, we recall the basis and amplitude encodings, which we use frequently in subsequent sections. Section~\ref{sec:cascade_rotations} introduces the concept of cascade of controlled rotations such to enable the decomposition of a single-qubit rotation with a parametric angle into a sequence of controlled rotations with fixed angles. Eventually, Section~\ref{sec:u2cr} describes the $U_{2CR}$ unitary, which plays a central role in the proposed state preparation algorithm. 

\subsection{Basis Encoding \& Amplitude Encoding}\label{sec:basis_and_amplitude}
With \emph{basis encoding} (also known as \emph{binary encoding}), we map a $t$-bit string to a computational basis state of a register of $t$ qubits. Formally, given a $t$-bit string $x = x_{t-1} \dots x_1 x_0$, where $x_i \in \{0,1\}$ for $ 0 \leq i < t$, basis encoding represents $x$ in a quantum register of $t$ qubits as $$\ket{x} = \ket{x_{t-1} \dots x_1 x_0},$$
where the leftmost qubit is the most significant. For example, the bit string $1011$ maps to the quantum state $\ket{1011}$. 

In \emph{amplitude encoding}, we embed classical data into the amplitudes of a quantum state. Given a vector $x = [x_0, \dots, x_{N-1}]$ of classical data, where $N$ can be padded with zeros to be a power of 2, we can encode this information in $n =  \log_2 N $ qubits as $$\ket{\psi} = \frac{1}{\lVert x \rVert_2}\sum_{j=0}^{N-1} x_j\ket{j}^n.$$
For example, consider the vector $x = [0.3, 0.4, 0.8]$ with $N=3$. Since $N$ is not a power of two, we pad the vector to $x=[0.3,0.4,0.8,0.0]$ so that $N=4$. Then, we normalize it to unit length by dividing it by $0.94$, resulting in $x = [0.32, 0.42, 0.85, 0.00]$, which we can now encode in the amplitudes of a $2$-qubit quantum register $\ket{\psi} = 0.32\ket{00} + 0.42\ket{01} + 0.85\ket{10}$. 

\subsection{Cascade of Controlled Rotations}\label{sec:cascade_rotations}

\begin{definition}[Cascade of Controlled Unitary Gates]
Consider a $t$-qubit register $a$ and a single target qubit $b$.
We define the cascade of controlled unitary gates, denoted by
$\mathrm{C}_{a}U_{\mapsto b} \in \mathbb{C}^{2^{t+1} \times 2^{t+1}}$,
with $U_{\mapsto b} \in \mathbb{C}^{2 \times 2}$, as

\begin{equation*}\label{eq:controlled_unitary}
\mathrm{C}_{a}U_{\mapsto b} = \prod_{i = 0}^{t-1} \mathrm{C}_{a_i}U_{\mapsto b},
\end{equation*}

where $\mathrm{C}_{a_i}$ denotes a single control qubit and the unitary $U_{\mapsto b}$ applies to the target qubit $b$ if the control qubit $a_i \in a$ is in the state $\ket{1}$.
For clarity, we omit tensor products with the identity operators $I_2$ acting on the other qubits $a_j \in a$, where $j \in [0, t)$ and $j \neq i$.
\end{definition}

In Lemma~\ref{lemma:controlled_rotations_equivalence}, we specialize Definition~\ref{eq:controlled_unitary} to the case of a \emph{cascade of controlled rotations}, showing that a single-qubit rotation with a parametric angle can be decomposed into a sequence of controlled rotations with fixed angles. 

\begin{lemma}[Cascade of Controlled Rotations Equivalence]\label{lemma:controlled_rotations_equivalence}
Let $a$ be a $t$-qubit register that encodes an angle $\theta \in (0, 2\pi]$ in fixed-point binary representation.
Let $b$ denote a target qubit, and let $R(\cdot) \in \mathbb{C}^{2 \times 2}$ denote a single-qubit rotation that satisfies the additive composition property $R(\phi_1 + \phi_2) = R(\phi_1) R(\phi_2)$. Then, 

\begin{equation*}
\mathrm{C}_{a}\mathrm{R}_{\mapsto b}(\theta) = \prod_{i = 0}^{t-1} \mathrm{C}_{a_i}\mathrm{R}_{\mapsto b}\left( 2^{\lfloor \log_2(\theta) \rfloor - i} \right) = \mathrm{R}_{\mapsto b}(\theta),
\end{equation*}
where $\mathrm{C}_{a_i}$ denotes a single control qubit and the unitary $R_{\mapsto b}$ applies to the target qubit $b$ if the control qubit $a_i \in a$ is in the state $\ket{1}$.
\end{lemma}

\begin{proof}
Let the binary expansion of $\theta$ with fixed precision $t$ be
\begin{equation*}
    \theta = \sum_{i = 0}^{t-1} a_i \, 2^{\lfloor \log_2(\theta) \rfloor - i},
\end{equation*}
where $a_i \in \{0,1\}$ denotes the $i$-th bit value of the fixed-point representation of $\theta$, and let
\begin{equation*}
    \mathrm{C}_{a}R_{\mapsto b}(\theta)
    = \prod_{i = 0}^{t-1} \mathrm{C}_{a_i} R_{\mapsto b}\!\left( 2^{\lfloor \log_2(\theta) \rfloor - i} \right),
\end{equation*}
where each control qubit $a_i$ triggers the rotation 
$R_{\mapsto b}\left( 2^{\lfloor \log_2(\theta)\rfloor - i} \right)$ on qubit $b$ if $a_i$ is in state $\ket{1}$. By making the control values $a_i$ explicit, we can rewrite the transformation as
\begin{equation*}
    \prod_{i = 0}^{t-1} R_{\mapsto b}\left( a_i \, 2^{\lfloor \log_2(\theta) \rfloor - i} \right).
\end{equation*}
\begin{figure*}[!t]
    \centering
    \begin{minipage}{0.65\textwidth}
        \centering
        \[
        \Qcircuit @C=0.8em @R=2em  {
            & \qw & \ctrl{6}  & \qw     &  \qw   & \qw    & \qw   & \qw  & \qw   & \qw   &  \ldots &  & \qw     & \qw \\
            &\qw & \qw  & \qw      & \ctrl{5}  & \qw    & \qw     & \qw  & \qw    & \qw   &  \ldots &  & \qw    & \qw \\
            &\qw & \qw  & \qw      & \qw   & \qw    & \ctrl{4}   & \qw  & \qw   & \qw   &  \ldots &  & \qw      & \qw \\
            &\qw & \qw  & \qw      & \qw   & \qw    & \qw    & \qw  & \ctrl{3}  & \qw   &  \ldots &  & \qw      & \qw &&& \rstick{=}\\
            & \smash[b]{\vdots}   &     &          &      &        &      &    &      &       &    \smash[b]{\vdots}     &  &         &  \\
            &\qw & \qw   & \qw      & \qw   & \qw    & \qw    & \qw  & \qw  & \qw     &\ldots &  & \ctrl{1}     & \qw 
                \inputgroupv{1}{6}{0.87em}{5.5em}{\ket{\theta}^t_a} \\
            \lstick{\ket{0}_b}&\qw & \gate{R_y  (2^2) } & \qw      & \gate{R_y (2^1) } & \qw    & \gate{R_y (2^0 )}  & \qw  & \gate{R_y(2^{-1})} & \qw   &  \ldots &  & \gate{R_y(2^{t-3})}& \qw 
        }
        \]
    \end{minipage}
    \begin{minipage}{0.30\textwidth}
        \centering
        \[
        \Qcircuit @C=1.5em @R=2.22em  {
            & \qw & \qw  & \qw \\
            & \qw & \qw  & \qw \\
            & \qw & \qw  & \qw \\
            & \qw & \qw  & \qw \\
            & & \smash[b]{\vdots}  & \\
            & \qw & \qw  & \qw \\
            & \qw & \gate{R_y(\theta)}  & \qw 
        }
        \]
    \end{minipage}
    \caption{The circuit on the left implements a cascade of controlled $R_y$ rotations that realize the single-qubit rotation $R_y(\theta)$ on the target qubit $\ket{0}_b$ shown on the right. The circuit encodes an angle $\theta$ in fixed-point representation within the quantum register $\ket{\theta}_a$ using basis encoding. It then applies a sequence of $t$ controlled $R_y$ rotations, each controlled by a qubit of $\ket{\theta}_a$ and associated with a fixed power of two.}
    \Description{} 
    \label{fig:cascade}
\end{figure*}
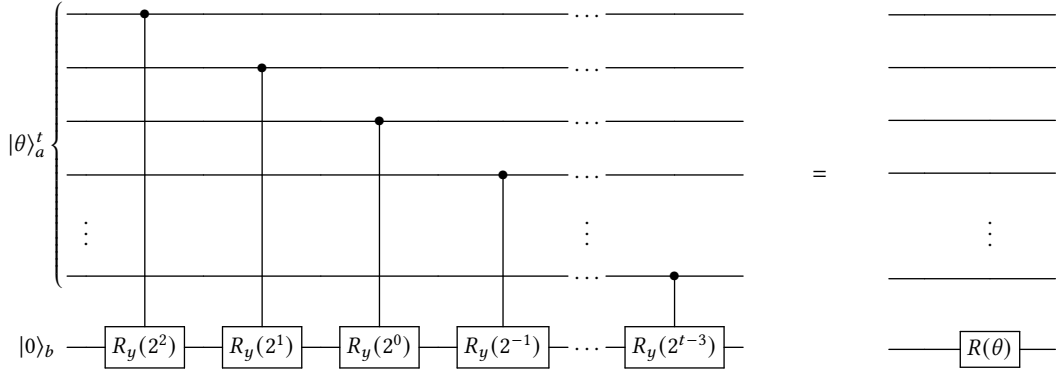

Then, the additive composition property of the rotation operator $R(\phi_1 + \phi_2) = R(\phi_1) R(\phi_2)$,
allows combining these rotations into a single operation:
\begin{equation*}
    \prod_{i = 0}^{t-1} R_{\mapsto b}\left(a_i \, 2^{\lfloor \log_2(\theta) \rfloor - i} \right)
    = R_{\mapsto b}\left( \sum_{i = 0}^{t-1} a_i \, 2^{\lfloor \log_2(\theta) \rfloor - i} \right)
    = R_{\mapsto b}(\theta).
\end{equation*}
Therefore, 
\begin{equation*}
    \mathrm{C}_{a}R_{\mapsto b}(\theta) = R_{\mapsto b}(\theta).
\end{equation*}
This result holds for any single-qubit rotation $R$ satisfying the additive composition property (e.g., $P$, $R_x$, $R_y$, and $R_z$).
\end{proof}

We highlight that, thanks to a \emph{cascade of controlled $R_y$ gates}, we can realize the transformation
\begin{equation*}
    \ket{\theta}_a^t\ket{0}_b \xmapsto{\textrm{Cascade }R_y} \ket{\theta}_a^t(\cos\frac{\theta}{2} \ket{0}_b + sin\frac{\theta}{2}\ket{1}_b),
\end{equation*}
where $\ket{\theta}_a^t$ denotes the basis encoding of the fixed-point representation of an angle $\theta \in (0, 2\pi]$, and $\ket{0}_b$ is the target qubit. Figure~\ref{fig:cascade} illustrates the circuit implementation of this cascade, which has depth $\mathcal{O}(t)$.

\subsection{From Basis to Amplitude}\label{sec:u2cr}
We employ a unitary operator, introduced in~\cite{chen_et_al:LIPIcs.ICALP.2023.38}, useful for the state preparation algorithm. This operator, denoted $U_{2CR}$, maps basis-encoded values to the amplitudes of an additional qubit, up to a normalization factor, through a sequence of reversible arithmetic operations followed by a cascade of controlled rotations. Formally,
\begin{equation}\label{eq:u2cr}
  U_{2CR}:\,\ket{a}\ket{b}\ket{0}=
    \begin{cases}
      \ket{a}\ket{b}\left( \frac{1}{\sqrt{2}} \ket{0} +\frac{1}{\sqrt{2}}\ket{1}\right) & \textrm {if } a = b = 0, \\
      \ket{a}\ket{b}\left( \sqrt{\frac{a}{a+b}} \ket{0} +\sqrt{\frac{b}{a+b}}\ket{1}\right) & \text{otherwise},
    \end{cases}
\end{equation}
where $a,b \in \{0,1\}^t$ represent two positive real values basis encoded with fixed precision $t$.

To implement $U_{2CR}$, we first check whether $a = b = 0$ by computing the logical OR over the $2t$ qubits. This operation produces a flag qubit. If the flag qubit equals $\ket{0}$, we apply a Hadamard gate on a target qubit, realizing the first case of Equation~\ref{eq:u2cr}.
If the flag qubit equals $\ket{1}$, the circuit computes the other transformation
$$
\ket{a}\ket{b}\ket{0}
\;\longmapsto\;
\ket{a}\ket{b}\!\left(
\sqrt{\frac{a}{a+b}}\ket{0} + \sqrt{\frac{b}{a+b}}\ket{1}
\right).
$$
Hereby, we list the reversible arithmetic subroutines needed for the implementation of $U_{2CR}$: 
\begin{itemize}
    \item $\ket{a}\ket{b}\ket{0} \xmapsto{\textrm{addition}} \ket{a}\ket{b}\ket{a+b}$,
    \item $\ket{b}\ket{a+b}\ket{0} \xmapsto{\textrm{division}} \ket{b}\ket{a+b}\ket{\frac{b}{a+b}}$,
    \item $\ket{\frac{b}{a+b}}\ket{0} \xmapsto{\textrm{square root}} \ket{\frac{b}{a+b}}\ket{\sqrt{\frac{b}{a+b}}}$,
    \item $\ket{\sqrt{\frac{b}{a+b}}}\ket{0} \xmapsto{\textrm{arcsine and left shift}} \ket{\sqrt{\frac{b}{a+b}}}\ket{2\cdot\arcsin (\sqrt{\frac{b}{a+b}})}$.
\end{itemize}

Eventually, we apply a cascade of controlled $R_y$ gates to map the angle $\ket{\theta} = \ket{2\cdot\arcsin (\sqrt{\frac{b}{a+b}})}$ to a target qubit (see Lemma~\ref{lemma:controlled_rotations_equivalence}). This procedure yields the state
\begin{equation*}
    \ket{2\cdot\arcsin (\sqrt{\frac{b}{a+b}})}\ket{0} \xmapsto{\textrm{Cascade }R_y} \ket{2\cdot\arcsin (\sqrt{\frac{b}{a+b}})}(\sqrt{\frac{a}{a+b}}\ket{0} + \sqrt{\frac{b}{a+b}}\ket{1}),
\end{equation*}
thereby completing the implementation of $U_{2CR}$. Finally, we discard the auxiliary qubits needed by these operations. 

{In general, given $t$-qubit precision for representing a value in basis encoding, the circuit depth of the cascade of controlled $R_y$ rotations is $\mathcal{O}(t)$. Focusing on asymptotic scaling, fast arithmetic algorithms for reversible addition, division, square root, and arcsine incur at most an $\tilde{O}(t)$ overhead in space and time}~\cite{bennett1989time}{; the reader can refer to}~\cite{10.1098/rsta.2023.0392} {for a detailed review of these implementations. Because this overhead depends strictly on the precision $t$, memory cells that contains the basis encoding of a value generally share a fixed word size that does not scale with the input size of the larger problem. Therefore, in this context, we can safely treat $t$ as a constant, meaning the overall cost to implement  $U_{2CR}$ reduces to $\tilde{O}(1)$.}

\section{Background}\label{sec:background}
In this section, we describe the main components we require for efficient quantum state preparation. Section~\ref{sec:QRAM} reviews the concept of QRAM, focusing on the Bucket Brigade architecture, and its role in enabling data retrieval in superposition. Section~\ref{sec:segment_tree} introduces the Segment Tree data structure, which we use to organize and preprocess classical data for quantum state preparation. 

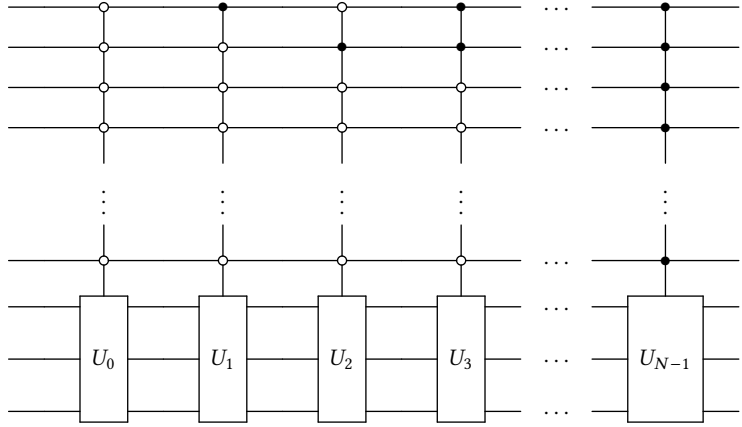
\begin{figure*}[t]
\[
\Qcircuit @C=1.5em @R=1.3em  {
    & \qw & \ctrlo{1}  & \qw      & \ctrl{1}   & \qw    & \ctrlo{1}   & \qw  & \ctrl{1}   & \qw   &  \ldots &  & \ctrl{1}      & \qw  \\
    &\qw & \ctrlo{1}  & \qw      & \ctrlo{1}  & \qw    & \ctrl{1}    & \qw  & \ctrl{1}   & \qw   &  \ldots &  & \ctrl{1}      & \qw \\
    &\qw & \ctrlo{1}  & \qw      & \ctrlo{1}  & \qw    & \ctrlo{1}   & \qw  & \ctrlo{1}  & \qw   &  \ldots &  & \ctrl{1}      & \qw \\
    &\qw & \ctrlo{1}  & \qw      & \ctrlo{1}  & \qw    & \ctrlo{1}   & \qw  & \ctrlo{1}  & \qw   &  \ldots &  & \ctrl{1}      & \qw \\
    &&&&&&&&&&&&& \\
    &    & \smash[b]{\vdots}     &          & \smash[b]{\vdots}     &        & \smash[b]{\vdots}      &      & \smash[b]{\vdots}     &       &         &  & \smash[b]{\vdots}        &  \\
    &&&&&&&&&&&&& \\
    &\qw & \ctrlo{-1}\qwx[1]  & \qw      & \ctrlo{-1}\qwx[1]  & \qw    & \ctrlo{-1}\qwx[1]   & \qw  & \ctrlo{-1}\qwx[1]  & \qw     &\ldots &  & \ctrl{-1}\qwx[1]      & \qw \\
    & \qw & \multigate{2}{U_0} & \qw & \multigate{2}{U_1} & \qw & \multigate{2}{U_2} & \qw & \multigate{2}{U_3} & \qw & \ldots & & \multigate{2}{U_{N-1}} & \qw \\
    & \qw & \ghost{U_0}        & \qw & \ghost{U_1}        & \qw & \ghost{U_2}        & \qw & \ghost{U_3}        & \qw & \ldots & & \ghost{U_{N-1}}        & \qw \\
    & \qw & \ghost{U_0}        & \qw & \ghost{U_1}        & \qw & \ghost{U_2}        & \qw & \ghost{U_3}        & \qw & \ldots & & \ghost{U_{N-1}}        & \qw \\
}
\]
\caption{A circuit multiplexer implementing $\sum_{i=0}^{N-1} \ket{i}\bra{i} U_i$ that prepares $N$ values in the form $\ket{x_i}^t$.}
\Description{} 
    \label{fig:multiplexer}
\end{figure*}

\subsection{Quantum Random Access Memory}\label{sec:QRAM}
A QRAM is a device designed to store data, which can be retrieved in 
superposition~\cite{giovannetti2008quantum, di2020fault, arunachalam2015robustness, allcock2023constant, hann2021resilience, PhysRevA.102.032608, Jaques2025qramsurveycritique} 
Specifically, a QRAM allows queries of the form:
\begin{equation}
    \mathsf{QRAM:}\ket{i}^{n}_{\mathrm{a}} \ket{b}^{t}_{\mathrm{d}} \mapsto  \ket{i}^{n}_{\mathrm{a}} \ket{b \oplus x_i}^{t}_{\mathrm{d}},
\label{eq:QRAM_access_transformation}\end{equation}
where $\ket{i}^{n}_{\mathrm{a}}$ for $i \in \mathbb{N}$ denotes an address register of size $n$, thus indexing $N=2^n$ values, and $\ket{b \oplus x_i}^{t}_{\mathrm{d}}$ denotes the value register where $b$ is any bit string and $x_i \in \{0,1\}^t$, where $t \in \mathbb{N}$, represents the value associated with the address $\ket{i}_{\mathrm{a}}$. The key feature of this type of memory model is its ability to retrieve a superposition of values when querying a superposition of addresses.
We observe that the linear transformation of the QRAM (i.e., $U_{\text{QRAM}}$) has the form of a block-diagonal unitary matrix, specifically:
\begin{align*}
    U_{\text{QRAM}} = \sum_{i=0}^{N-1} \ket{i}\bra{i}  U_i  = 
    \begin{bmatrix}
    U_0 &      &         & \\
        & U_1  &         &  \\
        &       & \ddots &  \\
        &       &        & U_{N-1}
\end{bmatrix},
\end{align*}

where $U_i\ket{0}^{t}=\ket{x_i}^{t}$. We can implement $U_{\mathsf{QRAM}}$ by means of a \textit{multiplexer} circuit~\cite{park2019circuit} (see Figure~\ref{fig:multiplexer}). However, this kind of circuit requires $\mathcal{O}(\log_2N)$ control qubits and its depth scales linearly with the number of values (i.e., $\mathcal{O}(N)$). This is evidently a non-viable solution for achieving a polylogarithmic complexity for state preparation.  

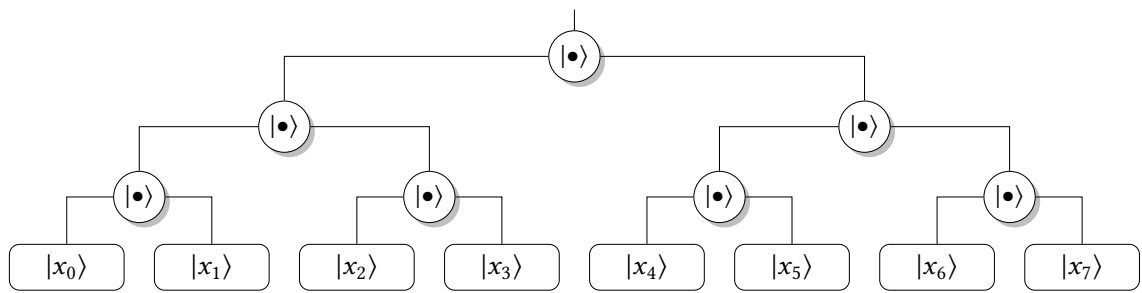
\begin{figure}[t]
  \centering
  \resizebox{\linewidth}{!}{%
    \begin{tikzpicture}[
        every label/.append style={font=\large},
        node/.style={
          draw, fill=white, circle, inner sep=0pt, font=\large,
          drop shadow, text width=2em, align=center
        },
        leaf/.style={
          shape=rectangle, rounded corners, draw, fill=white,
          inner sep=4pt, minimum width=5em, minimum height=2em,
          font=\Large, align=center
        }
      ]
      \def\dx{4} \def\dy{0.97} \def\dxii{2} \def\leafsep{1}

       \node[node] (root) at (0,0) {$\ket{\bullet}$};
        \coordinate (above) at ($(root)+(0,\dy/1.5)$);
        \draw (above) -- (root);

      \node[node] (L)  at ($(root)+(-\dx,-\dy)$) {$\ket{\bullet}$};
      \node[node] (R)  at ($(root)+(\dx,-\dy)$)  {$\ket{\bullet}$};
      \node[node] (LL) at ($(L)+(-\dxii,-\dy)$) {$\ket{\bullet}$};
      \node[node] (LR) at ($(L)+(\dxii,-\dy)$)  {$\ket{\bullet}$};
      \node[node] (RL) at ($(R)+(-\dxii,-\dy)$) {$\ket{\bullet}$};
      \node[node] (RR) at ($(R)+(\dxii,-\dy)$)  {$\ket{\bullet}$};

      \node[leaf] (LLL) at ($(LL)+(-\leafsep,-\dy)$) {$\ket{x_0}$};
      \node[leaf] (LLR) at ($(LL)+(\leafsep,-\dy)$)  {$\ket{x_1}$};
      \node[leaf] (LRL) at ($(LR)+(-\leafsep,-\dy)$) {$\ket{x_2}$};
      \node[leaf] (LRR) at ($(LR)+(\leafsep,-\dy)$)  {$\ket{x_3}$};
      \node[leaf] (RLL) at ($(RL)+(-\leafsep,-\dy)$) {$\ket{x_4}$};
      \node[leaf] (RLR) at ($(RL)+(\leafsep,-\dy)$)  {$\ket{x_5}$};
      \node[leaf] (RRL) at ($(RR)+(-\leafsep,-\dy)$) {$\ket{x_6}$};
      \node[leaf] (RRR) at ($(RR)+(\leafsep,-\dy)$)  {$\ket{x_7}$};

      \draw (root) -- ++(-\dx,0) coordinate(auxL) -- (L);
      \draw (root) -- ++(\dx,0)  coordinate(auxR) -- (R);
      \draw (L)    -- ++(-\dxii,0) coordinate(auxLL) -- (LL);
      \draw (L)    -- ++(\dxii,0)  coordinate(auxLR) -- (LR);
      \draw (R)    -- ++(-\dxii,0) coordinate(auxRL) -- (RL);
      \draw (R)    -- ++(\dxii,0)  coordinate(auxRR) -- (RR);
      \draw (LL)   -- ++(-\leafsep,0) coordinate(auxLLL) -- (LLL);
      \draw (LL)   -- ++(\leafsep,0)  coordinate(auxLLR) -- (LLR);
      \draw (LR)   -- ++(-\leafsep,0) coordinate(auxLRL) -- (LRL);
      \draw (LR)   -- ++(\leafsep,0)  coordinate(auxLRR) -- (LRR);
      \draw (RL)   -- ++(-\leafsep,0) coordinate(auxRLL) -- (RLL);
      \draw (RL)   -- ++(\leafsep,0)  coordinate(auxRLR) -- (RLR);
      \draw (RR)   -- ++(-\leafsep,0) coordinate(auxRRL) -- (RRL);
      \draw (RR)   -- ++(\leafsep,0)  coordinate(auxRRR) -- (RRR);
    \end{tikzpicture}
  }
  \caption{Bucket Brigade QRAM architecture (BBQRAM). Internal nodes act as switches that route address qubits, while leaf nodes represent memory cells storing the values $\ket{x_i}$.}
  \label{fig:bbqram}
  \Description{} 
\end{figure}

In this work, we adopt the hardware model of the \emph{Bucket Brigade QRAM} (BBQRAM)~\cite{giovannetti2008quantum}. Unlike other solutions~\cite{phalak2023quantum, xu2023systems}, the BBQRAM is arranged as a binary tree (see Figure~\ref{fig:bbqram}), thus exhibiting logarithmic depth in the number of indexed values, and also showing to be more robust to errors~\cite{hann2021resilience}. In particular, in this architecture:
\begin{itemize}
    \item the leaves represent \textit{memory cells};
    \item the internal nodes serve as \textit{switches}, routing the access towards memory cells.  
\end{itemize}
Specifically, given $N$ memory cells (assume that $N$ is a power of $2$), the depth of the binary tree is $n = \log_2 N$, and the total number of switches is $N-1 = \mathcal{O}(N)$.

\subsubsection{The Routing algorithm.} 
Let us provide some intuition about the routing algorithm used when querying a value associated with an address register $\ket{i}^{n}_{\mathrm{a}} = \ket{i_{n-1} \dots i_1 i_0}_{\mathrm{a}}$, where the leftmost qubit is the most significant.
The routing algorithm involves the use of switches implemented through qutrits whose state can be $\ket{0}$, $\ket{1}$, or the wait state $\ket{\bullet}$. Initially, each switch is in the $\ket{\bullet}$ state. Then, the algorithm iteratively routes each qubit of address $\ket{i}_{\mathrm{a}}$ through the tree by starting with the most significant qubit. It is worth noting that accessing a memory cell or a superposition of memory cells requires the activation of only the switches along the routing path in the tree.
When routing a qubit $\ket{i_j}_{\mathrm{a}}$, the algorithm works as follows upon encountering a switch in the tree. If the switch is in the state:
\begin{itemize}
    \item $\ket{0}$: route the qubit to left;
    \item $\ket{1}$: route the qubit to right;
    \item $\ket{\bullet}$: switch the state of the qutrit to the state of $\ket{i_j}_{\mathrm{a}}$.
\end{itemize}

\begin{figure*}[t]
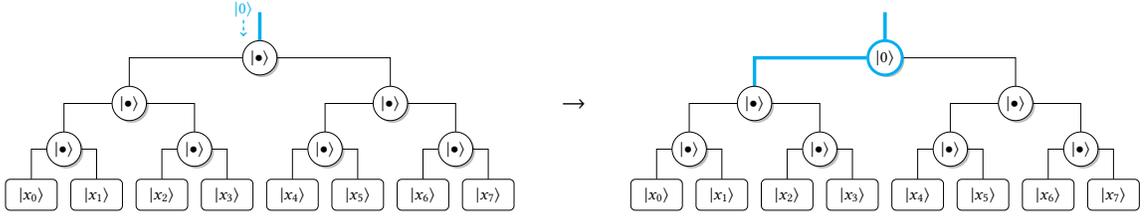
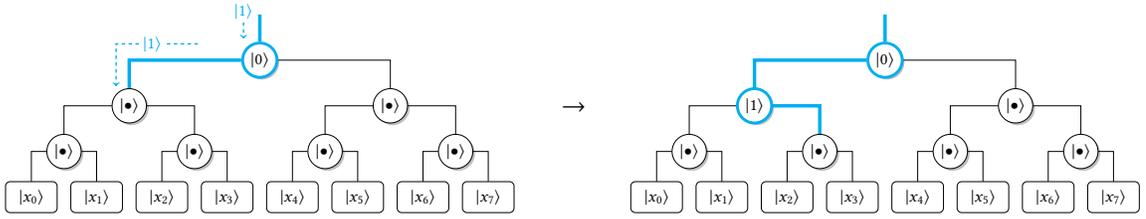
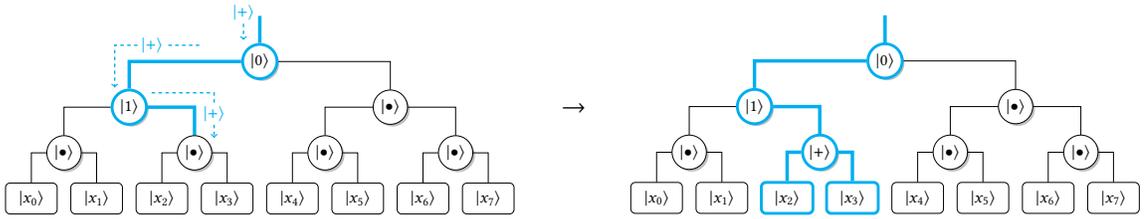

  \centering

  \begin{subfigure}[t]{\textwidth}
    \centering
    \begin{minipage}{0.45\linewidth}
      \resizebox{\linewidth}{!}{%

      }
    \end{minipage}
    \caption{The left figure shows the first address qubit $\ket{0}$ reaching the root of the BBQRAM. The right figure illustrates the BBQRAM state after setting the first encountered switch $\ket{\bullet}$ to $\ket{0}$, resulting in routing subsequent address qubits to the left subtree.}\Description{}
    \label{fig:routing_1}
  \end{subfigure}

  \vspace{2em}

  \begin{subfigure}[t]{\textwidth}
    \centering
    \begin{minipage}{0.45\linewidth}
      \resizebox{\linewidth}{!}{%
        %
      }
    \end{minipage}
    \caption{The figure shows how the second address qubit $\ket{1}$ propagates from the root to the first level of BBQRAM. The right figure illustrates the BBQRAM state after setting the second encountered switch $\ket{\bullet}$ to $\ket{1}$, therefore routing subsequent address qubits to the right subtree.}\Description{}
    \label{fig:routing_2}
  \end{subfigure}

  \vspace{2em}

  \begin{subfigure}[t]{\textwidth}
    \centering
    \begin{minipage}{0.45\linewidth}
      \resizebox{\linewidth}{!}{%
        %
      }
    \end{minipage}
    \caption{The left figure shows how the $\ket{+}$ state propagates through the three levels of BBQRAM. The right figure illustrates the BBQRAM state after the address qubit $\ket{+}$ reaches the next $\ket{\bullet}$ switch, displaying equal superposition access to the memory cells $\ket{x_2}$ and $\ket{x_3}$.}\Description{}
    \label{fig:routing_3}
  \end{subfigure}
  \caption{QRAM routing algorithm for the address register $\ket{01+}$. (Figure~\ref{fig:routing_1}) The algorithm routes the first qubit $\ket{0}$ through the tree until it reaches the first switch in the $\ket{\bullet}$ state, which corresponds to the root. Then, the switch updates its state $\ket{\bullet}$ to match the address incoming qubit $\ket{0}$. As a result, any subsequent qubits that encounter this switch is routed to the left subtree. The same logic applies also for the next address qubit $\ket{1}$ (Figure~\ref{fig:routing_2}), but routing subsequent qubits to the right subtree. The last qubit register is in the superposition state $\ket{+}$ (Figure~\ref{fig:routing_3}), and linearity ensures that the switch routes the access coherently to both left and right memory cells, enabling equal superposition access to $\ket{x_2}$ and $\ket{x_3}$.}
  \Description{} 
  \label{fig:qram_routing_algorithm}
\end{figure*}

We observe that if the state of the switch is $\ket{\gamma} = \alpha\ket{0} + \beta\ket{1}$ where $\alpha, \beta \neq 0$, the switch acts as a quantum switch due to the linearity of quantum mechanics, thus routing the subsequent qubits to the left and right subtrees in superposition.
In Figure~\ref{fig:qram_routing_algorithm}, we illustrate the routing algorithm for querying the values associated with the address $\ket{01+}= \frac{1}{\sqrt{2}}(\ket{010} +\ket{011})$, which retrieves the memory location of the addresses, $\ket{010}$ and $\ket{011}$ in equal superposition.

Since each qubit of the register $\ket{i}_{\mathrm{a}}$ is routed starting from the root to the next $\ket{\bullet}$, the total number of levels of the binary tree traversed to access a memory cell is $$\sum_{k=0}^{n-1} (n-k) = \frac{n(n-1)}{2} = \mathcal{O}(n^2) = \mathcal{O}(\log_2^2N).$$ 
\begin{figure*}[t]
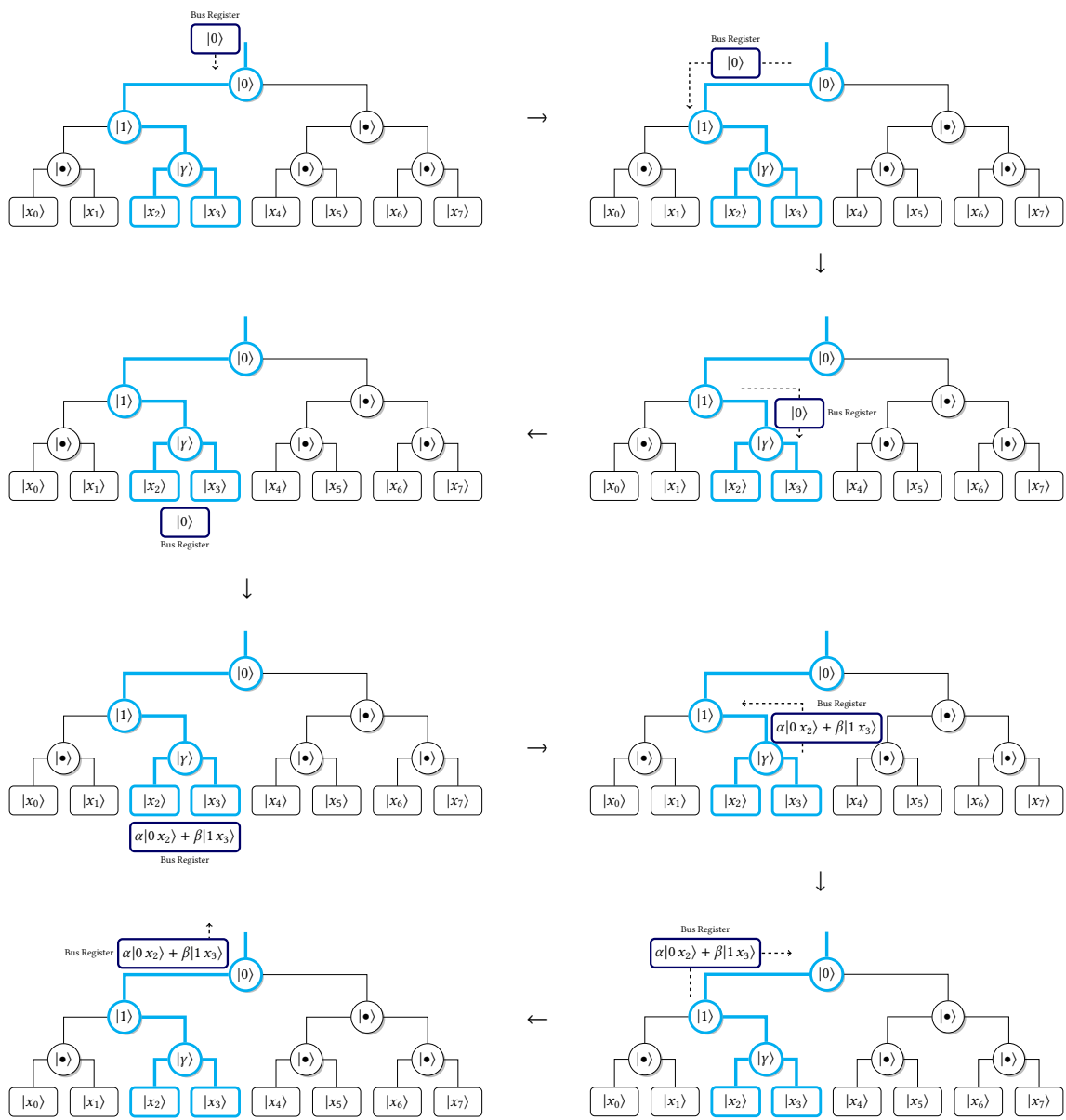

  \centering

  \begin{subfigure}[t]{\textwidth}
    \centering
    \begin{minipage}{0.45\linewidth}
      \resizebox{\linewidth}{!}{%

      }
    \end{minipage}
  \end{subfigure}
  \caption{
  Bus register traversal in BBQRAM. After establishing the access path for the address register $\ket{01\gamma}$, where $\ket{\gamma} = \alpha\ket{0} + \beta\ket{1}$ with $\alpha, \beta \neq 0$, the bus register follows this path to reach the target memory cell. Because the memory data is encoded in the computational basis, we use CNOT gates to copy the memory contents into the bus register. This process does not violate the no-cloning theorem, as the CNOT gate copy only classical information encoded in the computational basis. Finally, the bus register returns to the root node along the same path.}
  \Description{} 
  \label{fig:qram_bus_register_algorithm}
\end{figure*}

In this approach, each address qubit waits until the previous one reaches a switch in state $\ket{\bullet}$, resulting in $\mathcal{O}(\log_2^2N)$ routing steps. However, in~\cite{hann2021resilience, xu2023systems}, the authors note that an address qubit need not reach its destination before routing the subsequent one. Therefore, they propose a routing optimization called \emph{pipelining}, in which the $(j+1)$-th qubit begins routing as soon as the $j$-th moves one level down, thus avoiding waits in a pipeline-fashion. This optimization reduces the routing complexity to $\mathcal{O}(\log_2N)$ steps for both a single and superposed memory accesses. We adopt the pipelined routing throughout this work. 

\begin{algorithm2e}[t]
\caption{Retrieval from Bucket Brigade QRAM with Pipelined Routing}
\label{alg:bbqram_retrieval}
\SetKwInput{KwPre}{Precondition}
\KwIn{Address register $\ket{i}^{n}_{\mathrm{a}} = \ket{i_{n-1} \dots i_1 i_0}_{\mathrm{a}}$}
\KwOut{Data load into the working register, and all switches reset to the wait state $\ket{\bullet}$}
\KwPre{All switches in the BBQRAM are in the wait state $\ket{\bullet}$}
\BlankLine
\ForEach{qubit in the address register}{
    Route the qubit in a pipeline-fashion, updating switches as needed \tcp*[r]{$\mathcal{O}(\log_2 N)$}
}
Send the bus register along the established access path to the target memory cell(s) \tcp*[r]{$\mathcal{O}(\log_2 N)$}
Copy the data from the accessed memory cell(s) into the bus register using CNOT gates \tcp*[r]{$\mathcal{O}(1)$}
Route the bus register back to the root along the same path \tcp*[r]{$\mathcal{O}(\log_2 N)$}
Copy the bus register's contents into the working register using CNOT gates \tcp*[r]{$\mathcal{O}(1)$}
Uncompute the access path by reversing the routing operations, restoring all switches to $\ket{\bullet}$ \tcp*[r]{$\mathcal{O}(\log_2 N)$}
\end{algorithm2e}

Once we establish the access path to the target memory cell, we retrieve the data from BBQRAM to a working register as follows. First, we send the bus register down the access path, traversing $\mathcal{O}(\log_2 N)$ levels. Because the memory cells store data in the computational basis (i.e., $\ket{0}$ and $\ket{1}$), we use Controlled-NOT (CNOT) gates to copy the data to the bus register in constant time, which does not violate the no-cloning theorem~\cite{wootters1982single}. Next, we route the bus register back to the root in $\mathcal{O}(\log_2 N)$ steps (see Figure~\ref{fig:qram_bus_register_algorithm}) and again use CNOT gates to transfer the bus register's contents to a working register. Finally, we disentangle all activated switches by uncomputing the routing operations, thereby restoring the internal nodes (switches) to the wait state $\ket{\bullet}$. Algorithm~\ref{alg:bbqram_retrieval} summarizes the retrieval procedure for accessing either a single memory cell or a superposition of cells in the BBQRAM architecture. Lemma~\ref{lemma:retrieval_cost} establishes the total retrieval cost.

\begin{lemma}[Bucket Brigade QRAM Retrieval cost]\label{lemma:retrieval_cost}
Let $N = 2^n$ denote the number of indexed values in a BBQRAM. With pipelined routing, the total retrieval cost for either a single memory cell or a superposition of memory cells is $\mathcal{O}(\log_2 N)$ in time. This cost includes access path setup, bus traversal to memory cells, copying data into the bus register, returning the bus to the root, transferring its contents to the working register, and uncomputation of the access path.
\end{lemma}

\subsubsection{BBQRAM Memory cells initialization}

{Up to this point, we assume that the BBQRAM memory cells already store the data to be queried. We now describe the initialization step that writes such data into the memory cells. Since the data to store are classical bitstrings encoded in the computational basis, the framework does not require an inherently quantum memory. Two scenarios arise depending on the hardware implementation of the memory architecture.}

\textit{Scenario 1: Classical memory cells.} {In this scenario, the BBQRAM directly accesses a classical memory storage, where each memory cell is a classical register holding the bitstring $x_i$. Since the BBQRAM interfaces directly with classical storage, no quantum initialization step is necessary: a classical computer writes the data to the memory cells through standard classical operations. The routing structure of the BBQRAM addresses the classical cells, and the leaf nodes do not require quantum registers. We note that the transfer of the data from the classical memory cells would require a classically controlled NOT operation on the bus. More details on this passage can be found in}~\cite{Hann_2019}. 
\begin{figure}[t]
  \centering
  \resizebox{\linewidth}{!}{%
\begin{tikzpicture}[
    every label/.append style={font=\large},
    node/.style={
      draw=black, fill=white, circle, inner sep=0pt, font=\large,
      drop shadow, text width=2em, align=center, thick
    },
    q_leaf/.style={
      shape=rectangle, rounded corners, draw=cyan, fill=cyan!10,
      inner sep=4pt, minimum width=5em, minimum height=2em,
      font=\Large, align=center, drop shadow, thick
    },
    c_leaf/.style={
      shape=rectangle, draw=red!80!black, fill=red!10,
      inner sep=4pt, minimum width=5em, minimum height=2em,
      font=\large, align=center, drop shadow, thick
    },
    tree link/.style={draw=black, thick},
    memory link/.style={draw=gray, solid, ->, >=Stealth, thick}
  ]

  \def\dx{4}
  \def\dy{1.1}
  \def\dxii{2}
  \def\leafsep{1}
  \def\classsep{1.6}

  \node[node] (root) at (0,0) {$\ket{\bullet}$};
  \coordinate (above) at ($(root)+(0,1)$);
  \draw[tree link] (above) -- (root);

  \node[node] (L)  at ($(root)+(-\dx,-\dy)$) {$\ket{\bullet}$};
  \node[node] (R)  at ($(root)+(\dx,-\dy)$)  {$\ket{\bullet}$};

  \node[node] (LL) at ($(L)+(-\dxii,-\dy)$) {$\ket{\bullet}$};
  \node[node] (LR) at ($(L)+(\dxii,-\dy)$)  {$\ket{\bullet}$};
  \node[node] (RL) at ($(R)+(-\dxii,-\dy)$) {$\ket{\bullet}$};
  \node[node] (RR) at ($(R)+(\dxii,-\dy)$)  {$\ket{\bullet}$};

  \node[q_leaf] (LLL) at ($(LL)+(-\leafsep,-\dy)$) {$\ket{000}$};
  \node[q_leaf] (LLR) at ($(LL)+(\leafsep,-\dy)$)  {$\ket{000}$};
  \node[q_leaf] (LRL) at ($(LR)+(-\leafsep,-\dy)$) {$\ket{000}$};
  \node[q_leaf] (LRR) at ($(LR)+(\leafsep,-\dy)$)  {$\ket{000}$};
  \node[q_leaf] (RLL) at ($(RL)+(-\leafsep,-\dy)$) {$\ket{000}$};
  \node[q_leaf] (RLR) at ($(RL)+(\leafsep,-\dy)$)  {$\ket{000}$};
  \node[q_leaf] (RRL) at ($(RR)+(-\leafsep,-\dy)$) {$\ket{000}$};
  \node[q_leaf] (RRR) at ($(RR)+(\leafsep,-\dy)$)  {$\ket{000}$};

  \foreach \parent/\child in {root/L, root/R, L/LL, L/LR, R/RL, R/RR, LL/LLL, LL/LLR, LR/LRL, LR/LRR, RL/RLL, RL/RLR, RR/RRL, RR/RRR}
      \draw[tree link] (\parent) -| (\child);

  \node[c_leaf, below=\classsep of LLL] (LLLC) {101};
  \node[c_leaf, below=\classsep of LLR] (LLRC) {011};
  \node[c_leaf, below=\classsep of LRL] (LRLC) {110};
  \node[c_leaf, below=\classsep of LRR] (LRRC) {000};
  \node[c_leaf, below=\classsep of RLL] (RLLC) {111};
  \node[c_leaf, below=\classsep of RLR] (RLRC) {010};
  \node[c_leaf, below=\classsep of RRL] (RRLC) {100};
  \node[c_leaf, below=\classsep of RRR] (RRRC) {001};

  \foreach \i in {LLL,LLR,LRL,LRR,RLL,RLR,RRL,RRR}
      \draw[memory link] (\i C) -- (\i) node[pos=0.6, right=1pt] {$cNOT$};

  \begin{scope}[on background layer]
      \coordinate (qrealm_top) at ($(above)+(0,0.6cm)$);
      \node[
        fill=cyan!5,
        rounded corners,
        inner xsep=10pt,
        inner ysep=10pt,
        fit=(qrealm_top) (LLL) (RRR)
      ] (qrealm) {};

      \coordinate (crealm_top) at ($(LLLC.north)+(0,0.2cm)$);
      \coordinate (crealm_bottom) at ($(RRRC.south)+(0,-0.9cm)$);
      \node[
        fill=red!5,
        rounded corners,
        inner xsep=10pt,
        fit=(crealm_top) (LLLC) (RRRC) (crealm_bottom)
      ] (crealm) {};

      \draw[thick, dashed, black!60]
        ($(qrealm.south west)!0.5!(crealm.north west)$) --
        ($(qrealm.south east)!0.5!(crealm.north east)$);
  \end{scope}

  \node[anchor=north, color=black] at ($(qrealm.north)+(0,-0.28cm)$)
    {Bucket Brigade QRAM};

  \node[anchor=center, color=black] at ($(crealm.south)!0.5!(crealm.center)$)
    {Classical Storage};

\end{tikzpicture}
  }
    \caption{{BBQRAM initialization scheme. The upper part of the figure shows the BBQRAM architecture, with bucket-brigade routing nodes arranged as a binary tree and quantum memory cells located at the leaves. Each memory cell initially stores the state $\ket{000}$ and, after initialization, stores the corresponding computational-basis state $\ket{x_i}$. The lower part of the figure shows the classical storage, which contains the bit strings $x_i \in \{0,1\}^3$ to be loaded into the BBQRAM. The vertical arrows between the two layers represent the bitwise writing operations that transfer each classical string to the corresponding quantum memory cell through classically controlled $NOT$ gates, (i.e., $cNOT$ gates).}}
  \label{fig:bbqram_initialization}
\end{figure}

\textit{Scenario 2: Quantum memory cells.} {If the hardware employs quantum registers at the leaf nodes, classically controlled $NOT$ ($cNOT$) gates transfer the classical data to the quantum memory cells. Consider a BBQRAM with $N$ memory cells, where each cell starts in state $\ket{0}^t$, with $t \in \mathbb{N}$. For $i \in \{0,\dots,N-1\}$, let $x_i \in \{0,1\}^t$ denote the binary string to map to the $i$-th memory cell. The initialization procedure maps each classical string $x_i$ to the computational-basis state $\ket{x_i}$ in the corresponding memory cell. Specifically, for every bit position $j \in [0,t)$, we apply an $X$ gate to the $j$-th qubit of memory cell $i$ whenever the $j$-th bit of $x_i$ equals $1$. The initialization of a single memory cell therefore requires $t$ $cNOT$ operations, one per bit of $x_i$, producing the desired basis state $\ket{x_i}$. Figure}~\ref{fig:bbqram_initialization} {illustrates this initialization scheme: classical storage provides the bitstrings, and the BBQRAM encodes them in the corresponding memory cells as computational-basis states. Since the $cNOT$ operations act on distinct qubits and distinct memory cells, they are mutually independent and can therefore run in parallel. As a result, this initialization strategy has time cost $\mathcal{O}(1)$, while the total number of $cNOT$ operations scales linearly in $t$.

We observe that quantum registers at the memory cells may suffer from decoherence, potentially requiring periodic re-initialization from the classical source. However, this does not affect the asymptotic cost of the framework, since re-initialization via $cNOT$ gates runs in $\mathcal{O}(1)$ time by exploiting the parallelism across independent memory cells.}

\subsection{Segment Tree data structure}\label{sec:segment_tree}
The segment tree is a classical data structure widely used in computer science for efficiently performing queries and updates over intervals of a vector~\cite{de2008computational}. A segment tree is a binary tree where each node represents a segment of a vector. The leaves store individual values, and internal nodes store the result of an associative operation (i.e., sum, minimum, or maximum) over their children. For a vector of $N$ values, the segment tree has the following properties:
\begin{itemize}
\item \textbf{Efficient queries:} It supports querying the result of an operation (e.g., sum, minimum, maximum) over any segment in $\mathcal{O}(\log N)$ time.
\item \textbf{Efficient updates:} Updating a single value and recalculating all affected nodes requires $\mathcal{O}(\log N)$ time.
\item \textbf{Space complexity:} The segment tree uses $\mathcal{O}(N)$ space.
\item \textbf{Construction cost:} Building the segment tree takes $\mathcal{O}(N)$ time.
\end{itemize}

In quantum state preparation, we can use the Segment Tree data structure to precompute amplitudes. This classical preprocessing of amplitudes in a segment tree is fundamental to achieving efficient state preparation. Specifically, we construct a segment tree such that each leaf stores the squared norm of a value to be encoded as an amplitude in the quantum state, and each internal node stores the sum of its children’s values. In Definition~\ref{def:segment_tree}, we formalize the Segment Tree data structure for encoding a matrix that can serve as input for a target quantum algorithm.

\begin{definition}[Segment Tree of Squared Norms]\label{def:segment_tree}
Let $A\in \mathbb{R}^{M\times N}$ be a matrix, where $M\,,N$ are powers of two, and adopt a row-major order indexing of the values of $A$ such that $a_z = a_{i,j}$ with $z = i \cdot N + j$, where $i \in \left[0, M\right)$ and $j \in \left[0, N\right)$.
We construct a segment tree of squared norms $T$ of $A$ such that an internal node $T_{h,p}$ at height $h \in  \left[0, \log_2K \right]$, where $K = MN$, and in position $p \in \left[0, 2^h \right)$ is defined as:
\begin{equation*}
T_{h,p} \;=\; T_{\left[  p\cdot\frac{K}{2^h} \;:\;  (p+1)\cdot \frac{K}{2^h} -1 \right]} \;=\; \sum_{z \;=\; p \cdot \frac{K}{2^h}}^{(p+1) \cdot \frac{K}{2^h} -1} \lvert a_z \rvert^2.
\end{equation*}
If $h = \log_2K$, then the node is a leaf and is denoted by the tuple $(\mathrm{s}(a_z), T_{h,p})$, where $T_{h,p}$ coincides with the value $\lvert a_p\rvert^2$. This data structure has the following properties:
\begin{enumerate}
    \item The total number of nodes is $2K-1$ and the depth is $\log_2K$;
    \item The time to query or update an entry of $A$ is $\mathcal{O}(\log_2 K)$.
\end{enumerate}
\end{definition}

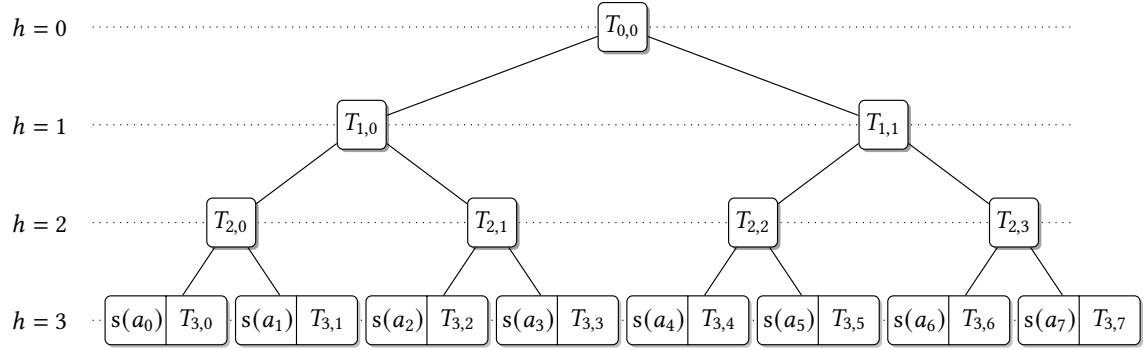
\begin{figure*}[t]
    \centering
    \resizebox{\linewidth}{!}{%
      \begin{tikzpicture}[
          treenode/.style={
            rectangle,
            rounded corners=2pt,            
            draw,
            fill=white,
            drop shadow={shadow xshift=1pt,shadow yshift=-1pt,opacity=0.5},
            minimum width=6mm,
            minimum height=6mm,
            inner sep=2pt,
            align=center,
          },
          leaf/.style={
            shape=rectangle split,
            rectangle split parts=2,
            rectangle split draw splits,
            rectangle split horizontal,
            fill={white!30},
            rounded corners=2pt,            
            draw,
            drop shadow={shadow xshift=1pt,shadow yshift=-1pt,opacity=0.5},
            minimum width=6mm,
            minimum height=6mm,
            inner sep=2pt,
            text width=6mm,
            align=center,
          },
          levelzero/.style={treenode}, 
          levelone/.style={treenode}, 
          leveltwo/.style={treenode}, 
          levelthree/.style={treenode},
          hlabel/.style={
            draw=none,
            fill=none,
            anchor=east,
          },
          level distance=10mm,
          level 1/.style={sibling distance=64mm},
          level 2/.style={sibling distance=32mm},
          level 3/.style={sibling distance=16mm},
        ]

        \begin{scope}[xshift=5mm]
          \node[treenode, levelzero] {$T_{0,0}$}
            child {
              node[treenode, levelone] {$T_{1,0}$}
              child {
                node[treenode, leveltwo] {$T_{2,0}$}
                child {
                  node[ leaf] {$\mathrm{s}(a_{0})$\nodepart{two}$T_{3,0}$}
                }
                child {
                  node[leaf] {$\mathrm{s}(a_{1})$\nodepart{two}$T_{3,1}$}
                }
              }
              child {
                node[treenode, leveltwo] {$T_{2,1}$}
                child {
                  node[leaf] {$\mathrm{s}(a_{2})$\nodepart{two}$T_{3,2}$}
                }
                child {
                  node[leaf] {$\mathrm{s}(a_{3})$\nodepart{two}$T_{3,3}$}
                }
              }
            }
            child {
              node[treenode, levelone] {$T_{1,1}$}
              child {
                node[treenode, leveltwo] {$T_{2,2}$}
                child {
                  node[leaf] {$\mathrm{s}(a_{4})$\nodepart{two}$T_{3,4}$}
                }
                child {
                  node[leaf] {$\mathrm{s}(a_{5})$\nodepart{two}$T_{3,5}$}
                }
              }
              child {
                node[treenode, leveltwo] {$T_{2,3}$}
                child {
                  node[leaf] {$\mathrm{s}(a_{6})$\nodepart{two}$T_{3,6}$}
                }
                child {
                  node[leaf] {$\mathrm{s}(a_{7})$\nodepart{two}$T_{3,7}$}
                }
              }
            };
        \end{scope}
      \end{tikzpicture}%
    }
    \caption{Segment tree of squared norms $T$ for a matrix $A \in \mathbb{R}^{2 \times 4}$. The segment tree has depth $\log_2(4 \cdot 2) = 3$ and contains $15$ nodes. Each leaf stores the squared norm $|a_{z}|^2 = T_{3,z}$ of a value of $A$ along with its sign $\mathrm{s}(a_z)$.}
    \Description{} 
    \label{fig:segment_tree}
  \end{figure*}

Figure~\ref{fig:segment_tree} depicts the segment tree of squared norms $T$ of the matrix
\begin{equation*}
A\in \mathbb{R}^{2\times4} = 
\begin{bmatrix}
a_{0,0} & a_{0,1} & a_{0,2} & a_{0,3} \\[6pt]
a_{1,0} & a_{1,1} & a_{1,2} & a_{1,3} 
\end{bmatrix}
\quad \xrightarrow{\text{row-major order indexing}} \quad
\begin{bmatrix}
a_{0} & a_{1} & a_{2} & a_{3} \\[6pt]
a_{4} & a_{5} & a_{6} & a_{7}
\end{bmatrix}.
\end{equation*}
We recall that the inverse mapping from an entry $a_z$ back to its matrix coordinates $a_{i,j}$ is immediate by observing that $i =  \left\lfloor \frac{z}{N} \right\rfloor$ and $j = z\ \mathrm{mod}\ N$.
Such segment tree $T$ has depth $k = \log_2 (4\cdot2) = 3$ and $15$ nodes. Furthermore, we observe that the square root of $T_{0,0}$ coincides with the Frobenius norm squared of the matrix $A$, denoted by ${\|A\|^2_F}$.

\section{Efficient state preparation using BBQRAM and Segment Tree} \label{sec:efficient_state_preparation}
In this section, we describe how to combine the BBQRAM from Section~\ref{sec:QRAM} with the Segment Tree of squared norms from Section~\ref{sec:segment_tree} to enable efficient state preparation of a matrix $A \in \mathbb{R}^{M \times N}$. Section~\ref{sec:memory_layout} illustrates the mapping of a segment tree of squared norms into the memory cells of a BBQRAM. Eventually, Section~\ref{sec:state_prep_alg} presents the algorithm that achieves state preparation in $\mathcal{O}(\log_2^2(MN))$ time.

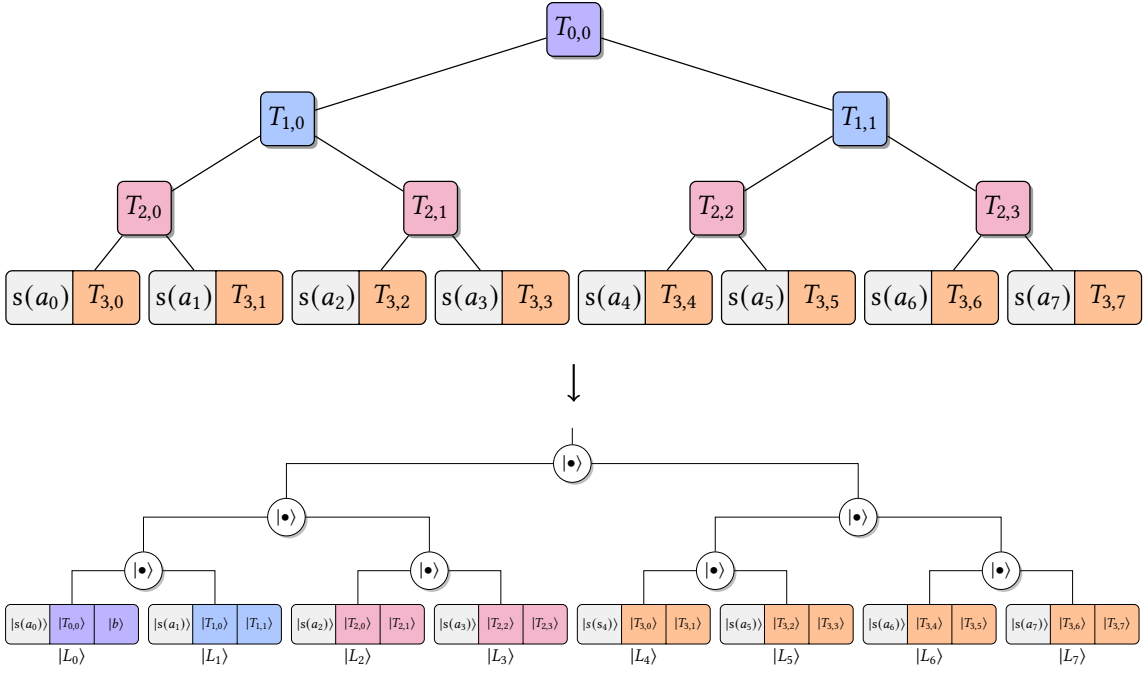
\begin{figure*}[t]
  \centering
  \begin{subfigure}[t]{\textwidth}
    \centering
    \resizebox{\linewidth}{!}{%
      \begin{tikzpicture}[
          treenode/.style={
            rectangle,
            rounded corners=2pt,            
            draw,
            fill=white,
            drop shadow={shadow xshift=1pt,shadow yshift=-1pt,opacity=0.5},
            minimum width=6mm,
            minimum height=6mm,
            inner sep=2pt,
            align=center,
          },
          leaf/.style={
            shape=rectangle split,
            rectangle split parts=2,
            rectangle split draw splits,
            rectangle split horizontal,
            rectangle split part fill={neutral-left,PastelIBMOrange},
            rounded corners=2pt,            
            draw,
            minimum width=6mm,
            minimum height=6mm,
            inner sep=2pt,
            text width=6mm,
            align=center,
          },
          levelzero/.style={treenode, fill=PastelIBMPurple},
          levelone/.style={treenode, fill=PastelIBMBlue},
          leveltwo/.style={treenode, fill=PastelIBMPink},
          levelthree/.style={treenode},
          hlabel/.style={
            draw=none,
            fill=none,
            anchor=east,
          },
          level distance=10mm,
          level 1/.style={sibling distance=64mm},
          level 2/.style={sibling distance=32mm},
          level 3/.style={sibling distance=16mm},
        ]

        \foreach \lvl in {0,1,2,3}{
          \draw[dotted] (-60mm,-\lvl*10mm) -- ( 60mm,-\lvl*10mm);
          \node[hlabel] at (-62mm,-\lvl*10mm) {$h=\lvl$};
        }

        \begin{scope}[xshift=5mm]
          \node[treenode, levelzero] {$T_{0,0}$}
            child {
              node[treenode, levelone] {$T_{1,0}$}
              child {
                node[treenode, leveltwo] {$T_{2,0}$}
                child {
                  node[ leaf] {$\mathrm{s}(a_{0})$\nodepart{two}$T_{3,0}$}
                }
                child {
                  node[leaf] {$\mathrm{s}(a_{1})$\nodepart{two}$T_{3,1}$}
                }
              }
              child {
                node[treenode, leveltwo] {$T_{2,1}$}
                child {
                  node[leaf] {$\mathrm{s}(a_{2})$\nodepart{two}$T_{3,2}$}
                }
                child {
                  node[leaf] {$\mathrm{s}(a_{3})$\nodepart{two}$T_{3,3}$}
                }
              }
            }
            child {
              node[treenode, levelone] {$T_{1,1}$}
              child {
                node[treenode, leveltwo] {$T_{2,2}$}
                child {
                  node[leaf] {$\mathrm{s}(a_{4})$\nodepart{two}$T_{3,4}$}
                }
                child {
                  node[leaf] {$\mathrm{s}(a_{5})$\nodepart{two}$T_{3,5}$}
                }
              }
              child {
                node[treenode, leveltwo] {$T_{2,3}$}
                child {
                  node[leaf] {$\mathrm{s}(a_{6})$\nodepart{two}$T_{3,6}$}
                }
                child {
                  node[leaf] {$\mathrm{s}(a_{7})$\nodepart{two}$T_{3,7}$}
                }
              }
            };
        \end{scope}
      \end{tikzpicture}%
    }
    \label{fig:segment_tree_colored}
  \end{subfigure}
  
  \begin{center}
    \begin{tikzpicture}
      \draw[->, thick, color=gray] (0,0.5) -- (0,0) node[below]{}; 
    \end{tikzpicture}
  \end{center}

  \begin{subfigure}[t]{\textwidth}
    \centering
    \resizebox{\linewidth}{!}{%

      \begin{tikzpicture}[
          every label/.append style={font=\huge},
          node/.style={
            draw, fill=white, circle, inner sep=0pt, font=\huge,
            drop shadow, text width=3em, align=center
          },
          levelzeroq/.style={rectangle split part fill={neutral-left,PastelIBMPurple,PastelIBMPurple}},
          leveloneq/.style={rectangle split part fill={neutral-left,PastelIBMBlue,PastelIBMBlue}},
          leveltwoq/.style={rectangle split part fill={neutral-left,PastelIBMPink,PastelIBMPink}},
          levelthreeq/.style={rectangle split part fill={neutral-left,PastelIBMOrange,PastelIBMOrange}},
          leaf/.style={
            shape=rectangle split, rectangle split parts=3,
            rectangle split draw splits, rectangle split horizontal,
            rounded corners, draw, inner sep=4pt,
            minimum width=5em, minimum height=3.5em,
            font=\Large, text width=3em, align=center
          }
        ]
        \def\dx{8} \def\dy{1.5} \def\dxii{4} \def\leafsep{2}

        \node[node] (root) at (0,0) {$\ket{\bullet}$};
        \coordinate (above) at ($(root)+(0,\dy/1.5)$);
        \draw (above) -- (root);

        \node[node] (L)  at ($(root)+(-\dx,-\dy)$) {$\ket{\bullet}$};
        \node[node] (R)  at ($(root)+(\dx,-\dy)$)  {$\ket{\bullet}$};

        \node[node] (LL) at ($(L)+(-\dxii,-\dy)$) {$\ket{\bullet}$};
        \node[node] (LR) at ($(L)+(\dxii,-\dy)$)  {$\ket{\bullet}$};
        \node[node] (RL) at ($(R)+(-\dxii,-\dy)$) {$\ket{\bullet}$};
        \node[node] (RR) at ($(R)+(\dxii,-\dy)$)  {$\ket{\bullet}$};

        \node[leaf, levelzeroq, label=below:{$\ket{L_0}$}] (LLL) at ($(LL)+(-\leafsep,-\dy)$)
          {$\ket{\mathrm{s}(a_{0})}$\nodepart{two}$\ket{T_{0,0}}$\nodepart{three}$\ket{b}$};
        \node[leaf, leveloneq,  label=below:{$\ket{L_1}$}] (LLR) at ($(LL)+(\leafsep,-\dy)$)
          {$\ket{\mathrm{s}(a_{1})}$\nodepart{two}$\ket{T_{1,0}}$\nodepart{three}$\ket{T_{1,1}}$};

        \node[leaf, leveltwoq,  label=below:{$\ket{L_2}$}] (LRL) at ($(LR)+(-\leafsep,-\dy)$)
          {$\ket{\mathrm{s}(a_{2})}$\nodepart{two}$\ket{T_{2,0}}$\nodepart{three}$\ket{T_{2,1}}$};
        \node[leaf, leveltwoq,  label=below:{$\ket{L_3}$}] (LRR) at ($(LR)+(\leafsep,-\dy)$)
          {$\ket{\mathrm{s}(a_{3})}$\nodepart{two}$\ket{T_{2,2}}$\nodepart{three}$\ket{T_{2,3}}$};

        \node[leaf, levelthreeq,label=below:{$\ket{L_4}$}] (RLL) at ($(RL)+(-\leafsep,-\dy)$)
          {$\ket{s(\mathrm{s}_{4})}$\nodepart{two}$\ket{T_{3,0}}$\nodepart{three}$\ket{T_{3,1}}$};
        \node[leaf, levelthreeq,label=below:{$\ket{L_5}$}] (RLR) at ($(RL)+(\leafsep,-\dy)$)
          {$\ket{\mathrm{s}(a_{5})}$\nodepart{two}$\ket{T_{3,2}}$\nodepart{three}$\ket{T_{3,3}}$};

        \node[leaf, levelthreeq,label=below:{$\ket{L_6}$}] (RRL) at ($(RR)+(-\leafsep,-\dy)$)
          {$\ket{\mathrm{s}(a_{6})}$\nodepart{two}$\ket{T_{3,4}}$\nodepart{three}$\ket{T_{3,5}}$};
        \node[leaf, levelthreeq,label=below:{$\ket{L_7}$}] (RRR) at ($(RR)+(\leafsep,-\dy)$)
          {$\ket{\mathrm{s}(a_{7})}$\nodepart{two}$\ket{T_{3,6}}$\nodepart{three}$\ket{T_{3,7}}$};

        \draw (root) -- ++(-\dx,0) coordinate(auxL) -- (L);
        \draw (root) -- ++(\dx,0)  coordinate(auxR) -- (R);
        \draw (L)  -- ++(-\dxii,0) coordinate(auxLL) -- (LL);
        \draw (L)  -- ++(\dxii,0)  coordinate(auxLR) -- (LR);
        \draw (R)  -- ++(-\dxii,0) coordinate(auxRL) -- (RL);
        \draw (R)  -- ++(\dxii,0)  coordinate(auxRR) -- (RR);
        \draw (LL) -- ++(-\leafsep,0) coordinate(auxLLL) -- (LLL);
        \draw (LL) -- ++(\leafsep,0)  coordinate(auxLLR) -- (LLR);
        \draw (LR) -- ++(-\leafsep,0) coordinate(auxLRL) -- (LRL);
        \draw (LR) -- ++(\leafsep,0)  coordinate(auxLRR) -- (LRR);
        \draw (RL) -- ++(-\leafsep,0) coordinate(auxRLL) -- (RLL);
        \draw (RL) -- ++(\leafsep,0)  coordinate(auxRLR) -- (RLR);
        \draw (RR) -- ++(-\leafsep,0) coordinate(auxRRL) -- (RRL);
        \draw (RR) -- ++(\leafsep,0)  coordinate(auxRRR) -- (RRR);

        \def\braceYshift{-3.2em}
        \def\braceAmp{8pt}
        \def\braceSep{12pt}
        
        \draw[decorate, decoration={brace, mirror, amplitude=\braceAmp}, very thick]
          ([yshift=\braceYshift]LLL.south west) -- ([yshift=\braceYshift]LLL.south east)
          node[midway, below=\braceSep, font=\fontsize{16}{18}\selectfont] {$h = 0$};
        
        \draw[decorate, decoration={brace, mirror, amplitude=\braceAmp}, very thick]
          ([yshift=\braceYshift]LLR.south west) -- ([yshift=\braceYshift]LLR.south east)
          node[midway, below=\braceSep, font=\fontsize{16}{18}\selectfont] {$h = 1$};
        
        \draw[decorate, decoration={brace, mirror, amplitude=\braceAmp}, very thick]
          ([yshift=\braceYshift]LRL.south west) -- ([yshift=\braceYshift]LRR.south east)
          node[midway, below=\braceSep, font=\fontsize{16}{18}\selectfont] {$h = 2$};
        
        \draw[decorate, decoration={brace, mirror, amplitude=\braceAmp}, very thick]
          ([yshift=\braceYshift]RLL.south west) -- ([yshift=\braceYshift]RRR.south east)
          node[midway, below=\braceSep, font=\fontsize{16}{18}\selectfont] {$h = 3$};
      \end{tikzpicture}
    }
    \label{fig:bbqram_mapping_colored}
  \end{subfigure}
  \caption{Illustration of the memory layout of a segment tree in a BBQRAM. The top figure shows the segment tree $T$ constructed from a matrix $A \in \mathbb{R}^{2 \times 4}$. The bottom figure visualizes how the nodes of the segment tree $T$ map to the memory cells of the BBQRAM. The figures use matching colors to highlight the correspondence between each specific node in the segment tree and its assigned location in the memory cells $\{\ket{L_z}\}_{z=0}^{K-1}$.}
  \Description{} 
  \label{fig:segment_tree_mapping_to_bbqram}
\end{figure*}

\subsection{Memory Layout of Segment Tree in BBQRAM}\label{sec:memory_layout}
Efficient quantum state preparation requires a precise memory layout that maps the nodes of the segment tree of squared norms onto the memory cells of the BBQRAM. {Since the state of the address register determines which superposition of memory cells the BBQRAM retrieves, the allocation of these nodes must allow the address register to be prepared efficiently at each step of the state preparation algorithm. Specifically, our goal is therefore to propose a memory layout that enables efficient preparation of the address register, so that each query retrieves, in superposition, all sibling nodes at a given level of the tree.}
Proposition~\ref{prop:mapping} formalizes the content of each memory cell, and Figure~\ref{fig:segment_tree_mapping_to_bbqram} illustrates how the segment tree nodes map to the BBQRAM memory cells. 

\begin{proposition}[Memory Layout of Segment Tree in Bucket Brigade QRAM]\label{prop:mapping}
Let $T$ be the segment tree of squared norms from a matrix $A\in \mathbb{R}^{M\times N}$, where $M\,,N$ are powers of two as defined in Definition~\ref{def:segment_tree}, with $MN=K$ leaves. We map $T$ into a BBQRAM with $K$ memory cells $\{\ket{L_z}\}_{z=0}^{K-1}$, where each $\ket{L_z}$ is a quantum register of $1+2t$ qubits such that
\begin{equation}\label{eq:mapping}
\ket{L_z}
=
\begin{cases}
\ket{ \mathrm{s}(a_{0})}^{1}\,\ket{T_{0,0}}^{t}\,\ket{b}^{t}
& \textrm{if } z = 0, \\[1ex]
\ket{\mathrm{s}(a_{z})}^{1}\,
           \ket{T_{l(z),\,2 d(z)}}^{t}\,
           \ket{T_{l(z),\,2 d(z)+1}}^{t}
& \textrm{otherwise,}
\end{cases}
\end{equation}

where:
\begin{itemize}
\item $\ket{\mathrm{s}(a_z)}$ is a single-qubit register encoding the sign of $a_z \in A $ using row-major indexing
\begin{equation}
    \mathrm{s}(a_z) =
        \begin{cases}
            0 & \text{if } a_z \geq 0, \\
            1 & \text{if } a_z < 0.
        \end{cases}
\end{equation}

\item $\ket{T_{l(z),\,2 d(z)}}^{t}$ and $\ket{T_{l(z),\,2 d(z)+1}}^{ t}$ are $t$-qubit registers each that basis encode the values of two sibling nodes of $\textrm{ }T$, with $l(z) = \lfloor \log_2 z \rfloor + 1$ and $d(z) = z - 2^{\lfloor \log_2 z \rfloor}$ with $t$-bit precision.
\item $\ket{b}$ is an arbitrary string used to standardize the length of the bitstrings stored in the memory cells.
\end{itemize}
Each memory cell $\ket{L_z}$ contains both the sign and the data of two sibling nodes from the segment tree $T$, except for $\ket{L_0}$, which instead stores the root $T_{0,0}$ along with an arbitrary string $b$ and the sign $\mathrm{s}(a_0)$. Under this construction, the BBQRAM has $K$ memory cells, $K-1$ switches, and depth $\log_2 K$.
\end{proposition}

\begin{proof}[Proposition~\ref{prop:mapping} Correctness]
Let $T$ be a segment tree constructed from a matrix $A \in \mathbb{R}^{M \times N}$, where $K = MN$ is the total number of leaves and $k = \log_2 K$ is the tree depth. We aim to map each node of $T$ to a corresponding memory cell in a BBQRAM, ensuring that both structures share the same depth.

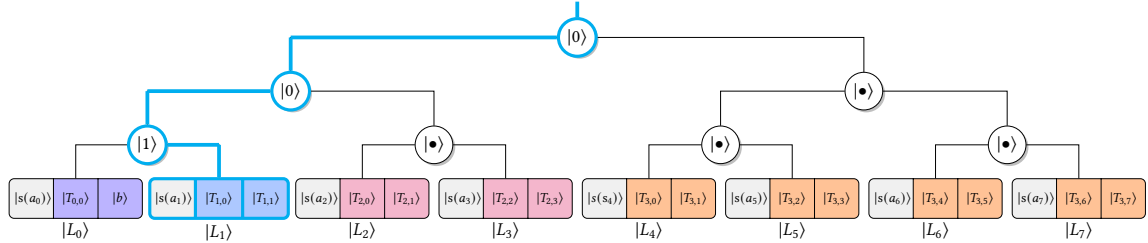
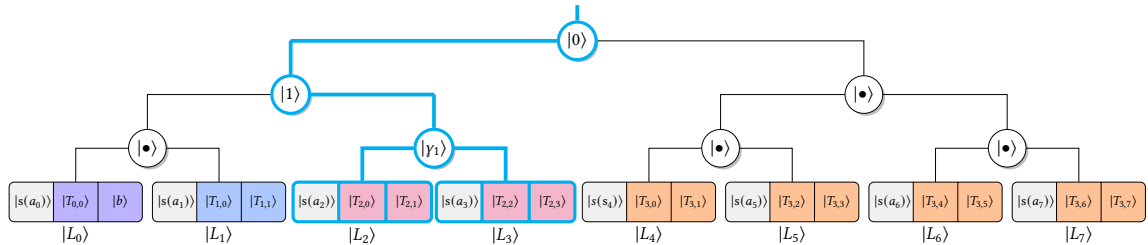
\begin{figure*}[b]
\centering
    \tikzset{
    line width/.initial=1.5pt,
    every label/.append style={font=\huge},
    node/.style={
      draw, fill=white, circle, inner sep=0pt, font=\huge,
      drop shadow, text width=3em, align=center
    },
    levelzeroq/.style={rectangle split part fill={neutral-left,PastelIBMPurple,PastelIBMPurple}},
    leveloneq/.style={rectangle split part fill={neutral-left,PastelIBMBlue,PastelIBMBlue}},
    leveltwoq/.style={rectangle split part fill={neutral-left,PastelIBMPink,PastelIBMPink}},
    levelthreeq/.style={rectangle split part fill={neutral-left,PastelIBMOrange,PastelIBMOrange}},
    leaf/.style={
      shape=rectangle split, rectangle split parts=3,
      rectangle split draw splits, rectangle split horizontal,
      rounded corners, draw, inner sep=4pt,
      minimum width=5em, minimum height=3.5em,
      font=\Large, text width=3em, align=center
    },
    hi/.style={draw=cyan,line width=3pt}  
  }
  \newcommand*\treedim{
    \def\dx{8}\def\dy{1.5}\def\dxii{4}\def\leafsep{2}}

  \begin{subfigure}[t]{\textwidth}
    \centering
    \resizebox{\linewidth}{!}{\treedim
      \begin{tikzpicture}
        \node[node, draw=cyan, line width=2.5pt] (root) at (0,0)             {$\ket{0}$};
        \coordinate  (above) at ($(root)+(0,\dy/1.5)$);
        \draw[hi] (above) -- (root);

        \node[node, draw=cyan, line width=2.5pt] (L)  at ($(root)+(-\dx,-\dy)$) {$\ket{0}$};
        \node[node] (R)  at ($(root)+(\dx,-\dy)$)  {$\ket{\bullet}$};

        \node[node, draw=cyan, line width=2.5pt] (LL) at ($(L)+(-\dxii,-\dy)$) {$\ket{1}$};
        \node[node] (LR) at ($(L)+(\dxii,-\dy)$)  {$\ket{\bullet}$};
        \node[node] (RL) at ($(R)+(-\dxii,-\dy)$) {$\ket{\bullet}$};
        \node[node] (RR) at ($(R)+(\dxii,-\dy)$)  {$\ket{\bullet}$};

        \node[leaf, levelzeroq, label=below:{$\ket{L_0}$}] (LLL) at
             ($(LL)+(-\leafsep,-\dy)$)
             {$\ket{\mathrm{s}(a_{0})}$\nodepart{two}$\ket{T_{0,0}}$\nodepart{three}$\ket{b}$};
        \node[leaf, draw=cyan, line width=2.5pt, leveloneq,  label=below:{$\ket{L_1}$}] (LLR) at
             ($(LL)+(\leafsep,-\dy)$)
             {$\ket{\mathrm{s}(a_{1})}$\nodepart{two}$\ket{T_{1,0}}$\nodepart{three}$\ket{T_{1,1}}$};

        \node[leaf, leveltwoq,  label=below:{$\ket{L_2}$}] (LRL) at
             ($(LR)+(-\leafsep,-\dy)$)
             {$\ket{\mathrm{s}(a_{2})}$\nodepart{two}$\ket{T_{2,0}}$\nodepart{three}$\ket{T_{2,1}}$};
        \node[leaf, leveltwoq,  label=below:{$\ket{L_3}$}] (LRR) at
             ($(LR)+(\leafsep,-\dy)$)
             {$\ket{\mathrm{s}(a_{3})}$\nodepart{two}$\ket{T_{2,2}}$\nodepart{three}$\ket{T_{2,3}}$};

        \node[leaf, levelthreeq,label=below:{$\ket{L_4}$}] (RLL) at
             ($(RL)+(-\leafsep,-\dy)$)
             {$\ket{s(\mathrm{s}_{4})}$\nodepart{two}$\ket{T_{3,0}}$\nodepart{three}$\ket{T_{3,1}}$};
        \node[leaf, levelthreeq,label=below:{$\ket{L_5}$}] (RLR) at
             ($(RL)+(\leafsep,-\dy)$)
             {$\ket{\mathrm{s}(a_{5})}$\nodepart{two}$\ket{T_{3,2}}$\nodepart{three}$\ket{T_{3,3}}$};

        \node[leaf, levelthreeq,label=below:{$\ket{L_6}$}] (RRL) at
             ($(RR)+(-\leafsep,-\dy)$)
             {$\ket{\mathrm{s}(a_{6})}$\nodepart{two}$\ket{T_{3,4}}$\nodepart{three}$\ket{T_{3,5}}$};
        \node[leaf, levelthreeq,label=below:{$\ket{L_7}$}] (RRR) at
             ($(RR)+(\leafsep,-\dy)$)
             {$\ket{\mathrm{s}(a_{7})}$\nodepart{two}$\ket{T_{3,6}}$\nodepart{three}$\ket{T_{3,7}}$};

        \draw (root)--++(-\dx,0) coordinate(auxL)--(L);
        \draw (root)--++(\dx,0)  coordinate(auxR)--(R);
        \draw (L)  --++(-\dxii,0) coordinate(auxLL)--(LL);
        \draw (L)  --++(\dxii,0)  coordinate(auxLR)--(LR);
        \draw (R)  --++(-\dxii,0) coordinate(auxRL)--(RL);
        \draw (R)  --++(\dxii,0)  coordinate(auxRR)--(RR);
        \draw (LL) --++(-\leafsep,0) coordinate(auxLLL)--(LLL);
        \draw (LL) --++(\leafsep,0)  coordinate(auxLLR)--(LLR);
        \draw (LR) --++(-\leafsep,0) coordinate(auxLRL)--(LRL);
        \draw (LR) --++(\leafsep,0)  coordinate(auxLRR)--(LRR);
        \draw (RL) --++(-\leafsep,0) coordinate(auxRLL)--(RLL);
        \draw (RL) --++(\leafsep,0)  coordinate(auxRLR)--(RLR);
        \draw (RR) --++(-\leafsep,0) coordinate(auxRRL)--(RRL);
        \draw (RR) --++(\leafsep,0)  coordinate(auxRRR)--(RRR);

        \draw[hi] (root)--(auxL);
        \draw[hi] (auxL)--(L);
        \draw[hi] (L)--(auxLL);
        \draw[hi] (auxLL)--(LL);
        \draw[hi] (LL)--(auxLLR);
        \draw[hi] (auxLLR)--(LLR);
      \end{tikzpicture}}
    \caption{Address register $\ket{0\,0\,1}$ traces a single path that accesses the memory cell $\ket{L_1}$ containing the sibling pair $\ket{T_{1,0}}\,\ket{T_{1,1}}$ at height $h=1$ in $T$.}
    \label{subfig:access_1}
  \end{subfigure}
  
  \vspace{1em}

  \begin{subfigure}[t]{\textwidth}
    \centering
    \resizebox{\linewidth}{!}{\treedim
      \begin{tikzpicture}
        \node[node, draw=cyan, line width=2.5pt] (root) at (0,0)             {$\ket{0}$};
        \coordinate  (above) at ($(root)+(0,\dy/1.5)$);
        \draw[hi] (above) -- (root);

        \node[node, draw=cyan, line width=2.5pt] (L)  at ($(root)+(-\dx,-\dy)$) {$\ket{1}$};
        \node[node] (R)  at ($(root)+(\dx,-\dy)$)  {$\ket{\bullet}$};

        \node[node] (LL) at ($(L)+(-\dxii,-\dy)$) {$\ket{\bullet}$};
        \node[node, draw=cyan, line width=2.5pt] (LR) at ($(L)+(\dxii,-\dy)$)  {$\ket{\gamma_1}$};
        \node[node] (RL) at ($(R)+(-\dxii,-\dy)$) {$\ket{\bullet}$};
        \node[node] (RR) at ($(R)+(\dxii,-\dy)$)  {$\ket{\bullet}$};

        \node[leaf, levelzeroq, label=below:{$\ket{L_0}$}] (LLL) at
             ($(LL)+(-\leafsep,-\dy)$)
             {$\ket{\mathrm{s}(a_{0})}$\nodepart{two}$\ket{T_{0,0}}$\nodepart{three}$\ket{b}$};
        \node[leaf, leveloneq,  label=below:{$\ket{L_1}$}] (LLR) at
             ($(LL)+(\leafsep,-\dy)$)
             {$\ket{\mathrm{s}(a_{1})}$\nodepart{two}$\ket{T_{1,0}}$\nodepart{three}$\ket{T_{1,1}}$};

        \node[leaf, draw=cyan, line width=2.5pt, leveltwoq,  label=below:{$\ket{L_2}$}] (LRL) at
             ($(LR)+(-\leafsep,-\dy)$)
             {$\ket{\mathrm{s}(a_{2})}$\nodepart{two}$\ket{T_{2,0}}$\nodepart{three}$\ket{T_{2,1}}$};
        \node[leaf, draw=cyan, line width=2.5pt, leveltwoq,  label=below:{$\ket{L_3}$}] (LRR) at
             ($(LR)+(\leafsep,-\dy)$)
             {$\ket{\mathrm{s}(a_{3})}$\nodepart{two}$\ket{T_{2,2}}$\nodepart{three}$\ket{T_{2,3}}$};

        \node[leaf, levelthreeq,label=below:{$\ket{L_4}$}] (RLL) at
             ($(RL)+(-\leafsep,-\dy)$)
             {$\ket{s(\mathrm{s}_{4})}$\nodepart{two}$\ket{T_{3,0}}$\nodepart{three}$\ket{T_{3,1}}$};
        \node[leaf, levelthreeq,label=below:{$\ket{L_5}$}] (RLR) at
             ($(RL)+(\leafsep,-\dy)$)
             {$\ket{\mathrm{s}(a_{5})}$\nodepart{two}$\ket{T_{3,2}}$\nodepart{three}$\ket{T_{3,3}}$};

        \node[leaf, levelthreeq,label=below:{$\ket{L_6}$}] (RRL) at
             ($(RR)+(-\leafsep,-\dy)$)
             {$\ket{\mathrm{s}(a_{6})}$\nodepart{two}$\ket{T_{3,4}}$\nodepart{three}$\ket{T_{3,5}}$};
        \node[leaf, levelthreeq,label=below:{$\ket{L_7}$}] (RRR) at
             ($(RR)+(\leafsep,-\dy)$)
             {$\ket{\mathrm{s}(a_{7})}$\nodepart{two}$\ket{T_{3,6}}$\nodepart{three}$\ket{T_{3,7}}$};

        \draw (root)--++(-\dx,0) coordinate(auxL)--(L);
        \draw (root)--++(\dx,0)  coordinate(auxR)--(R);
        \draw (L)  --++(-\dxii,0) coordinate(auxLL)--(LL);
        \draw (L)  --++(\dxii,0)  coordinate(auxLR)--(LR);
        \draw (R)  --++(-\dxii,0) coordinate(auxRL)--(RL);
        \draw (R)  --++(\dxii,0)  coordinate(auxRR)--(RR);
        \draw (LL) --++(-\leafsep,0) coordinate(auxLLL)--(LLL);
        \draw (LL) --++(\leafsep,0)  coordinate(auxLLR)--(LLR);
        \draw (LR) --++(-\leafsep,0) coordinate(auxLRL)--(LRL);
        \draw (LR) --++(\leafsep,0)  coordinate(auxLRR)--(LRR);
        \draw (RL) --++(-\leafsep,0) coordinate(auxRLL)--(RLL);
        \draw (RL) --++(\leafsep,0)  coordinate(auxRLR)--(RLR);
        \draw (RR) --++(-\leafsep,0) coordinate(auxRRL)--(RRL);
        \draw (RR) --++(\leafsep,0)  coordinate(auxRRR)--(RRR);

        \draw[hi] (root)--(auxL);
        \draw[hi] (auxL)--(L);
        \draw[hi] (L)--(auxLR);
        \draw[hi] (auxLR)--(LR);
        \draw[hi] (LR)--(auxLRL) -- (LRL);
        \draw[hi] (LR)--(auxLRR) -- (LRR);
      \end{tikzpicture}}
    \caption{Address registers $\ket{0\,1\,\gamma_1}$, where $\ket{\gamma_1} = \alpha_1 \ket{0} + \beta_1\ket{1}$ with $\alpha,\beta \neq 0$, branches once, accessing in superposition the memory cells $\ket{L_2}$ and $\ket{L_3}$ containing the two sibling pairs $\ket{T_{2,0}}\ket{T_{2,1}}$ and $\ket{T_{2,2}}\ket{T_{2,3}}$, respectively, that reside at height $h=2$ in $T$.}
    \label{subfig:access_2}
    \end{subfigure}
        \caption{Two representative access paths allowed by the retrieval primitives of Corollary~\ref{cor:structured_access}. In Figure~\ref{subfig:access_1}, the address register establishes the access path that targets the sibling pair at level $h=1$ in $T$, as specified by Equation~\eqref{eq:case_ret} for $h=1$. Figure~\ref{subfig:access_2} shows the path accessing in superposition the sibling pairs at level $h=2$ in $T$.}\Description{}
  \label{fig:access_patterns_1}
\end{figure*}
\begin{figure*}[t]
  \centering
  \begin{subfigure}[t]{\textwidth}
    \centering
    \tikzset{
    every label/.append style={font=\huge,  cm={0.8,-0.3,0,1,(0,0)}},
    node/.style={
      draw, fill=white, circle, inner sep=0pt, font=\Large,
      drop shadow, text width=2.5em, align=center,  cm={0.8,-0.3,0,1,(0,0)}
    },
    levelzeroq/.style={rectangle split part fill={neutral-left,PastelIBMPurple,PastelIBMPurple}},
    leveloneq/.style={rectangle split part fill={neutral-left,PastelIBMBlue,PastelIBMBlue}},
    leveltwoq/.style={rectangle split part fill={neutral-left,PastelIBMPink,PastelIBMPink}},
    levelthreeq/.style={rectangle split part fill={neutral-left,PastelIBMOrange,PastelIBMOrange}},
    leaf/.style={
      shape=rectangle split, rectangle split parts=3,
      rectangle split draw splits, rectangle split horizontal,
      rounded corners, draw, inner sep=4pt,
      minimum width=5em, minimum height=3.5em,
      font=\Large, text width=3em, align=center, cm={0.4,-0.13,0,0.6,(0,0)}
    },
    hi/.style={draw=cyan,line width=3pt} 
  }
  \resizebox{\linewidth}{!}{%
    \begin{tikzpicture}[yscale=0.5]
    \begin{scope}[cm={0.5,-0.3,0,1,(11,-1)}]
          \node[
            fill=teal,
            fill opacity=0.15,
            inner xsep=51em,
            inner ysep=18em,
            yshift=-2em,
            transform shape
          ]{};
        \end{scope}
        
        \begin{scope}[cm={0.5,-0.3,0,1,(11,-1)}]
          \node[
            draw=teal,
            line width=1pt,
            inner xsep=51em,
            inner ysep=18em,
            yshift=-2em,
            transform shape
          ]{};
        \end{scope}
       \begin{scope}[cm={0.5,-0.3,0,1,(4.4,6.4)}]
            \node[transform shape, text=teal, scale=6] at (0,0) {$\ket{11\gamma_3}$};
      \end{scope}
      \begin{scope}[cm={0.5,-0.3,0,1,(11,1)}]
        \def\dx{8} \def\dy{1.8} \def\dxii{4} \def\leafsep{2}
    
        \node[node, draw=cyan, line width=2.5pt] (root) at (0,0) {$\ket{1}$};
        \coordinate (above) at ($(root)+(0,2)$);
        \draw[hi] (above) -- (root);
    
        \node[node] (L)  at ($(root)+(-\dx,-\dy)$) {$\ket{\bullet}$};
        \node[node, draw=cyan, line width=2.5pt] (R)  at ($(root)+(\dx,-\dy)$)  {$\ket{1}$};
    
        \node[node] (LL) at ($(L)+(-\dxii,-\dy)$) {$\ket{\bullet}$};
        \node[node] (LR) at ($(L)+(\dxii,-\dy)$)  {$\ket{\bullet}$};
        \node[node,] (RL) at ($(R)+(-\dxii,-\dy)$) {$\ket{\bullet}$};
        \node[node, draw=cyan, line width=2.5pt] (RR) at ($(R)+(\dxii,-\dy)$)  {$\ket{\gamma_3}$};
    
        \node[leaf, levelzeroq, label=below:{$\ket{L_0}$}] (LLL) at ($(LL)+(-\leafsep,-\dy)$)
          {\nodepart{two}\nodepart{three}};
        \node[leaf, leveloneq,  label=below:{$\ket{L_1}$}] (LLR) at ($(LL)+(\leafsep,-\dy)$)
          {\nodepart{two}\nodepart{three}};
        \node[leaf, leveltwoq,  label=below:{$\ket{L_2}$}] (LRL) at ($(LR)+(-\leafsep,-\dy)$)
          {\nodepart{two}\nodepart{three}};
        \node[leaf, leveltwoq,  label=below:{$\ket{L_3}$}] (LRR) at ($(LR)+(\leafsep,-\dy)$)
          {\nodepart{two}\nodepart{three}};
        \node[leaf, levelthreeq,label=below:{$\ket{L_4}$}] (RLL) at ($(RL)+(-\leafsep,-\dy)$)
          {\nodepart{two}\nodepart{three}};
        \node[leaf, levelthreeq,label=below:{$\ket{L_5}$}] (RLR) at ($(RL)+(\leafsep,-\dy)$)
          {\nodepart{two}\nodepart{three}};
        \node[leaf, draw=cyan, line width=2.5pt, levelthreeq,label=below:{$\ket{L_6}$}] (RRL) at ($(RR)+(-\leafsep,-\dy)$)
          {\nodepart{two}\nodepart{three}};
        \node[leaf, draw=cyan, line width=2.5pt, levelthreeq,label=below:{$\ket{L_7}$}] (RRR) at ($(RR)+(\leafsep,-\dy)$)
          {\nodepart{two}\nodepart{three}};
    
        \draw (root) -- ++(-\dx,0) coordinate(auxL) -- (L);
        \draw (root) -- ++(\dx,0)  coordinate(auxR) -- (R);
        \draw (L)  -- ++(-\dxii,0) coordinate(auxLL) -- (LL);
        \draw (L)  -- ++(\dxii,0)  coordinate(auxLR) -- (LR);
        \draw (R)  -- ++(-\dxii,0) coordinate(auxRL) -- (RL);
        \draw (R)  -- ++(\dxii,0)  coordinate(auxRR) -- (RR);
        \draw (LL) -- ++(-\leafsep,0) coordinate(auxLLL) -- (LLL);
        \draw (LL) -- ++(\leafsep,0)  coordinate(auxLLR) -- (LLR);
        \draw (LR) -- ++(-\leafsep,0) coordinate(auxLRL) -- (LRL);
        \draw (LR) -- ++(\leafsep,0)  coordinate(auxLRR) -- (LRR);
        \draw (RL) -- ++(-\leafsep,0) coordinate(auxRLL) -- (RLL);
        \draw (RL) -- ++(\leafsep,0)  coordinate(auxRLR) -- (RLR);
        \draw (RR) -- ++(-\leafsep,0) coordinate(auxRRL) -- (RRL);
        \draw (RR) -- ++(\leafsep,0)  coordinate(auxRRR) -- (RRR);
    
        \draw[hi] (root)--(auxR)--(R)
                  (R)--(auxRR)--(RR)
                  (RR)--(auxRRL)--(RRL)
                  (RR)--(auxRRR)--(RRR);
      \end{scope}
      \begin{scope}[cm={0.5,-0.3,0,1,(0,-2)}]
          \node[
            fill=teal,
            fill opacity=0.15,
            inner xsep=51em,
            inner ysep=18em,
            yshift=-2em,
            transform shape
          ]{};
        \end{scope}
        
        \begin{scope}[cm={0.5,-0.3,0,1,(0,-2)}]
          \node[
            draw=teal,
            line width=1pt,
            inner xsep=51em,
            inner ysep=18em,
            yshift=-2em,
            transform shape
          ]{};
        \end{scope}
     \begin{scope}[cm={0.5,-0.3,0,1,(-6.6,5.4)}]
            \node[transform shape, text=teal, scale=6] at (0,0) {$\ket{10\gamma_2}$};
     \end{scope}
     \begin{scope}[cm={0.5,-0.3,0,1,(0,0)}]
        
        \def\dx{8} \def\dy{1.8} \def\dxii{4} \def\leafsep{2}
    
        \node[node, draw=cyan, line width=2.5pt] (root) at (0,0) {$\ket{1}$};
        \coordinate (above) at ($(root)+(0,2)$);
        \draw[hi] (above) -- (root);
    
        \node[node] (L)  at ($(root)+(-\dx,-\dy)$) {$\ket{\bullet}$};
        \node[node, draw=cyan, line width=2.5pt] (R)  at ($(root)+(\dx,-\dy)$)  {$\ket{0}$};
    
        \node[node] (LL) at ($(L)+(-\dxii,-\dy)$) {$\ket{\bullet}$};
        \node[node] (LR) at ($(L)+(\dxii,-\dy)$)  {$\ket{\bullet}$};
        \node[node, draw=cyan, line width=2.5pt] (RL) at ($(R)+(-\dxii,-\dy)$) {$\ket{\gamma_2}$};
        \node[node] (RR) at ($(R)+(\dxii,-\dy)$)  {$\ket{\bullet}$};
    
        \node[leaf, levelzeroq, label=below:{$\ket{L_0}$}] (LLL) at ($(LL)+(-\leafsep,-\dy)$)
          {\nodepart{two}\nodepart{three}};
        \node[leaf, leveloneq,  label=below:{$\ket{L_1}$}] (LLR) at ($(LL)+(\leafsep,-\dy)$)
          {\nodepart{two}\nodepart{three}};
        \node[leaf, leveltwoq,  label=below:{$\ket{L_2}$}] (LRL) at ($(LR)+(-\leafsep,-\dy)$)
          {\nodepart{two}\nodepart{three}};
        \node[leaf, leveltwoq,  label=below:{$\ket{L_3}$}] (LRR) at ($(LR)+(\leafsep,-\dy)$)
          {\nodepart{two}\nodepart{three}};
        \node[leaf, draw=cyan, line width=2.5pt, levelthreeq,label=below:{$\ket{L_4}$}] (RLL) at ($(RL)+(-\leafsep,-\dy)$)
          {\nodepart{two}\nodepart{three}};
        \node[leaf, draw=cyan, line width=2.5pt, levelthreeq,label=below:{$\ket{L_5}$}] (RLR) at ($(RL)+(\leafsep,-\dy)$)
          {\nodepart{two}\nodepart{three}};
        \node[leaf, levelthreeq,label=below:{$\ket{L_6}$}] (RRL) at ($(RR)+(-\leafsep,-\dy)$)
          {\nodepart{two}\nodepart{three}};
        \node[leaf, levelthreeq, label=below:{$\ket{L_7}$}] (RRR) at ($(RR)+(\leafsep,-\dy)$)
          {\nodepart{two}\nodepart{three}};
    
        \draw (root) -- ++(-\dx,0) coordinate(auxL) -- (L);
        \draw (root) -- ++(\dx,0)  coordinate(auxR) -- (R);
        \draw (L)  -- ++(-\dxii,0) coordinate(auxLL) -- (LL);
        \draw (L)  -- ++(\dxii,0)  coordinate(auxLR) -- (LR);
        \draw (R)  -- ++(-\dxii,0) coordinate(auxRL) -- (RL);
        \draw (R)  -- ++(\dxii,0)  coordinate(auxRR) -- (RR);
        \draw (LL) -- ++(-\leafsep,0) coordinate(auxLLL) -- (LLL);
        \draw (LL) -- ++(\leafsep,0)  coordinate(auxLLR) -- (LLR);
        \draw (LR) -- ++(-\leafsep,0) coordinate(auxLRL) -- (LRL);
        \draw (LR) -- ++(\leafsep,0)  coordinate(auxLRR) -- (LRR);
        \draw (RL) -- ++(-\leafsep,0) coordinate(auxRLL) -- (RLL);
        \draw (RL) -- ++(\leafsep,0)  coordinate(auxRLR) -- (RLR);
        \draw (RR) -- ++(-\leafsep,0) coordinate(auxRRL) -- (RRL);
        \draw (RR) -- ++(\leafsep,0)  coordinate(auxRRR) -- (RRR);
    
        \draw[hi] (root)--(auxR)--(R)
                  (R)--(auxRL)--(RL)
                  (RL)--(auxRLL)--(RLL)
                  (RL)--(auxRLR)--(RLR);
      \end{scope}
    \end{tikzpicture}}
    \caption{Address Register $\ket{1}(\ket{0\gamma_2}+\ket{1\gamma_3})$ branches twice, accessing in superposition the memory cells $\ket{L_4}$, $\ket{L_5}$, $\ket{L_6}$, and $\ket{L_7}$ containing the four sibling pairs $\ket{T_{3,0}}\ket{T_{3,1}}$, $\ket{T_{3,2}}\ket{T_{3,3}}$, $\ket{T_{3,4}}\ket{T_{3,5}}$, and $\ket{T_{3,6}}\ket{T_{3,7}}$ located at the leaf level $h=3$ in $T$.}
    \label{subfig:access_3}
  \end{subfigure}

  \vspace{1em}

  \begin{subfigure}[t]{\textwidth}
   \centering
    \tikzset{
    every label/.append style={font=\huge,  cm={0.8,-0.3,0,1,(0,0)}},
    node/.style={
      draw, fill=white, circle, inner sep=0pt, font=\Large,
      drop shadow, text width=2.5em, align=center,  cm={0.8,-0.3,0,1,(0,0)}
    },
    levelzeroq/.style={rectangle split part fill={neutral-left,PastelIBMPurple,PastelIBMPurple}},
    leveloneq/.style={rectangle split part fill={neutral-left,PastelIBMBlue,PastelIBMBlue}},
    leveltwoq/.style={rectangle split part fill={neutral-left,PastelIBMPink,PastelIBMPink}},
    levelthreeq/.style={rectangle split part fill={neutral-left,PastelIBMOrange,PastelIBMOrange}},
    leaf/.style={
      shape=rectangle split, rectangle split parts=3,
      rectangle split draw splits, rectangle split horizontal,
      rounded corners, draw, inner sep=4pt,
      minimum width=5em, minimum height=3.5em,
      font=\Large, text width=3em, align=center, cm={0.4,-0.13,0,0.6,(0,0)}
    },
    hi/.style={draw=cyan,line width=3pt} 
  }
  \resizebox{\linewidth}{!}{%
    \begin{tikzpicture}[yscale=0.5]

     \begin{scope}[cm={0.5,-0.3,0,1,(24,4)}]
          \node[
            fill=purple,
            fill opacity=0.15,
            inner xsep=51em,
            inner ysep=17em,
            transform shape
          ]{};
        \end{scope}
        
        \begin{scope}[cm={0.5,-0.3,0,1,(24,4)}]
          \node[
            draw=purple,
            line width=1pt,
            inner xsep=51em,
            inner ysep=17em,
            transform shape
          ]{};
        \end{scope}
        \begin{scope}[cm={0.5,-0.3,0,1,(17.2,12)}]
            \node[transform shape, text=purple, scale=5] at (0,0) {$\ket{11\gamma_3}$};
        \end{scope}
    \begin{scope}[cm={0.5,-0.3,0,1,(24,6)}]
        \def\dx{8} \def\dy{1.8} \def\dxii{4} \def\leafsep{2}
    
        \node[node, draw=cyan, line width=2.5pt] (root) at (0,0) {$\ket{1}$};
        \coordinate (above) at ($(root)+(0,2)$);
        \draw[hi] (above) -- (root);
    
        \node[node] (L)  at ($(root)+(-\dx,-\dy)$) {$\ket{\bullet}$};
        \node[node, draw=cyan, line width=2.5pt] (R)  at ($(root)+(\dx,-\dy)$)  {$\ket{1}$};
    
        \node[node] (LL) at ($(L)+(-\dxii,-\dy)$) {$\ket{\bullet}$};
        \node[node] (LR) at ($(L)+(\dxii,-\dy)$)  {$\ket{\bullet}$};
        \node[node,] (RL) at ($(R)+(-\dxii,-\dy)$) {$\ket{\bullet}$};
        \node[node, draw=cyan, line width=2.5pt] (RR) at ($(R)+(\dxii,-\dy)$)  {$\ket{\gamma_3}$};
    
        \node[leaf, levelzeroq, label=below:{}] (LLL) at ($(LL)+(-\leafsep,-\dy)$)
          {\nodepart{two}\nodepart{three}};
        \node[leaf, leveloneq,  label=below:{}] (LLR) at ($(LL)+(\leafsep,-\dy)$)
          {\nodepart{two}\nodepart{three}};
        \node[leaf, leveltwoq,  label=below:{}] (LRL) at ($(LR)+(-\leafsep,-\dy)$)
          {\nodepart{two}\nodepart{three}};
        \node[leaf, leveltwoq,  label=below:{}] (LRR) at ($(LR)+(\leafsep,-\dy)$)
          {\nodepart{two}\nodepart{three}};
        \node[leaf, levelthreeq,label=below:{}] (RLL) at ($(RL)+(-\leafsep,-\dy)$)
          {\nodepart{two}\nodepart{three}};
        \node[leaf, levelthreeq,label=below:{}] (RLR) at ($(RL)+(\leafsep,-\dy)$)
          {\nodepart{two}\nodepart{three}};
        \node[leaf, draw=cyan, line width=2.5pt, levelthreeq,label=below:{}] (RRL) at ($(RR)+(-\leafsep,-\dy)$)
          {\nodepart{two}\nodepart{three}};
        \node[leaf, draw=cyan, line width=2.5pt, levelthreeq,label=below:{}] (RRR) at ($(RR)+(\leafsep,-\dy)$)
          {\nodepart{two}\nodepart{three}};
    
        \draw (root) -- ++(-\dx,0) coordinate(auxL) -- (L);
        \draw (root) -- ++(\dx,0)  coordinate(auxR) -- (R);
        \draw (L)  -- ++(-\dxii,0) coordinate(auxLL) -- (LL);
        \draw (L)  -- ++(\dxii,0)  coordinate(auxLR) -- (LR);
        \draw (R)  -- ++(-\dxii,0) coordinate(auxRL) -- (RL);
        \draw (R)  -- ++(\dxii,0)  coordinate(auxRR) -- (RR);
        \draw (LL) -- ++(-\leafsep,0) coordinate(auxLLL) -- (LLL);
        \draw (LL) -- ++(\leafsep,0)  coordinate(auxLLR) -- (LLR);
        \draw (LR) -- ++(-\leafsep,0) coordinate(auxLRL) -- (LRL);
        \draw (LR) -- ++(\leafsep,0)  coordinate(auxLRR) -- (LRR);
        \draw (RL) -- ++(-\leafsep,0) coordinate(auxRLL) -- (RLL);
        \draw (RL) -- ++(\leafsep,0)  coordinate(auxRLR) -- (RLR);
        \draw (RR) -- ++(-\leafsep,0) coordinate(auxRRL) -- (RRL);
        \draw (RR) -- ++(\leafsep,0)  coordinate(auxRRR) -- (RRR);
    
        \draw[hi] (root)--(auxR)--(R)
                  (R)--(auxRR)--(RR)
                  (RR)--(auxRRL)--(RRL)
                  (RR)--(auxRRR)--(RRR);
    
      \end{scope}

     \begin{scope}[cm={0.5,-0.3,0,1,(16,2)}]
          \node[
            fill=purple,
            fill opacity=0.15,
            inner xsep=51em,
            inner ysep=17em,
            transform shape
          ]{};
        \end{scope}
        
        \begin{scope}[cm={0.5,-0.3,0,1,(16,2)}]
          \node[
            draw=purple,
            line width=1pt,
            inner xsep=51em,
            inner ysep=17em,
            transform shape
          ]{};
        \end{scope}
        \begin{scope}[cm={0.5,-0.3,0,1,(9.2,10)}]
            \node[transform shape, text=purple, scale=5] at (0,0) {$\ket{10\gamma_2}$};
        \end{scope}
      \begin{scope}[cm={0.5,-0.3,0,1,(16,4)}]
        
        \def\dx{8} \def\dy{1.8} \def\dxii{4} \def\leafsep{2}
    
        \node[node, draw=cyan, line width=2.5pt] (root) at (0,0) {$\ket{1}$};
        \coordinate (above) at ($(root)+(0,2)$);
        \draw[hi] (above) -- (root);
    
        \node[node] (L)  at ($(root)+(-\dx,-\dy)$) {$\ket{\bullet}$};
        \node[node, draw=cyan, line width=2.5pt] (R)  at ($(root)+(\dx,-\dy)$)  {$\ket{0}$};
    
        \node[node] (LL) at ($(L)+(-\dxii,-\dy)$) {$\ket{\bullet}$};
        \node[node] (LR) at ($(L)+(\dxii,-\dy)$)  {$\ket{\bullet}$};
        \node[node, draw=cyan, line width=2.5pt] (RL) at ($(R)+(-\dxii,-\dy)$) {$\ket{\gamma_2}$};
        \node[node] (RR) at ($(R)+(\dxii,-\dy)$)  {$\ket{\bullet}$};
    
        \node[leaf, levelzeroq, label=below:{}] (LLL) at ($(LL)+(-\leafsep,-\dy)$)
          {\nodepart{two}\nodepart{three}};
        \node[leaf, leveloneq,  label=below:{}] (LLR) at ($(LL)+(\leafsep,-\dy)$)
          {\nodepart{two}\nodepart{three}};
        \node[leaf, leveltwoq,  label=below:{}] (LRL) at ($(LR)+(-\leafsep,-\dy)$)
          {\nodepart{two}\nodepart{three}};
        \node[leaf, leveltwoq,  label=below:{}] (LRR) at ($(LR)+(\leafsep,-\dy)$)
          {\nodepart{two}\nodepart{three}};
        \node[leaf, draw=cyan, line width=2.5pt, levelthreeq,label=below:{}] (RLL) at ($(RL)+(-\leafsep,-\dy)$)
          {\nodepart{two}\nodepart{three}};
        \node[leaf, draw=cyan, line width=2.5pt, levelthreeq,label=below:{}] (RLR) at ($(RL)+(\leafsep,-\dy)$)
          {\nodepart{two}\nodepart{three}};
        \node[leaf, levelthreeq,label=below:{}] (RRL) at ($(RR)+(-\leafsep,-\dy)$)
          {\nodepart{two}\nodepart{three}};
        \node[leaf, levelthreeq, label=below:{}] (RRR) at ($(RR)+(\leafsep,-\dy)$)
          {\nodepart{two}\nodepart{three}};
    
        \draw (root) -- ++(-\dx,0) coordinate(auxL) -- (L);
        \draw (root) -- ++(\dx,0)  coordinate(auxR) -- (R);
        \draw (L)  -- ++(-\dxii,0) coordinate(auxLL) -- (LL);
        \draw (L)  -- ++(\dxii,0)  coordinate(auxLR) -- (LR);
        \draw (R)  -- ++(-\dxii,0) coordinate(auxRL) -- (RL);
        \draw (R)  -- ++(\dxii,0)  coordinate(auxRR) -- (RR);
        \draw (LL) -- ++(-\leafsep,0) coordinate(auxLLL) -- (LLL);
        \draw (LL) -- ++(\leafsep,0)  coordinate(auxLLR) -- (LLR);
        \draw (LR) -- ++(-\leafsep,0) coordinate(auxLRL) -- (LRL);
        \draw (LR) -- ++(\leafsep,0)  coordinate(auxLRR) -- (LRR);
        \draw (RL) -- ++(-\leafsep,0) coordinate(auxRLL) -- (RLL);
        \draw (RL) -- ++(\leafsep,0)  coordinate(auxRLR) -- (RLR);
        \draw (RR) -- ++(-\leafsep,0) coordinate(auxRRL) -- (RRL);
        \draw (RR) -- ++(\leafsep,0)  coordinate(auxRRR) -- (RRR);
    
        \draw[hi] (root)--(auxR)--(R)
                  (R)--(auxRL)--(RL)
                  (RL)--(auxRLL)--(RLL)
                  (RL)--(auxRLR)--(RLR);
      \end{scope}
    
    \begin{scope}[cm={0.5,-0.3,0,1,(8,0)}]
      \node[
        fill=purple,
        fill opacity=0.15,
        inner xsep=51em,
        inner ysep=17em,
        transform shape
      ]{};
    \end{scope}
    
    \begin{scope}[cm={0.5,-0.3,0,1,(8,0)}]
      \node[
        draw=purple,
        line width=1pt,
        inner xsep=51em,
        inner ysep=17em,
        transform shape
      ]{};
    \end{scope}
    \begin{scope}[cm={0.5,-0.3,0,1,(1.2,8)}]
        \node[transform shape, text=purple, scale=5] at (0,0) {$\ket{01\gamma_1}$};
    \end{scope}
    \begin{scope}[cm={0.5,-0.3,0,1,(8,2)}]
\def\dx{8} \def\dy{1.8} \def\dxii{4} \def\leafsep{2}
    
        \node[node, draw=cyan, line width=2.5pt] (root) at (0,0) {$\ket{0}$};
        \coordinate (above) at ($(root)+(0,2)$);
        \draw[hi] (above) -- (root);
    
        \node[node, draw=cyan, line width=2.5pt] (L)  at ($(root)+(-\dx,-\dy)$) {$\ket{1}$};
        \node[node] (R)  at ($(root)+(\dx,-\dy)$)  {$\ket{\bullet}$};
    
        \node[node] (LL) at ($(L)+(-\dxii,-\dy)$) {$\ket{\bullet}$};
        \node[node, draw=cyan, line width=2.5pt] (LR) at ($(L)+(\dxii,-\dy)$)  {$\ket{\gamma_1}$};
        \node[node] (RL) at ($(R)+(-\dxii,-\dy)$) {$\ket{\bullet}$};
        \node[node] (RR) at ($(R)+(\dxii,-\dy)$)  {$\ket{\bullet}$};
    
        \node[leaf, levelzeroq, label=below:{}] (LLL) at ($(LL)+(-\leafsep,-\dy)$)
          {\nodepart{two}\nodepart{three}};
        \node[leaf, leveloneq,  label=below:{}] (LLR) at ($(LL)+(\leafsep,-\dy)$)
          {\nodepart{two}\nodepart{three}};
        \node[leaf, draw=cyan, line width=2.5pt, leveltwoq,  label=below:{}] (LRL) at ($(LR)+(-\leafsep,-\dy)$)
          {\nodepart{two}\nodepart{three}};
        \node[leaf, draw=cyan, line width=2.5pt, leveltwoq,  label=below:{}] (LRR) at ($(LR)+(\leafsep,-\dy)$)
          {\nodepart{two}\nodepart{three}};
        \node[leaf, levelthreeq,label=below:{}] (RLL) at ($(RL)+(-\leafsep,-\dy)$)
          {\nodepart{two}\nodepart{three}};
        \node[leaf, levelthreeq,label=below:{}] (RLR) at ($(RL)+(\leafsep,-\dy)$)
          {\nodepart{two}\nodepart{three}};
        \node[leaf, levelthreeq,label=below:{}] (RRL) at ($(RR)+(-\leafsep,-\dy)$)
          {\nodepart{two}\nodepart{three}};
        \node[leaf, levelthreeq,label=below:{}] (RRR) at ($(RR)+(\leafsep,-\dy)$)
          {\nodepart{two}\nodepart{three}};
    
        \draw (root) -- ++(-\dx,0) coordinate(auxL) -- (L);
        \draw (root) -- ++(\dx,0)  coordinate(auxR) -- (R);
        \draw (L)  -- ++(-\dxii,0) coordinate(auxLL) -- (LL);
        \draw (L)  -- ++(\dxii,0)  coordinate(auxLR) -- (LR);
        \draw (R)  -- ++(-\dxii,0) coordinate(auxRL) -- (RL);
        \draw (R)  -- ++(\dxii,0)  coordinate(auxRR) -- (RR);
        \draw (LL) -- ++(-\leafsep,0) coordinate(auxLLL) -- (LLL);
        \draw (LL) -- ++(\leafsep,0)  coordinate(auxLLR) -- (LLR);
        \draw (LR) -- ++(-\leafsep,0) coordinate(auxLRL) -- (LRL);
        \draw (LR) -- ++(\leafsep,0)  coordinate(auxLRR) -- (LRR);
        \draw (RL) -- ++(-\leafsep,0) coordinate(auxRLL) -- (RLL);
        \draw (RL) -- ++(\leafsep,0)  coordinate(auxRLR) -- (RLR);
        \draw (RR) -- ++(-\leafsep,0) coordinate(auxRRL) -- (RRL);
        \draw (RR) -- ++(\leafsep,0)  coordinate(auxRRR) -- (RRR);
    
        \draw[hi] (root)--(auxL)--(L)
                  (L)--(auxLR)--(LR)  
                  (LR)--(auxLRL)--(LRL) 
                  (LR)--(auxLRR)--(LRR);   
      \end{scope}

      \begin{scope}[cm={0.5,-0.3,0,1,(0,-2)}]
      \node[
        fill=purple,
        fill opacity=0.15,
        inner xsep=51em,
        inner ysep=17em,
        transform shape
      ]{};
    \end{scope}
    
    \begin{scope}[cm={0.5,-0.3,0,1,(0,-2)}]
      \node[
        draw=purple,
        line width=1pt,
        inner xsep=51em,
        inner ysep=17em,
        transform shape
      ]{};
    \end{scope}
    \begin{scope}[cm={0.5,-0.3,0,1,(-6.8,6)}]
        \node[transform shape, text=purple, scale=5] at (0,0) {$\ket{00\gamma_0}$};
      \end{scope}
     \begin{scope}[cm={0.5,-0.3,0,1,(0,0)}]
        \def\dx{8} \def\dy{1.8} \def\dxii{4} \def\leafsep{2}
    
        \node[node, draw=cyan, line width=2.5pt] (root) at (0,0) {$\ket{0}$};
        \coordinate (above) at ($(root)+(0,2)$);
        \draw[hi] (above) -- (root);
    
        \node[node, draw=cyan, line width=2.5pt] (L)  at ($(root)+(-\dx,-\dy)$) {$\ket{0}$};
        \node[node] (R)  at ($(root)+(\dx,-\dy)$)  {$\ket{\bullet}$};
    
        \node[node,draw=cyan, line width=2.5pt] (LL) at ($(L)+(-\dxii,-\dy)$) {$\ket{\gamma_0}$};
        \node[node] (LR) at ($(L)+(\dxii,-\dy)$)  {$\ket{\bullet}$};
        \node[node] (RL) at ($(R)+(-\dxii,-\dy)$) {$\ket{\bullet}$};
        \node[node] (RR) at ($(R)+(\dxii,-\dy)$)  {$\ket{\bullet}$};
    
        \node[leaf, draw=cyan, line width=2.5pt, levelzeroq, label=below:{}] (LLL) at ($(LL)+(-\leafsep,-\dy)$)
          {\nodepart{two}\nodepart{three}};
        \node[leaf, draw=cyan, line width=2.5pt, leveloneq,  label=below:{}] (LLR) at ($(LL)+(\leafsep,-\dy)$)
          {\nodepart{two}\nodepart{three}};
        \node[leaf, leveltwoq,  label=below:{}] (LRL) at ($(LR)+(-\leafsep,-\dy)$)
          {\nodepart{two}\nodepart{three}};
        \node[leaf, leveltwoq,  label=below:{}] (LRR) at ($(LR)+(\leafsep,-\dy)$)
          {\nodepart{two}\nodepart{three}};
        \node[leaf, levelthreeq,label=below:{}] (RLL) at ($(RL)+(-\leafsep,-\dy)$)
          {\nodepart{two}\nodepart{three}};
        \node[leaf, levelthreeq,label=below:{}] (RLR) at ($(RL)+(\leafsep,-\dy)$)
          {\nodepart{two}\nodepart{three}};
        \node[leaf, levelthreeq,label=below:{}] (RRL) at ($(RR)+(-\leafsep,-\dy)$)
          {\nodepart{two}\nodepart{three}};
        \node[leaf, levelthreeq,label=below:{}] (RRR) at ($(RR)+(\leafsep,-\dy)$)
          {\nodepart{two}\nodepart{three}};
    
        \draw (root) -- ++(-\dx,0) coordinate(auxL) -- (L);
        \draw (root) -- ++(\dx,0)  coordinate(auxR) -- (R);
        \draw (L)  -- ++(-\dxii,0) coordinate(auxLL) -- (LL);
        \draw (L)  -- ++(\dxii,0)  coordinate(auxLR) -- (LR);
        \draw (R)  -- ++(-\dxii,0) coordinate(auxRL) -- (RL);
        \draw (R)  -- ++(\dxii,0)  coordinate(auxRR) -- (RR);
        \draw (LL) -- ++(-\leafsep,0) coordinate(auxLLL) -- (LLL);
        \draw (LL) -- ++(\leafsep,0)  coordinate(auxLLR) -- (LLR);
        \draw (LR) -- ++(-\leafsep,0) coordinate(auxLRL) -- (LRL);
        \draw (LR) -- ++(\leafsep,0)  coordinate(auxLRR) -- (LRR);
        \draw (RL) -- ++(-\leafsep,0) coordinate(auxRLL) -- (RLL);
        \draw (RL) -- ++(\leafsep,0)  coordinate(auxRLR) -- (RLR);
        \draw (RR) -- ++(-\leafsep,0) coordinate(auxRRL) -- (RRL);
        \draw (RR) -- ++(\leafsep,0)  coordinate(auxRRR) -- (RRR);
    
        \draw[hi] (root)--(auxL)--(L)
                  (L)--(auxLL)--(LL)  
                  (LL)--(auxLLL)--(LLL)
                  (LL)--(auxLLR)--(LLR); 
      \end{scope}
    \end{tikzpicture}}
    \caption{Address register $\ket{00}_{\mathrm{a}}\ket{\gamma_0}_{\mathrm{a}} + \ket{01}_{\mathrm{a}}\ket{\gamma_1}_{\mathrm{a}} + \ket{10}_{\mathrm{a}}\ket{\gamma_2}_{\mathrm{a}} + \ket{11}_{\mathrm{a}}\ket{\gamma_3}_{\mathrm{a}}$ activates every path towards all $K$ memory cells and such to access the sign bits in superposition.}
    \label{subfig:access_4}
  \end{subfigure}
  \caption{Two representative access paths defined by the retrieval primitives of Corollary~\ref{cor:structured_access}. Figure~\ref{subfig:access_3} illustrates the path accessing in superposition the sibling pairs at level $h=3$ in $T$. Figure~\ref{subfig:access_4} depicts the configuration where all access paths are active, enabling retrieval of the sign bits from every memory cell $\ket{L_z}$ in superposition.}\Description{}
  \label{fig:access_patterns_2}
\end{figure*}
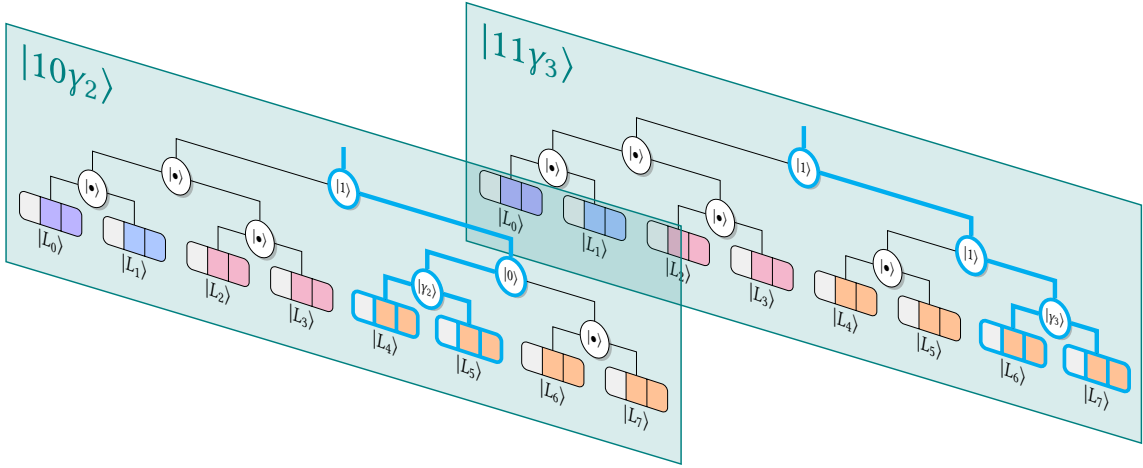
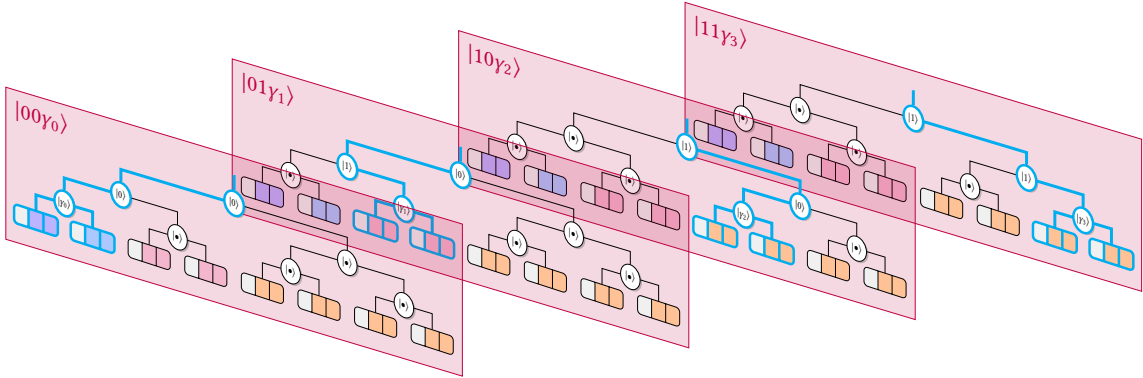

At each level $h \in [1, k]$, the segment tree contains $2^h$ nodes. We organize these nodes into sibling pairs, specifically $(T_{l(z),\,2 d(z)}, T_{l(z),\,2 d(z)+1})$, where $l(z) = \lfloor \log_2 z \rfloor + 1$ denotes the tree level and $d(z) = z - 2^{\lfloor \log_2 z \rfloor}$ specifies the position within that level. Then, we encode each pair into a quantum register $\ket{L_z}$. Since each quantum register encodes two sibling nodes, the total number of registers is $\sum_{h=1}^{k} \frac{2^h}{2} = 2^k - 1 = K - 1$.

The root node $T_{0,0}$ is unique and does not have a sibling. We store it separately in the register $\ket{L_0}$, along with an arbitrary string $b$. Therefore, the BBQRAM contains $K$ memory cells: $K - 1$ for sibling pairs and one for the root. This mapping guarantees that the number of BBQRAM memory cells equals the number of leaves in the segment tree, and both structures maintain identical depth.

Eventually, since $K = MN$, the number of BBQRAM memory cells matches the number of entries $a_z$ in the matrix $A$. To store the sign of each value $a_z$, we prepend a single qubit to the leftmost position of each quantum register $\ket{L_z}$.
\end{proof}

In Corollary~\ref{cor:structured_access}, we describe three retrieval primitives that leverage the memory layout of Proposition~\ref{prop:mapping} to enable efficient retrieval of precomputed amplitudes in superposition. Figure~\ref{fig:access_patterns_1} and Figure~\ref{fig:access_patterns_2} display four representative access paths, each corresponding to one of the retrieval scenarios outlined in the corollary.

\begin{corollary}[Quantum Retrieval Primitives]\label{cor:structured_access}
Let $\mathsf{BBQRAM}$ denote a Bucket Brigade QRAM that stores the nodes of a segment tree $T$ with depth $k = \log_2 K$ in memory cells $\{\ket{L_z}\}_{z=0}^{K-1}$. Each $\ket{L_z}$ is a $(1+2t)$-qubit register according to Proposition~\ref{prop:mapping}. Given a $k$-qubit address register $\ket{z}^{k}$, where $0 \leq z < K$, the retrieval of the memory cell $\ket{L_z}$ into a $(1+2t)$-qubit working register is defined by the following mapping:
$$   
\mathsf{BBQRAM}:\;
   \ket{z}^{k}\ket{0}^{(1+2t)}
   \longmapsto
   \ket{z}^{k}\ket{L_z}^{1+2t}.
$$ 
\noindent
The memory layout enables the following primitives for efficient retrieval from the $\mathsf{BBQRAM}$ of the states:

\begin{equation}\label{eq:case_ret}
\mathsf{PRIMITIVES:}
\begin{cases}
    \ket{0}^{k}\ket{0}^{1}\ket{0}^{t}\ket{0}^{t} \longmapsto \ket{0}^{ k}\ket{0}^{1}\ket{T_{0,0}}^{t}\ket{0}^{t}&\textrm{if }  h = 0, \\[1ex]
    \ket{0}^{k - h}\sum_{z = 2^{h-1}}^{2^{h-1}-1}\alpha_z\ket{z}^{h}\ket{0}^{1}\ket{0}^{2t} \longmapsto  \sum_{z = 2^{h-1}}^{2^{h}-1}\alpha_z\ket{z}^{k}\ket{0}^{1}\ket{T_{l(z),\,2\dot d(z)}}^{t}\,
           \ket{T_{l(z),\,2\dot d(z)+1}}^{t} &\textrm{else if }  1 \leq h \leq k, \\[1ex]
    \sum_{z = 0}^{K}\alpha_z\ket{z}^{k}\ket{0}^{1}\ket{0}^{t}\ket{0}^{t} \longmapsto \sum_{z = 0}^{K}\alpha_z\ket{z}^{k}\ket{\mathrm{s}(a_{z})}^{1}\ket{0}^{t}\ket{0}^{t} & \textrm{otherwise, }  \\[1ex]
\end{cases}
\end{equation}
where $$ 
\alpha_z \in \mathbb{C} \textrm{ and } \alpha_z \neq 0, \quad
l(z) = \lfloor \log_2 z \rfloor + 1, \qquad
d(z) = z - 2^{\lfloor \log_2 z \rfloor},\textrm{ and} \qquad
\mathrm{s}(a_z) =
    \begin{cases}
        0 & \text{if } a_z \geq 0, \\
        1 & \text{if } a_z < 0.
    \end{cases}
$$
 
\end{corollary}
\begin{proof}[Corollary~\ref{cor:structured_access} Correctness]
We now verify the correctness of the three retrieval primitives described in Equation~\eqref{eq:case_ret} that correspond to the following scenarios:
\begin{enumerate}
    \item retrieval of the root $\ket{T_{0,0}}$ \label{item:root};
    \item retrieval of the superposition of every sibling nodes pairs (i.e., $\ket{T_{l(z),\,2\dot d(z)}}^{t}\, \ket{T_{l(z),\,2\dot d(z)+1}}^{t}$) at a given height $h$ in $T$ for $1 \leq h \leq k$ \label{item:sibling_pairs};
    \item retrieval of the superposition of the signs $\ket{\mathrm{s}(a_{z})}$ for $0 \leq z < K$\label{item:signs}.
\end{enumerate}

\paragraph{(\ref{item:root}) \textbf{Root retrieval}.}
With the address register set to $\ket{0}^{ k}$, we retrieve from BBQRAM the memory cell
$\ket{L_0}=\ket{\mathrm{s}(a_0)}\ket{T_{0,0}}^{t}\ket{b}^{t}$. Since the root $\ket{T_{0,0}}^{t}$ occupies the $t$ middle qubits of $\ket{L_0}$, copying only these $t$ qubits from the memory cell to the working register yields the state $\ket{0}^{ k}\ket{0}^{1}\ket{T_{0,0}}^{t}\ket{0}^{t}$.

\paragraph{(\ref{item:sibling_pairs})  \textbf{Superposition of sibling nodes pairs}.} We prove by induction that the hypothesis $P(h)$ below holds for $1\le h\le k$, given $k \in \mathbf{N}$.

$$
P(h) : \
\ket{0}^{k - h}\sum_{z = 2^{h-1}}^{2^{h-1}-1}\alpha_z\ket{z}^{h}\ket{0}^{1}\ket{0}^{2t} \longmapsto \sum_{z = 2^{h-1}}^{2^{h}-1}\alpha_z\ket{z}^{k}\ket{0}^{1}\ket{T_{l(z),\,2\dot d(z)}}^{t}\,\ket{T_{l(z),\,2\dot d(z)+1}}^{t},
$$
recalling that $l(z) = \lfloor \log_2 z\rfloor + 1 $ $(\clubsuit)$ and $d(z) = z - 2^{\lfloor \log_2 z\rfloor}$ $(\spadesuit)$ .

\begin{itemize} 
\item \textit{Base Case:} $h = 1$.  
\begin{equation*}
    P(1):  \ket{0}^{k - 1}\ket{1}\ket{0}^{1}\ket{0}^{ 2t}\longmapsto \ket{0}^{k - 1}\ket{1}\ket{0}^{1}\ket{T_{l(1),2d(1)}}^{t}\ket{T_{l(1),2d(1)+1}}^{t}=  \ket{0}^{ k - 1}\ket{1}\ket{0}^{1}\ket{T_{1,0}}^{t}\ket{T_{1,1}}^{t}.
\end{equation*}

We observe that the base case $P(1)$ holds, since setting the address register to $\ket{0}^{k - 1}\ket{1} = \ket{1}^k$ allows the retrieval from the memory cell $\ket{L_1}$ of the only two sibling nodes (i.e., $\ket{T_{1,0}}^{t}\ket{T_{1,1}}^{t}$) at height $h=1$ of segment tree $T$.

\item \textit{Inductive step}. Assume $P(h)$ holds for some $ 1 \leq h  < k$, we prove it for $h+1$.
\begin{equation}\label{eq:h+1}
    P(h+1): \ket{0}^{ k - (h+1)}\sum_{f = 2^{(h+1)-1}}^{2^{(h+1)}-1}\alpha_f\ket{f}^{h+1}\ket{0}^{1}\ket{0}^{2t}\longmapsto  \sum_{f = 2^{h}}^{2^{(h+1)}-1}\alpha_f\ket{f}^{h+1}\ket{0}^{1}\ket{T_{l(f),\,2\dot d(f)}}^{t}\,\ket{T_{l(f),\,2\dot d(f)+1}}^{t}.
\end{equation}

The induction step results immediately by observing that we can write any index $f \in [2^h, 2^{(h+1)}-1]$ as $f = 2z + g$, where $z \in \left[2^{h-1}, 2^h-1\right]$ and $g \in [0,1]$. At level $h+1$, this decomposition corresponds to shifting the binary representation of $z$ left by one bit (i.e., multiplying $z$ by $2$), with the least significant bit determined by $g$, which indexes each superposition. The mapping $(z,g)\mapsto f$ defines a bijection from $[2^{h-1}, 2^h-1] \times \{0, 1\}$ onto $[2^{h},\,2^{h+1}-1]$, therefore, we can rewrite Equation~\eqref{eq:h+1} as:
\begin{align*}
    & P(h+1): \ket{0}^{ k-(h+1)}\sum_{f = 2^{h}}^{2^{(h+1)}-1}\alpha_f\ket{f}^{h+1}\ket{0}^{1}\ket{0}^{2t} \longmapsto \\ & \qquad \qquad \longmapsto \sum_{z = 2^{h-1}}^{2^{h}-1}\ket{z}^h \left(\sum_{g=0}^1\alpha_{(2z+g)}\ket{g}^1\ket{0}^{1}\ket{T_{l(2z+g),\,2\dot d(2z+g)}}^{t}\,\ket{T_{l(2z+g),\,2\dot d(2z+g)+1}}^{t}\right).
\end{align*}

Since $2^{h-1} \leq i \leq 2^h -1$ and $g \in [0,1]$, we have that $ \lfloor\log_2(2z+g)\rfloor = \log_2(2z)$, so:
\begin{equation*}
   l(f) =  l(2z+g) = \lfloor\log_2(2z+g)\rfloor + 1 = \log_2(2z) + 1  = 1 + \underbrace{ \log_2(z) + 1}_{(\clubsuit)} = 1 + l(z).
\end{equation*}
From the induction hypothesis $P(h)$, we have that $l(z)= h$, and thus $l(2z+g) = l(z) +1 = h + 1 = l(f)$.

Next, we observe that:
\begin{equation*}
    d(2z+g) = 2z + g - 2^{\lfloor\log_2(2z+g)\rfloor} = 2\underbrace{(z - 2^{\lfloor\log_2(z)\rfloor})}_{(\spadesuit)}+g  = 2d(z)+g = f - 2^{\lfloor\log_2(f)\rfloor}.
\end{equation*}
Thus, 
\begin{equation*}
    \ket{T_{l(f),\,2\dot d(f)}}^{t}\,\ket{T_{l(f),\,2\dot d(f)+1}}^{t} = \ket{T_{l(2z+g),\,2\dot d(2z+g)}}^{t}\,\ket{T_{l(2z+g),\,2\dot d(2z+g)+1}}^{t}
\end{equation*}
for both the superposition indexed by $g = 0$ and $g = 1$. 
Therefore $P(h+1)$ holds, and by induction $P(h)$ is true for every $h\le k$.
\end{itemize}

\paragraph{(\ref{item:signs})  \textbf{Superposition of signs}.}
With the address register set to superposition $\sum_{z = 0}^{K}\alpha_z\ket{z}^{k}$, where $\alpha_z \neq 0$, we access in superposition every memory cell $\{\ket{L_z}\}_{z=0}^{K-1}$ in the BBQRAM. Since the leftmost qubit of each $\ket{L_z}$ encodes the sign $\mathrm{s}(a_z)$, the retrieval of only this qubit into the working register yields the state $\frac{1}{\sqrt{2^k}}\sum_{z=0}^{K-1} \ket{z}^{k}\ket{\mathrm{s}(a_{z})}^{1}\ket{0}^{t}\ket{0}^{t}$.
\end{proof}

{We observe that, in Proposition}~\ref{prop:mapping} {, we assign all sibling pairs at height $h$ of the segment tree to the contiguous BBQRAM address range $[2^{h-1},\,2^{h}-1]$, forming a sequence of memory partitions that double in size at each level, where each
partition contains exactly the sibling nodes at height $h$ of the segment tree. These memory addresses share the common binary prefix $0^{k-h}1$, where $k = \log_2 K$ denotes the depth of the BBQRAM, so that all addresses within each partition share a common prefix. This property is part of what enables the state preparation algorithm to prepare the address register efficiently for the next retrieval, as we show in Section}~\ref{sec:state_prep_alg}{. Additionally, packing two sibling nodes into a single memory cell allows the retrieval of both operands that $U_{2CR}$ (Section}~\ref{sec:u2cr}{) requires in a single query, halving the number of BBQRAM queries per level.}

\subsection{Efficient State Preparation Algorithm}\label{sec:state_prep_alg}
We present an algorithm for efficiently preparing a quantum state that encodes a real-valued matrix $A \in \mathbb{R}^{M \times N}$ using the memory layout of the segment tree in BBQRAM introduced in Section~\ref{sec:memory_layout}. We describe the algorithm step-by-step, highlight the sequence of retrievals and unitary operations, and analyze the computational cost of each stage. Theorem~\ref{theo:efficient_state_prep} demonstrates how classical preprocessing, hierarchically organized quantum memory, and recursive amplitude encoding enable polylogarithmic time state preparation for a matrix.

\begin{theorem}[Efficient State preparation with BBQRAM]\label{theo:efficient_state_prep}
Let $A\in \mathbb{R}^{M\times N}$, where $M = 2^m, N=2^n$ and $m, n \in \mathbb{N}$. Let $T$ be the segment tree of squared norms of $A$ with depth $k = \log_2 K$, where $K = MN$. Consider the memory layout of $T$ in a BBQRAM as described in Proposition~\ref{prop:mapping}, such that each node of $T$ is basis encoded in the memory cells $\{\ket{L_z}\}_{z=0}^{K-1}$ using $t$ bits for fixed-point representation. Then, there exists a unitary operator $E_A$ that maps the matrix $A$ into a quantum register of $\Theta(\log_2(MN))$ qubits in $\mathcal{O}(\log_2^2(MN))$ time:
\begin{equation*}
    E_A: \ket{0}^{m+n} \longmapsto \frac{1}{\|A\|_F}\sum_{i=0}^{M-1}\sum_{j=0}^{N-1} a_{i,j} \ket{i}^m \ket{j}^n ,
\end{equation*}
where $\|A\|_F$ is the Frobenius norm of $A$, and $\ket{i}^m$ and $\ket{j}^n$ denote the basis encodings of the row and column indices $i$ and $j$ of the entry  $a_{i,j} \in A$, respectively.
\end{theorem}
\begin{proof}[Theorem~\ref{theo:efficient_state_prep} Correctness and Complexity Analysis]

Given a BBQRAM with $K$ memory cells that encode the segment tree $T$ of depth $k = \log_2 K$, where each internal node of $T$ is represented using $t$ bits, the initial state is:
$$
\ket{0}^{1}_{\mathrm{s}} \ket{0}^{t}_{\mathrm{l}} \ket{0}^{t}_{\mathrm{r}} \ket{0}^{1}_{\mathrm{v}} \ket{0}^{k}_{\mathrm{a}},
$$
where:
\begin{itemize}
    \item the $1$-qubit register $\ket{0}^{1}_{\mathrm{s}}$ is for the sign bit;
    \item the $t$-qubit registers, respectively $\ket{0}^{t}_{\mathrm{l}}$ and $\ket{0}_{\mathrm{r}}^{t}$, are working registers that store the retrieved values from the memory cells $\{\ket{L_z}\}_{z=0}^{K-1}$ of the BBQRAM; 
    \item the $1$-qubit registers $\ket{0}^{1}_{\mathrm{v}}$ is another working register;
    \item the $k$-qubit address register $\ket{0}^{k}_{\mathrm{a}}$ sets the access path to retrieve the needed memory cells and it will encode the matrix $A\in \mathbb{R}^{M\times N}$ in the desidered quantum state a the end of the procedure.
\end{itemize}

For improved readability, we omit superscripts indicating the size of quantum registers, except where ambiguity may arise. Then, we write:
\begin{align*}
\ket{0}_{\mathrm{s}}\ket{0}_{\mathrm{l}}\ket{0}_{\mathrm{r}}\ket{0}_{\mathrm{v}}\ket{0}_{\mathrm{a}}.
\end{align*}

\noindent
As first step, we retrieve the sibling nodes $\ket{T_{1,0}}\ket{T_{1,1}}$ at level $h=1$ of $T$ from BBQRAM. Thus, we set the address register to state $\ket{0}^{ k-1}_{\mathrm{a}}\ket{1}_{\mathrm{a}}$ such to access the memory cell $\ket{L_1}$ (see Proposition~\ref{prop:mapping}):
\begin{align*}
\ket{0}_{\mathrm{s}}\ket{0}_{\mathrm{l}}\ket{0}_{\mathrm{r}}\ket{0}_{\mathrm{v}}\ket{0}^{ k-1}_{\mathrm{a}}\ket{1}_{\mathrm{a}}.
\end{align*}

\noindent
After the retrieval, we yield the following state:
\begin{align*}
\ket{0}_{\mathrm{s}}\ket{T_{1,0}}_{\mathrm{l}}\ket{T_{1,1}}_{\mathrm{r}}\ket{0}_{\mathrm{v}}\ket{0}^{ k-1}_{\mathrm{a}}\ket{1}_{\mathrm{a}}.
\end{align*}

\noindent
Then, we apply the unitary $U_{2CR}$ on $\ket{T_{1,0}}_{\mathrm{l}}\ket{T_{1,1}}_{\mathrm{r}}$ and targeting $\ket{0}_{\mathrm{v}}$ (see Equation~\ref{eq:u2cr}):
\begin{align*}
\ket{0}_{\mathrm{s}}\ket{T_{1,0}}_{\mathrm{l}}\ket{T_{1,1}}_{\mathrm{r}}\left(\sqrt{\frac{T_{1,0}}{T_{1,0}+T_{1,1}}}\ket{0}_{\mathrm{v}} +\sqrt{\frac{T_{1,1}}{T_{1,0}+T_{1,1}}}\ket{1}_{\mathrm{v}} \right)\ket{0}^{ k-1}_{\mathrm{a}}\ket{1}_{\mathrm{a}}.
\end{align*}

\noindent
By recalling Definition~\ref{def:segment_tree}, we observe that $T_{1,0}+T_{1,1} = T_{0,0}$, and  $\sqrt{T_{0,0}} = \|A\|_F$. Therefore, we rewrite the state as follows:
\begin{align*}
    \frac{1}{\|A\|_F}\ket{0}_{\mathrm{s}}\ket{T_{1,0}}_{\mathrm{l}}\ket{T_{1,1}}_{\mathrm{r}} \left(\sqrt{T_{1,0}}\ket{0}_{\mathrm{v}} +\sqrt{T_{1,1}}\ket{1}_{\mathrm{v}} \right)\ket{0}^{ k-1}_{\mathrm{a}}\ket{1}_{\mathrm{a}}.
\end{align*}

\noindent
Then, we uncompute the registers $\ket{T_{1,0}}_{\mathrm{l}}\ket{T_{1,1}}_{\mathrm{r}}$:
\begin{align*}
    \frac{1}{\|A\|_F}\ket{0}_{\mathrm{s}}\ket{0}_{\mathrm{l}}\ket{0}_{\mathrm{r}} \left(\sqrt{T_{1,0}}\ket{0}_{\mathrm{v}} +\sqrt{T_{1,1}}\ket{1}_{\mathrm{v}} \right)\ket{0}^{ k-1}_{\mathrm{a}}\ket{1}_{\mathrm{a}}.
\end{align*}

\noindent
Next step consists of retrieving the pairs of sibling nodes at level $h=2$ of $T$ in superposition. To accomplish this, we first prepare the address register in the appropriate superposition, as described in Proposition~\ref{prop:mapping}. Specifically, we perform a left circular shift on the working register $\mathrm{v}$ and the address register $\mathrm{a}$:
\begin{align*}
    \frac{1}{\|A\|_F} \ket{0}_{\mathrm{s}}\ket{0}_{\mathrm{l}}\ket{0}_{\mathrm{r}} \ket{0}_{\mathrm{v}}\ket{0}^{ k-2}_{\mathrm{a}}\ket{1}_{\mathrm{a}}\left(\sqrt{T_{1,0}}\ket{0}_{\mathrm{a}} +\sqrt{T_{1,1}}\ket{1}_{\mathrm{a}} \right),
\end{align*}
which we rewrite as
\begin{align*}
    \frac{1}{\|A\|_F}  \ket{0}_{\mathrm{s}} 
\left(\sqrt{T_{1,0}} \ket{0}_{\mathrm{l}}\ket{0}_{\mathrm{r}} \ket{0}_{\mathrm{v}} \ket{0}^{ k-2}_{\mathrm{a}}\ket{1}_{\mathrm{a}}\ket{0}_{\mathrm{a}} +\sqrt{T_{1,1}}\ket{T_{0}}_{\mathrm{l}}\ket{T_{0}}_{\mathrm{r}} \ket{0}_{\mathrm{v}} \ket{0}^{ k-2}_{\mathrm{a}}\ket{1}_{\mathrm{a}}\ket{1}_{\mathrm{a}}
\right).
\end{align*}

{To access the memory cells at the next tree level in superposition, the algorithm must prepare the address register $a$ in a specific coherent superposition. A naive approach treats this as an independent arbitrary state preparation problem at each of the $\log_2 K$ levels, incurring an $O(K)$ cost per level that nullifies the polylogarithmic advantage of the QRAM. Our memory layout avoids this bottleneck through the combination of $U_{2CR}$ and the left circular shift. The $U_{2CR}$ operator computes the amplitudes from the retrieved sibling nodes and encodes them into the single working qubit $v$. The left circular shift then moves $v$ into the address register $a$, setting it in the correct superposition for querying the subsequent level. This reduces what would otherwise be an arbitrary state preparation problem into an operation based on SWAP gates, whose exact complexity we analyze at the end of this section.}

Given that the address register $\mathrm{a}$ is now in superposition, we can retrieve from the BBQRAM the pairs $\ket{T_{2,0}}\ket{T_{2,1}}$ and $\ket{T_{2,2}}\ket{T_{2,3}}$ in superposition and load them into the working registers.
The quantum state after this step is:
\noindent
\begin{align*}
    \frac{1}{\|A\|_F} \ket{0}_{\mathrm{s}} 
\left(\sqrt{T_{1,0}} \ket{T_{2,0}}_{\mathrm{l}}\ket{T_{2,1}}_{\mathrm{r}} \ket{0}_{\mathrm{v}} \ket{0}^{ k-2}_{\mathrm{a}}\ket{1}_{\mathrm{a}}\ket{0}_{\mathrm{a}} +\sqrt{T_{1,1}}\ket{T_{2,2}}_{\mathrm{l}}\ket{T_{2,3}}_{\mathrm{r}} \ket{0}_{\mathrm{v}} \ket{0}^{ k-2}_{\mathrm{a}}\ket{1}_{\mathrm{a}}\ket{1}_{\mathrm{a}}
\right).
\end{align*}

\noindent
As before, we apply $U_{2CR}$ to the superpositions of the quantum registers $\mathrm{l}$, and $\mathrm{r}$, and $\mathrm{v}$:
\begin{align*}
& \frac{1}{\|A\|_F} \ket{0}_{\mathrm{s}} \Bigg( 
    \sqrt{T_{1,0}}\, \ket{T_{2,0}}_{\mathrm{l}} \ket{T_{2,1}}_{\mathrm{r}}
    \left(
        \sqrt{\frac{T_{2,0}}{T_{2,0} + T_{2,1}}}\ket{0}_{\mathrm{v}}
        + \sqrt{\frac{T_{2,1}}{T_{2,0} + T_{2,1}}}\ket{1}_{\mathrm{v}}
    \right)
    \ket{0}^{ k-2}_{\mathrm{a}} \ket{1}_{\mathrm{a}} \ket{0}_{\mathrm{a}} +
    \\
& \hspace{4em}
    + \sqrt{T_{1,1}}\, \ket{T_{2,2}}_{\mathrm{l}} \ket{T_{2,3}}_{\mathrm{r}}
    \left(
        \sqrt{\frac{T_{2,2}}{T_{2,2} + T_{2,3}}}\ket{0}_{\mathrm{v}}
        + \sqrt{\frac{T_{2,3}}{T_{2,2} + T_{2,3}}}\ket{1}_{\mathrm{v}}
    \right)
    \ket{0}^{ k-2}_{\mathrm{a}} \ket{1}_{\mathrm{a}} \ket{1}_{\mathrm{a}}
    \Bigg).
\end{align*}

\noindent
Recalling that $T_{1,0} = T_{2,0} + T_{2,1}$ and $T_{1,1} = T_{2,2} + T_{2,3}$, we rewrite the state as
\begin{align*}
\frac{1}{\|A\|_F} \ket{0}_{\mathrm{s}} 
\left( \ket{T_{2,0}}_{\mathrm{l}}\ket{T_{2,1}}_{\mathrm{r}} \left(\sqrt{T_{2,0}}\ket{0}_{\mathrm{v}} +\sqrt{T_{2,1}}\ket{1}_{\mathrm{v}} \right) \ket{0}^{ k-2}_{\mathrm{a}}\ket{1}_{\mathrm{a}}\ket{0}_{\mathrm{a}} +\ket{T_{2,2}}_{\mathrm{l}}\ket{T_{2,3}}_{\mathrm{r}} \left(\sqrt{T_{2,2}}\ket{0}_{\mathrm{v}} +\sqrt{T_{2,3}}\ket{1}_{\mathrm{v}} \right) \ket{0}^{ k-2}_{\mathrm{a}}\ket{1}_{\mathrm{a}}\ket{1}_{\mathrm{a}}
\right).
\end{align*}

\noindent
Then, we uncompute the registers $\mathrm{l}$, and $\mathrm{r}$:
\begin{align*}
\frac{1}{\|A\|_F}\ket{0}_{\mathrm{s}} \ket{0}_{\mathrm{l}}\ket{0}_{\mathrm{r}}\left( \left(\sqrt{T_{2,0}}\ket{0}_{\mathrm{v}} +\sqrt{T_{2,1}}\ket{1}_{\mathrm{v}} \right) \ket{0}^{ k-2}_{\mathrm{a}}\ket{1}_{\mathrm{a}}\ket{0}_{\mathrm{a}} +\left(\sqrt{T_{2,2}}\ket{0}_{\mathrm{v}} +\sqrt{T_{2,3}}\ket{1}_{\mathrm{v}} \right) \ket{0}^{ k-2}_{\mathrm{a}}\ket{1}_{\mathrm{a}}\ket{1}_{\mathrm{a}}
\right).
\end{align*}

The same sequence of steps applies for each subsequent retrieval. Before the last $k$-th retrieval, the quantum state is
\begin{align*}
\frac{1}{\|A\|_F} \ket{0}_{\mathrm{s}} \ket{0}_{\mathrm{l}}\ket{0}_{\mathrm{r}}\ket{0}_{\mathrm{v}}
\sum_{z=0}^{2^{k-1}-1} \sqrt{T_{k-1,z}}
\ket{1}_{\mathrm{a}}\ket{z}_{\mathrm{a}}^{k-1}.
\end{align*}

\noindent
After the retrieval, the quantum state becomes 
\begin{align*}
\frac{1}{\|A\|_F} \ket{0}_{\mathrm{s}} \sum_{z=0}^{2^{k-1}-1}\sqrt{T_{k-1,z}}\ket{T_{k, 2d(z)}}_{\mathrm{l}}\ket{T_{k, 2d(z)+1}}_{\mathrm{r}}\ket{0}_{\mathrm{v}}
\ket{1}_{\mathrm{a}}\ket{z}_{\mathrm{a}}.
\end{align*}

\noindent
As previously, we apply the unitary $U_{2CR}$ to the registers $\mathrm{l}$, $\mathrm{r}$, and $\mathrm{v}$ in superposition, yielding the state
\begin{align*}
\frac{1}{\|A\|_F} \ket{0}_{\mathrm{s}} \sum_{z=0}^{2^{k-1}-1}\sqrt{T_{k-1,z}}\ket{T_{k, 2d(z)}}_{\mathrm{l}}\ket{T_{k, 2d(z)+1}}_{\mathrm{r}}
\left(\sqrt{\frac{T_{k, 2d(z)}}{T_{k-1,z}}}\ket{0}_{\mathrm{v}} +\sqrt{\frac{T_{k, 2d(z)+1}}{T_{k-1,z}}}\ket{1}_{\mathrm{v}} \right) 
\ket{1}_{\mathrm{a}}\ket{z}_{\mathrm{a}}.
\end{align*}

\noindent
Next, we uncompute the registers $\mathrm{l}$ and $\mathrm{r}$:
\begin{align*}
\frac{1}{\|A\|_F} \ket{0}_{\mathrm{s}} \sum_{z=0}^{2^{k-1}-1}\ket{0}_{\mathrm{l}}\ket{0}_{\mathrm{r}}
\left(\sqrt{T_{k, 2d(z)}}\ket{0}_{\mathrm{v}} +\sqrt{T_{k, 2d(z)+1}}\ket{1}_{\mathrm{v}} \right) 
\ket{1}_{\mathrm{a}}\ket{z}_{\mathrm{a}} ,
\end{align*}
\begin{algorithm2e}[!t]
\caption{Efficient Quantum State Preparation via BBQRAM}
\label{alg:state_prep}
\KwIn{
    $A \in \mathbb{R}^{M \times N}$, a real-value matrix to encode in a quantum state.
}
\KwOut{
    $\displaystyle \frac{1}{\|A\|_F} \sum_{i=0}^{M-1} \sum_{j=0}^{N-1} a_{i,j} \ket{i}^m \ket{j}^n$, where $a_{i,j} \in A$.
}
\BlankLine

\textbf{Build the Segment Tree:} Construct the segment tree $T$ from $A$, with depth $k = \log_2 K$, where $K = MN$\;

\textbf{Map to BBQRAM:} Store $T$ in BBQRAM using the memory layout described in Proposition~\ref{prop:mapping}\;

\textbf{Initialize Quantum Registers:} Prepare the quantum registers in the state $\ket{0}_{\mathrm{s}} \ket{0}_{\mathrm{l}} \ket{0}_{\mathrm{r}} \ket{0}_{\mathrm{v}} \ket{0}^{k}_{\mathrm{a}}$, where:\;
\begin{itemize}
    \item $\mathrm{s}$: 1 qubit for the sign of $a_{i,j}$,
    \item $\mathrm{l}, \mathrm{r}$: $t$ qubits each for left/right child values,
    \item $\mathrm{v}$: 1 qubit for amplitude encoding,
    \item $\mathrm{a}$: $k$ address qubits.
\end{itemize}
\nl \textbf{Set Address Register:} Set the address register $\mathrm{a}$ to $\ket{0}^{k-1}\ket{1}$\;
\BlankLine
\nl\For{$h = 1$ \KwTo $k$}{
    \nl\textbf{Retrieve Sibling Nodes:} In superposition, retrieve from BBQRAM the pairs of sibling nodes at level $h$ of $T$ (memory cells $\{\ket{L_z}\}_{z=2^{h-1}}^{2^h-1}$) into registers $\mathrm{l}$ and $\mathrm{r}$.
    
    \nl\textbf{Amplitude Encoding:} Apply the unitary $U_{2CR}$ to registers $\mathrm{l}$, $\mathrm{r}$, and $\mathrm{v}$, producing the superposition:
    
    $$
    \sqrt{\frac{T_{l(z),\,2\dot d(z)}}{T_{l(z),\,2\dot d(z)}+T_{l(z),\,2\dot d(z)+1}}} \ket{0}_{\mathrm{v}} +
    \sqrt{\frac{T_{l(z),\,2\dot d(z)+1}}{T_{l(z),\,2\dot d(z)}+T_{l(z),\,2\dot d(z)+1}}} \ket{1}_{\mathrm{v}},
    $$
    where $T_{l(z),\,2\dot d(z)}$ and $T_{l(z),\,2\dot d(z)+1}$ are the values of the left and right children, respectively\;
    \nl\textbf{Uncompute Working Registers:} Uncompute the registers $\mathrm{l}$ and $\mathrm{r}$\;
    \nl\textbf{Circular Shift:} Perform a left circular shift on registers $\mathrm{v}$ and $\mathrm{a}$ to prepare the address for the next iteration\;
}
\BlankLine
\nl\textbf{Sign Retrieval:} Retrieve in superposition the sign qubits $\ket{\mathrm{s}(a_z)}_{\mathrm{s}}$ from memory cells $\{\ket{L_z}\}_{z=0}^{K-1}$ into register $\mathrm{s}$\;
\nl\textbf{Phase Correction:} Apply a controlled-$Z$ gate with register $\mathrm{s}$ as the control and register $\mathrm{v}$ as the target qubit\;
\nl\textbf{Clean Up:} Trace out the working registers $\mathrm{s}$, $\mathrm{l}$, $\mathrm{r}$, and $\mathrm{v}$\;
\nl\Return{$\displaystyle  \frac{1}{\|A\|_F}\sum_{i=0}^{M-1}\sum_{j=0}^{N-1} a_{i,j} \ket{i}_{\mathrm{a}}^{m} \ket{j}_{\mathrm{a}}^{n}$}
\end{algorithm2e}
\noindent
and we perform the left circular shift on the working register $\mathrm{v}$ and the address register $\mathrm{a}$:
\begin{align*}
\frac{1}{\|A\|_F}\ket{0}_{\mathrm{s}} \sum_{z=0}^{2^{k-1}-1}\ket{0}_{\mathrm{l}}\ket{0}_{\mathrm{r}}
\ket{1}_{\mathrm{v}}\ket{z}_{\mathrm{a}} \left(\sqrt{T_{k, 2d(z)}}\ket{0}_{\mathrm{a}} +\sqrt{T_{k, 2d(z)+1}}\ket{1}_{\mathrm{a}}
\right) = \frac{1}{\|A\|_F} \sum_{z=0}^{2^{k}-1}\ket{0}_{\mathrm{s}}\ket{0}_{\mathrm{l}}\ket{0}_{\mathrm{r}}
\ket{1}_{\mathrm{v}}\sqrt{T_{k, z}}\ket{z}_{\mathrm{a}}, 
\end{align*}
and since $\sqrt{T_{k,z}} = |a_z|$, the state corresponds to:
\begin{align*}
\frac{1}{\|A\|_F}  \sum_{z=0}^{2^{k}-1}\ket{0}_{\mathrm{s}}\ket{0}_{\mathrm{l}}\ket{0}_{\mathrm{r}}
\ket{1}_{\mathrm{v}}\,\lvert a_z\rvert \,\ket{z}_{\mathrm{a}}.
\end{align*}
Then, we retrieve in superposition the sign qubits $\ket{\mathrm{s}(a_{z})}$ from each memory cell $\ket{L_z}$, for $0 \leq z < K$, where each sign is associated with the corresponding amplitude $a_z$:
\begin{align*}
\frac{1}{{\|A\|_F}} \sum_{z=0}^{2^{k}-1}\ket{\mathrm{s}(a_{z})}_{\mathrm{s}}\ket{0}_{\mathrm{l}}\ket{0}_{\mathrm{r}}
\ket{1}_{\mathrm{v}}\,\lvert a_z\rvert \,\ket{z}_{\mathrm{a}}^k.
\end{align*}

At this point, the register $\mathrm{v}$ is already in the state $\ket{1}_{\mathrm{v}}$. We now apply a controlled-$Z$ ($CZ$) gate, where the sign qubit $\mathrm{s}$ acts as the control qubit and $\mathrm{v}$ as the target qubit. This operation multiplies by $-1$ the amplitude $\vert a_z \rvert$ whenever $\ket{\mathrm{s}(a_{z})}_{\mathrm{s}} = \ket{1}$. The quantum state after this step becomes:
\begin{align*}
\frac{1}{{\|A\|_F}} \sum_{z=0}^{2^{k}-1}\ket{\mathrm{s}(a_{z})}_{\mathrm{s}}\ket{0}_{\mathrm{l}}\ket{0}_{\mathrm{r}}
\ket{1}_{\mathrm{v}}\, \mathrm{s}(a_{z})\cdot \vert a_z \rvert \,\ket{z}_{\mathrm{a}}^k = \frac{1}{{\|A\|_F}} \sum_{z=0}^{2^{k}-1}\ket{\mathrm{s}(a_{z})}_{\mathrm{s}}\ket{0}_{\mathrm{l}}\ket{0}_{\mathrm{r}}
\ket{1}_{\mathrm{v}}\,a_z\,\ket{z}_{\mathrm{a}}^k.
\end{align*}

\noindent
Finally, we uncompute the working registers $\mathrm{s}$, $\mathrm{l}$, $\mathrm{r}$, and $\mathrm{v}$ and we observe that $2^k$ equals $MN$ and $k = m + n = \log_2(MN)$:
\begin{align*}
\frac{1}{{\|A\|_F}} \sum_{z=0}^{MN-1}\ket{0}_{\mathrm{s}}\ket{0}_{\mathrm{l}}\ket{0}_{\mathrm{r}}
\ket{0}_{\mathrm{v}}\, a_z\,\ket{z}^k_{\mathrm{a}} =  \frac{1}{\|A\|_F}\sum_{i=0}^{M-1}\sum_{j=0}^{N-1} \ket{0}_{\mathrm{s}}\ket{0}_{\mathrm{l}}\ket{0}_{\mathrm{r}}
\ket{0}_{\mathrm{v}}\,  a_{i,j} \,\ket{i}_{\mathrm{a}}^{m} \ket{j}_{\mathrm{a}}^{n},
\end{align*}

Algorithm~\ref{alg:state_prep} summarizes these steps. We analyze the total cost of the quantum state preparation algorithm by counting both the number of quantum memory retrievals and the complexity of the involved unitary operations at each level of the segment tree $T$. Given that $T$ has depth $k = \log_2 K$, the algorithm performs exactly $k$ retrievals (i.e., one for each level $h \in [1, k]$) to access all pairs of sibling nodes in superposition. As defined by Lemma~\ref{lemma:retrieval_cost}, each retrieval from BBQRAM requires $\mathcal{O}(\log_2 K)$ time.
\begin{figure*}[t]
    \centering
    \[
    \Qcircuit @C=2.5em @R=2.0em {
        & \lstick{\ket{\gamma_z}_{\mathrm{v}}} & \qswap               & \qw                & \qw                & \qw \\
        & \lstick{\ket{0}}        & \qw                  & \qw                & \qw                & \qw \\
        & \lstick{\ket{0}}        & \qw                  & \qw                & \qw                & \qw \\
        & \smash[b]{\vdots}       &                      &                    &                    & \smash[b]{\vdots}   \\
        & \lstick{\ket{0}}        & \qswap \qwx[-4]      & \qswap             & \qw                & \qw \inputgroupv{2}{5}{0.1em}{2.8em}{\ket{0}_{\mathrm{a}}^{k-h}}\\
        & \lstick{\ket{z_5}}      & \qswap               & \qw                & \qw                & \qw \\
        & \lstick{\ket{z_4}}      & \qswap \qwx          & \qswap \qwx[-2]    & \qswap             & \qw \\
        & \lstick{\ket{z_3}}      & \qswap               & \qw                & \qw                & \qw \inputgroupv{6}{11}{1em}{4.8em}{\ket{z}_{\mathrm{a}}^{h}}\\
        & \lstick{\ket{z_2}}      & \qswap \qwx          & \qswap             & \qw                & \qw \\
        & \lstick{\ket{z_1}}      & \qswap               & \qw                & \qw                & \qw \\
        & \lstick{\ket{z_0}}      & \qswap \qwx          & \qswap \qwx[-2]    & \qswap \qwx[-4]    & \qw \\
    }
    \]
    \caption{Tree of SWAP gates quantum circuit that performs a left circular shift on the working register $\mathrm{v}$ and the address register $\mathrm{a}$, as used in the state preparation algorithm. The circuit achieves a depth of $\mathcal{O}(\log_2 h)$ for $h$ qubits. For simplicity, we assume working with a number of qubits equal to a power of two.}\Description{}
    \label{fig:swap_circuit}
\end{figure*}
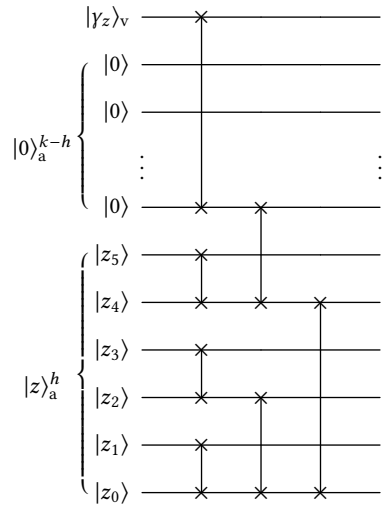
and, we trace out the working qubits, yielding the final state encoding the matrix $A\in \mathbb{R}^{M\times N}$:
\begin{align*}
 \frac{1}{\|A\|_F}\sum_{i=0}^{M-1}\sum_{j=0}^{N-1} a_{i,j} \ket{i}_{\mathrm{a}}^{m} \ket{j}_{\mathrm{a}}^{n}.
\end{align*}

After each retrieval, the algorithm applies the unitary $U_{2CR}$. {As discussed in Section}~\ref{sec:u2cr} {, the cost introduced by $U_{2CR}$ strictly depends on the fixed precision $t$, which is independent of the size of the matrix $A$. Therefore, the overhead of $U_{2CR}$ is  $\tilde{O}(1)$, leaving the overall asymptotic cost unaffected.} Subsequently, we uncompute the working registers $\mathrm{l}$ and $\mathrm{r}$ by applying the conjugate transpose of the unitary implementing the retrieval operation, which also requires $\mathcal{O}(\log_2 K)$ time. The algorithm then updates the address and working registers for the next iteration using a left circular shift. In general, before each shift, the state is:
$$\frac{1}{\|A\|_F}\ket{0}_{\mathrm{s}}\ket{0}_{\mathrm{l}}\ket{0}_{\mathrm{r}} \sum^{2^h-1}_{z=2^{h-1}}\ket{\gamma_z}_{\mathrm{v}}\ket{0}_{\mathrm{a}}^{k-h}\ket{z}^{h}_{\mathrm{a}},$$ for $h \in [1,k]$, where $\ket{\gamma_z} =  \sqrt{T_{l(z), 2d(z)}}\ket{0}_{\mathrm{a}} +\sqrt{T_{l(z), 2d(z)+1}}\ket{1}_{\mathrm{a}}$. After the shift, the state becomes 
\begin{equation*}
\begin{cases}
    \frac{1}{\|A\|_F}\ket{0}_{\mathrm{s}}\ket{0}_{\mathrm{l}}\ket{0}_{\mathrm{r}} \sum^{2^h-1}_{z=2^{h-1}}\ket{0}_{\mathrm{v}}\ket{0}_{\mathrm{a}}^{k-h-1}\ket{z}^{h}_{\mathrm{a}}\ket{\gamma_z}^1_{\mathrm{a}} &\textrm{if }  1 \leq h < k .\\[1ex]
    \frac{1}{\|A\|_F}\ket{0}_{\mathrm{s}}\ket{0}_{\mathrm{l}}\ket{0}_{\mathrm{r}} \sum^{2^{(k-1)}-1}_{z=2^{k}}\ket{1}_{\mathrm{v}}\ket{z}^{k-1}_{\mathrm{a}}\ket{\gamma_z}^1_{\mathrm{a}} &\textrm{ otherwise .}\\[1ex]
\end{cases}
\end{equation*}

We can implement this shift either with a quantum circuit of depth $\mathcal{O}(h)$ using a sequence of SWAP gates, or with a tree-like arrangement of SWAP gates with depth $\mathcal{O}(\log_2 h)$, as depicted in Figure~\ref{fig:swap_circuit}. 

Therefore, for an iteration $h \in [1, k]$, the costs consist of: QRAM retrieval ($\mathcal{O}(\log_2 K)$), uncomputation of the retrieved data ($\mathcal{O}(\log_2 K)$), and preparation of the address register via left circular shift ($\mathcal{O}(\log_2 h)$). Since $\mathcal{O}(\log_2 h) < \mathcal{O}(\log_2\log_2 K) = \mathcal{O}(\log_2 k)$, and the address register preparation can be performed in parallel with the uncomputation step --- because they act on disjoint sets of qubits --- the total cost per iteration at level $h$ is $\mathcal{O}(\log_2 K)$. With such iterations $k = \log_2 K$, the total cost of retrieving all sibling nodes at all levels of $T$ is $\mathcal{O}(\log_2^2 K)$.
{This overall complexity relies on the memory layout of Proposition}~\ref{prop:mapping}{. An arbitrary mapping incurs an $O(K)$ overhead at each tree level, since it requires an independent address preparation with no exploitable structure. The proposed layout avoids this by placing all sibling pairs at a given height in a contiguous address range, which allows the $U_{2CR}$ and left circular shift to prepare the address register incrementally at a cost of $O(\log_2 h)$ per level. Simultaneously, packing two sibling nodes into a single memory cell ensures that the algorithm retrieves both operands in a single query.}

Finally, the algorithm includes a single retrieval step to access the sign qubits in superposition, which also costs $\mathcal{O}(\log_2 K)$. In this step, the $U_{2CR}$ operation is not required; instead, a single controlled-$Z$ gate is applied, which has constant cost $\mathcal{O}(1)$. Thus, the sign retrieval does not change the overall asymptotic complexity.

{Regarding classical computation costs, constructing the segment tree requires $O(MN)$ space and $O(MN)$ time. Since this construction occurs only once for the BBQRAM, the time cost of initialization amortizes over multiple state preparations of the same input matrix and becomes negligible.}

In summary, the dominant contribution to the total cost comes from the $k$ rounds of sibling node retrievals, resulting in an overall state preparation complexity of $\mathcal{O}(\log_2^2 K) = \mathcal{O}(\log^2_2(MN))$ for a matrix $A \in \mathbb{R}^{M \times N}$, where $K = MN$. 
\end{proof}

\begin{figure*}[!t]
    \centering
    \begin{subfigure}[t]{\textwidth}
        \centering
        \resizebox{\linewidth}{!}{%
          \begin{tikzpicture}[
              treenode/.style={
                rectangle,
                rounded corners=2pt,            
                draw,
                fill=white,
                drop shadow={shadow xshift=1pt,shadow yshift=-1pt,opacity=0.5},
                minimum width=6mm,
                minimum height=6mm,
                inner sep=2pt,
                align=center,
              },
              leaf/.style={
                shape=rectangle split,
                rectangle split parts=2,
                rectangle split draw splits,
                rectangle split horizontal,
                rectangle split part fill={neutral-left,PastelIBMOrange},
                rounded corners=2pt,            
                draw,
                minimum width=6mm,
                minimum height=6mm,
                inner sep=2pt,
                text width=6mm,
                align=center,
              },
              levelzero/.style={treenode, fill=PastelIBMPurple},
              levelone/.style={treenode, fill=PastelIBMBlue},
              leveltwo/.style={treenode, fill=PastelIBMPink},
              levelthree/.style={treenode},
              hlabel/.style={
                draw=none,
                fill=none,
                anchor=east,
              },
              level distance=12mm,
              level 1/.style={sibling distance=64mm},
              level 2/.style={sibling distance=32mm},
              level 3/.style={sibling distance=16mm},
            ]
            \foreach \lvl in {0,1,2,3}{
              \draw[dotted] (-60mm,-\lvl*12mm) -- ( 60mm,-\lvl*12mm);
              \node[hlabel] at (-62mm,-\lvl*12mm) {$h=\lvl$};
            }
            \begin{scope}[xshift=5mm]
              \node[treenode, levelzero] {$32.48$}
                child {
                  node[treenode, levelone] {$24.89$}
                  child {
                    node[treenode, leveltwo] {$14.45$}
                    child { node[leaf] {$0$\nodepart{two}$4.84$} }
                    child { node[leaf] {$0$\nodepart{two}$9.61$} }
                  }
                  child {
                    node[treenode, leveltwo] {$10.44$}
                    child { node[leaf] {$1$\nodepart{two}$9.00$} }
                    child { node[leaf] {$0$\nodepart{two}$1.44$} }
                  }
                }
                child {
                  node[treenode, levelone] {$7.59$}
                  child {
                    node[treenode, leveltwo] {$1.09$}
                    child { node[leaf] {$0$\nodepart{two}$0.09$} }
                    child { node[leaf] {$0$\nodepart{two}$1.00$} }
                  }
                  child {
                    node[treenode, leveltwo] {$6.50$}
                    child { node[leaf] {$0$\nodepart{two}$0.25$} }
                    child { node[leaf] {$1$\nodepart{two}$6.25$} }
                  }
                };
            \end{scope}
          \end{tikzpicture}%
        }
        \caption{}\Description{}
        \label{fig:segment_tree_colored_example}
    \end{subfigure}

  \vspace{0.25cm}

  \begin{subfigure}[t]{\textwidth}
    \centering
    \resizebox{\linewidth}{!}{%
      \begin{tikzpicture}[
          every label/.append style={font=\huge},
          node/.style={
            draw, fill=white, circle, inner sep=0pt, font=\huge,
            drop shadow, text width=3em, align=center
          },
          levelzeroq/.style={rectangle split part fill={neutral-left,PastelIBMPurple,PastelIBMPurple}},
          leveloneq/.style={rectangle split part fill={neutral-left,PastelIBMBlue,PastelIBMBlue}},
          leveltwoq/.style={rectangle split part fill={neutral-left,PastelIBMPink,PastelIBMPink}},
          levelthreeq/.style={rectangle split part fill={neutral-left,PastelIBMOrange,PastelIBMOrange}},
          leaf/.style={
            shape=rectangle split, rectangle split parts=3,
            rectangle split draw splits, rectangle split horizontal,
            rounded corners, draw, inner sep=4pt,
            minimum width=5em, minimum height=3.5em,
            font=\Large, text width=3em, align=center
          }
        ]
        \def\dx{8} \def\dy{1.5} \def\dxii{4} \def\leafsep{2}

        \node[node] (root) at (0,0) {$\ket{\bullet}$};
        \coordinate (above) at ($(root)+(0,\dy/1.5)$);
        \draw (above) -- (root);

        \node[node] (L)  at ($(root)+(-\dx,-\dy)$) {$\ket{\bullet}$};
        \node[node] (R)  at ($(root)+(\dx,-\dy)$)  {$\ket{\bullet}$};

        \node[node] (LL) at ($(L)+(-\dxii,-\dy)$) {$\ket{\bullet}$};
        \node[node] (LR) at ($(L)+(\dxii,-\dy)$)  {$\ket{\bullet}$};
        \node[node] (RL) at ($(R)+(-\dxii,-\dy)$) {$\ket{\bullet}$};
        \node[node] (RR) at ($(R)+(\dxii,-\dy)$)  {$\ket{\bullet}$};

        \node[leaf, levelzeroq, label=below:{$\ket{L_0}$}] (LLL) at ($(LL)+(-\leafsep,-\dy)$)
          {$\ket{0}$\nodepart{two}$\ket{32.48}$\nodepart{three}$\ket{b}$};

        \node[leaf, leveloneq,  label=below:{$\ket{L_1}$}] (LLR) at ($(LL)+(\leafsep,-\dy)$)
          {$\ket{0}$\nodepart{two}$\ket{24.89}$\nodepart{three}$\ket{7.59}$};

        \node[leaf, leveltwoq,  label=below:{$\ket{L_2}$}] (LRL) at ($(LR)+(-\leafsep,-\dy)$)
          {$\ket{1}$\nodepart{two}$\ket{14.45}$\nodepart{three}$\ket{10.44}$};

        \node[leaf, leveltwoq,  label=below:{$\ket{L_3}$}] (LRR) at ($(LR)+(\leafsep,-\dy)$)
          {$\ket{0}$\nodepart{two}$\ket{1.09}$\nodepart{three}$\ket{6.50}$};

        \node[leaf, levelthreeq,label=below:{$\ket{L_4}$}] (RLL) at ($(RL)+(-\leafsep,-\dy)$)
          {$\ket{0}$\nodepart{two}$\ket{4.84}$\nodepart{three}$\ket{9.61}$};

        \node[leaf, levelthreeq,label=below:{$\ket{L_5}$}] (RLR) at ($(RL)+(\leafsep,-\dy)$)
          {$\ket{0}$\nodepart{two}$\ket{9.00}$\nodepart{three}$\ket{1.44}$};

        \node[leaf, levelthreeq,label=below:{$\ket{L_6}$}] (RRL) at ($(RR)+(-\leafsep,-\dy)$)
          {$\ket{0}$\nodepart{two}$\ket{0.09}$\nodepart{three}$\ket{1.00}$};

        \node[leaf, levelthreeq,label=below:{$\ket{L_7}$}] (RRR) at ($(RR)+(\leafsep,-\dy)$)
          {$\ket{1}$\nodepart{two}$\ket{0.25}$\nodepart{three}$\ket{6.25}$};

        \draw (root) -- ++(-\dx,0) coordinate(auxL) -- (L);
        \draw (root) -- ++(\dx,0)  coordinate(auxR) -- (R);
        \draw (L)  -- ++(-\dxii,0) coordinate(auxLL) -- (LL);
        \draw (L)  -- ++(\dxii,0)  coordinate(auxLR) -- (LR);
        \draw (R)  -- ++(-\dxii,0) coordinate(auxRL) -- (RL);
        \draw (R)  -- ++(\dxii,0)  coordinate(auxRR) -- (RR);
        \draw (LL) -- ++(-\leafsep,0) coordinate(auxLLL) -- (LLL);
        \draw (LL) -- ++(\leafsep,0)  coordinate(auxLLR) -- (LLR);
        \draw (LR) -- ++(-\leafsep,0) coordinate(auxLRL) -- (LRL);
        \draw (LR) -- ++(\leafsep,0)  coordinate(auxLRR) -- (LRR);
        \draw (RL) -- ++(-\leafsep,0) coordinate(auxRLL) -- (RLL);
        \draw (RL) -- ++(\leafsep,0)  coordinate(auxRLR) -- (RLR);
        \draw (RR) -- ++(-\leafsep,0) coordinate(auxRRL) -- (RRL);
        \draw (RR) -- ++(\leafsep,0)  coordinate(auxRRR) -- (RRR);
        
        \def\braceYshift{-3.2em}
        \def\braceAmp{8pt}
        \def\braceSep{12pt}
        
        \draw[decorate, decoration={brace, mirror, amplitude=\braceAmp}, very thick]
          ([yshift=\braceYshift]LLL.south west) -- ([yshift=\braceYshift]LLL.south east)
          node[midway, below=\braceSep, font=\fontsize{16}{18}\selectfont] {$h = 0$};
        
        \draw[decorate, decoration={brace, mirror, amplitude=\braceAmp}, very thick]
          ([yshift=\braceYshift]LLR.south west) -- ([yshift=\braceYshift]LLR.south east)
          node[midway, below=\braceSep, font=\fontsize{16}{18}\selectfont] {$h = 1$};
        
        \draw[decorate, decoration={brace, mirror, amplitude=\braceAmp}, very thick]
          ([yshift=\braceYshift]LRL.south west) -- ([yshift=\braceYshift]LRR.south east)
          node[midway, below=\braceSep, font=\fontsize{16}{18}\selectfont] {$h = 2$};
        
        \draw[decorate, decoration={brace, mirror, amplitude=\braceAmp}, very thick]
          ([yshift=\braceYshift]RLL.south west) -- ([yshift=\braceYshift]RRR.south east)
          node[midway, below=\braceSep, font=\fontsize{16}{18}\selectfont] {$h = 3$};
      \end{tikzpicture}
    }
    \caption{}\Description{}
    \label{fig:bbqram_mapping_colored_example}
   \end{subfigure}
  \caption{Figure~\ref{fig:segment_tree_colored_example} depicts the segment tree $T$ obtained from matrix $A \in \mathbb{R}^{2 \times 4}$ of the numerical example accoding to Definition~\ref{def:segment_tree}. Figure~\ref{fig:bbqram_mapping_colored_example} illustrates how $T$ maps into the memory cells of a BBQRAM architecture according to Proposition~\ref{prop:mapping}.}\Description{}
  \label{fig:segment_tree_mapping_to_bbqram_example}
\end{figure*}
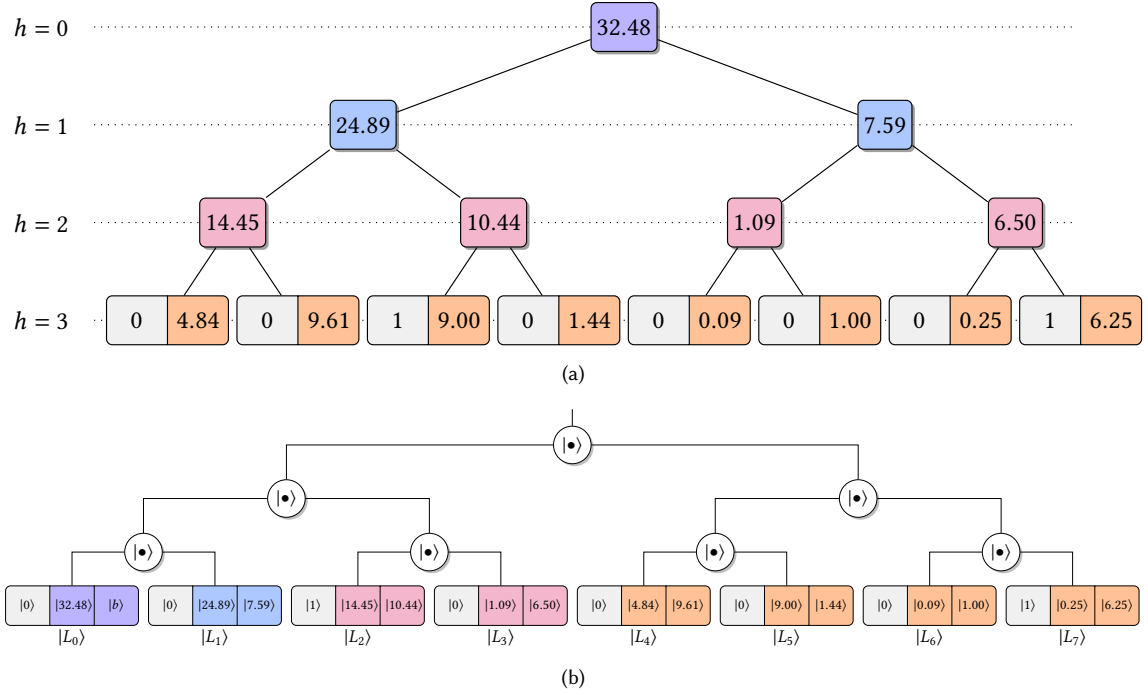

\section{Numerical Example}\label{sec:numerical_example}

In this section, we present a numerical example that showcases how to encode the real-valued matrix 
\begin{equation*}
A \in \mathbb{R}^{2 \times 4} = 
    \begin{bmatrix}
    2.2 & 3.1 & -3.0 & 1.2 \\
    0.3 & 1.0 & 0.5 & -2.5
    \end{bmatrix}
\end{equation*}
into a quantum state using the state preparation procedure of Algorithm~\ref{alg:state_prep}. In this example, we adopt the same variable names introduced in Section~\ref{sec:state_prep_alg} to maintain a correspondence between the two settings:
\begin{equation*}
    M = 2, \quad N = 4, \quad K = M \times N = 8, \quad m = \log_2 M = 1, \quad n = \log_2 N = 2, \quad k = \log_2 K = 3,
\end{equation*}
where $M$ and $N$ denote the number of rows and columns of $A$, respectively; $K$ is the total number of entries in $A$, indexed in row-major order; and $m$, $n$, and $k$ represent the number of qubits required to address rows, columns, and the full set of entries, respectively.

\begin{figure*}[t]
  \centering
  \tikzset{
    every label/.append style={font=\huge},
    node/.style={
      draw, fill=white, circle, inner sep=0pt, font=\huge,
      drop shadow, text width=3em, align=center
    },
    levelzeroq/.style={rectangle split part fill={neutral-left,PastelIBMPurple,PastelIBMPurple}},
    leveloneq/.style={rectangle split part fill={neutral-left,PastelIBMBlue,PastelIBMBlue}},
    leveltwoq/.style={rectangle split part fill={neutral-left,PastelIBMPink,PastelIBMPink}},
    levelthreeq/.style={rectangle split part fill={neutral-left,PastelIBMOrange,PastelIBMOrange}},
    leaf/.style={
      shape=rectangle split, rectangle split parts=3,
      rectangle split draw splits, rectangle split horizontal,
      rounded corners, draw, inner sep=4pt,
      minimum width=5em, minimum height=3.5em,
      font=\Large, text width=3em, align=center
    },
    hi/.style={draw=cyan,line width=3pt} 
  }
  \resizebox{\linewidth}{!}{%
    \begin{tikzpicture}
      \def\dx{8} \def\dy{1.5} \def\dxii{4} \def\leafsep{2}

      \node[node, draw=cyan, line width=2.5pt] (root) at (0,0) {$\ket{0}$};
      \coordinate (above) at ($(root)+(0,\dy/1.5)$);
      \draw[hi] (above) -- (root);        

      \node[node, draw=cyan, line width=2.5pt] (L)  at ($(root)+(-\dx,-\dy)$) {$\ket{0}$};
      \node[node] (R)  at ($(root)+(\dx,-\dy)$)  {$\ket{\bullet}$};

      \node[node, draw=cyan, line width=2.5pt] (LL) at ($(L)+(-\dxii,-\dy)$) {$\ket{1}$};
      \node[node] (LR) at ($(L)+(\dxii,-\dy)$)  {$\ket{\bullet}$};
      \node[node] (RL) at ($(R)+(-\dxii,-\dy)$) {$\ket{\bullet}$};
      \node[node] (RR) at ($(R)+(\dxii,-\dy)$)  {$\ket{\bullet}$};

      \node[leaf, levelzeroq, label=below:{$\ket{L_0}$}] (LLL) at ($(LL)+(-\leafsep,-\dy)$)
        {$\ket{0}$\nodepart{two}$\ket{32.48}$\nodepart{three}$\ket{b}$};

      \node[leaf, draw=cyan, line width=2.5pt, leveloneq,  label=below:{$\ket{L_1}$}] (LLR) at ($(LL)+(\leafsep,-\dy)$)
        {$\ket{0}$\nodepart{two}$\ket{24.89}$\nodepart{three}$\ket{7.59}$};

      \node[leaf, leveltwoq,  label=below:{$\ket{L_2}$}] (LRL) at ($(LR)+(-\leafsep,-\dy)$)
        {$\ket{1}$\nodepart{two}$\ket{14.45}$\nodepart{three}$\ket{10.44}$};

      \node[leaf, leveltwoq,  label=below:{$\ket{L_3}$}] (LRR) at ($(LR)+(\leafsep,-\dy)$)
        {$\ket{0}$\nodepart{two}$\ket{1.09}$\nodepart{three}$\ket{6.50}$};

      \node[leaf, levelthreeq,label=below:{$\ket{L_4}$}] (RLL) at ($(RL)+(-\leafsep,-\dy)$)
        {$\ket{0}$\nodepart{two}$\ket{4.84}$\nodepart{three}$\ket{9.61}$};

      \node[leaf, levelthreeq,label=below:{$\ket{L_5}$}] (RLR) at ($(RL)+(\leafsep,-\dy)$)
        {$\ket{0}$\nodepart{two}$\ket{9.00}$\nodepart{three}$\ket{1.44}$};

      \node[leaf, levelthreeq,label=below:{$\ket{L_6}$}] (RRL) at ($(RR)+(-\leafsep,-\dy)$)
        {$\ket{0}$\nodepart{two}$\ket{0.09}$\nodepart{three}$\ket{1.00}$};

      \node[leaf, levelthreeq,label=below:{$\ket{L_7}$}] (RRR) at ($(RR)+(\leafsep,-\dy)$)
        {$\ket{1}$\nodepart{two}$\ket{0.25}$\nodepart{three}$\ket{6.25}$};

      \draw (root) -- ++(-\dx,0) coordinate(auxL) -- (L);
      \draw (root) -- ++(\dx,0)  coordinate(auxR) -- (R);
      \draw (L)  -- ++(-\dxii,0) coordinate(auxLL) -- (LL);
      \draw (L)  -- ++(\dxii,0)  coordinate(auxLR) -- (LR);
      \draw (R)  -- ++(-\dxii,0) coordinate(auxRL) -- (RL);
      \draw (R)  -- ++(\dxii,0)  coordinate(auxRR) -- (RR);
      \draw (LL) -- ++(-\leafsep,0) coordinate(auxLLL) -- (LLL);
      \draw (LL) -- ++(\leafsep,0)  coordinate(auxLLR) -- (LLR);
      \draw (LR) -- ++(-\leafsep,0) coordinate(auxLRL) -- (LRL);
      \draw (LR) -- ++(\leafsep,0)  coordinate(auxLRR) -- (LRR);
      \draw (RL) -- ++(-\leafsep,0) coordinate(auxRLL) -- (RLL);
      \draw (RL) -- ++(\leafsep,0)  coordinate(auxRLR) -- (RLR);
      \draw (RR) -- ++(-\leafsep,0) coordinate(auxRRL) -- (RRL);
      \draw (RR) -- ++(\leafsep,0)  coordinate(auxRRR) -- (RRR);

      \draw[hi] (root) -- (auxL);
      \draw[hi] (auxL) -- (L);
      \draw[hi] (L)    -- (auxLL);
      \draw[hi] (auxLL) -- (LL);
      \draw[hi] (LL)   -- (auxLLR);
      \draw[hi] (auxLLR) -- (LLR);
    \end{tikzpicture}}
  \caption{Address register $\ket{0\,0\,1}_{\mathrm{a}}$ traces a single path that accesses the memory cell $\ket{L_1}$ containing the sibling pair $\ket{24.89}\ket{7.59}$ at height $h=1$ in $T$.}
  \label{fig:bbqram_path_001_example}\Description{}
\end{figure*}
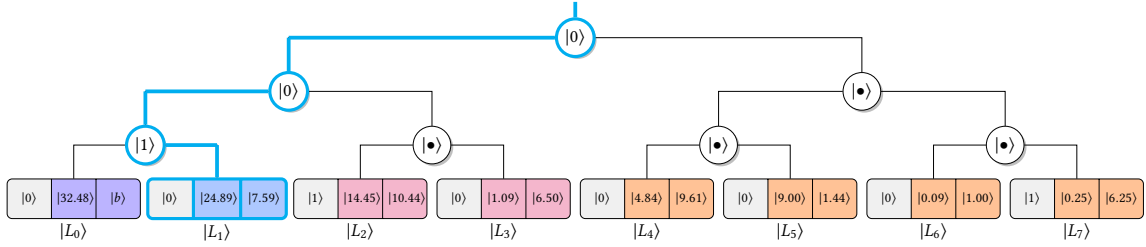

Figure~\ref{fig:segment_tree_mapping_to_bbqram_example} illustrates the memory layout of the segment tree $T$ constructed from the example matrix $A$ and its mapping onto the memory cells of a BBQRAM.
With all the necessary components in place, we now proceed to describe, step by step, how to encode the matrix $A$ into a quantum state. Let us start with the initial state:
\begin{align*}
\ket{0}_{\mathrm{s}}\ket{0}_{\mathrm{l}}\ket{0}_{\mathrm{r}}\ket{0}_{\mathrm{v}}\ket{0}^{ k=3}_{\mathrm{a}} = \ket{0}_{\mathrm{s}}\ket{0}_{\mathrm{l}}\ket{0}_{\mathrm{r}}\ket{0}_{\mathrm{v}}\ket{000}_{\mathrm{a}}.
\end{align*}
The next step is to set the address register $\mathrm{a}$ to the state $\ket{001}$, which allows access to the memory cell $\ket{L_1}$ (see Figure~\ref{fig:bbqram_path_001_example}):
\begin{align*}
\ket{0}\ket{0}_{\mathrm{l}}\ket{0}_{\mathrm{r}}\ket{0}\ket{001}.
\end{align*}
After retrieving the sibling nodes at the first level of $T$ (i.e., $\ket{24.89}\ket{7.59}$), we obtain the state
\begin{align*}
\ket{0}_{\mathrm{s}}\ket{24.89}_{\mathrm{l}}\ket{7.59}_{\mathrm{r}}\ket{0}_{\mathrm{v}}\ket{001}_{\mathrm{a}}.
\end{align*}
Then, we apply $U_{2CR}$ to the registers $\mathrm{l}$, $\mathrm{r}$, and $\mathrm{v}$
\begin{align*}
\ket{0}_{\mathrm{s}}\ket{24.89}_{\mathrm{l}}\ket{7.59}_{\mathrm{r}} \left(\sqrt\frac{24.89}{32.48}\ket{0}_{\mathrm{v}}+ \sqrt\frac{7.59}{32.48}\ket{1}_{\mathrm{v}}\right) \ket{001}_{\mathrm{a}},
\end{align*}
and we uncompute $\ket{24.89}_{\mathrm{l}}\ket{7.59}_{\mathrm{r}}$:
\begin{align*}
\ket{0}_{\mathrm{s}}\ket{0}_{\mathrm{l}}\ket{0}_{\mathrm{r}} \left(\sqrt\frac{24.89}{32.48}\ket{0}_{\mathrm{v}}+ \sqrt\frac{7.59}{32.48}\ket{1}_{\mathrm{v}}\right) \ket{001}_{\mathrm{a}} = \frac{1}{\sqrt{32.48}}\ket{0}_{\mathrm{s}}\ket{0}_{\mathrm{l}}\ket{0}_{\mathrm{r}} \left(\sqrt{24.89}\ket{0}_{\mathrm{v}}+ \sqrt{7.59}\ket{1}_{\mathrm{v}}\right) \ket{001}_{\mathrm{a}}.
\end{align*}
Subsequently, we perform a left circular shift over the quantum register $\mathrm{v}$ and $\mathrm{a}$:
\begin{align*}
\frac{1}{\sqrt{32.48}}\ket{0}_{\mathrm{s}}\ket{0}_{\mathrm{l}}\ket{0}_{\mathrm{r}}\ket{0}_{\mathrm{v}} \ket{01}_{\mathrm{a}}\left(\sqrt{24.89}\ket{0}_{\mathrm{a}}+ \sqrt{7.59}\ket{1}_{\mathrm{a}}\right)
\end{align*}
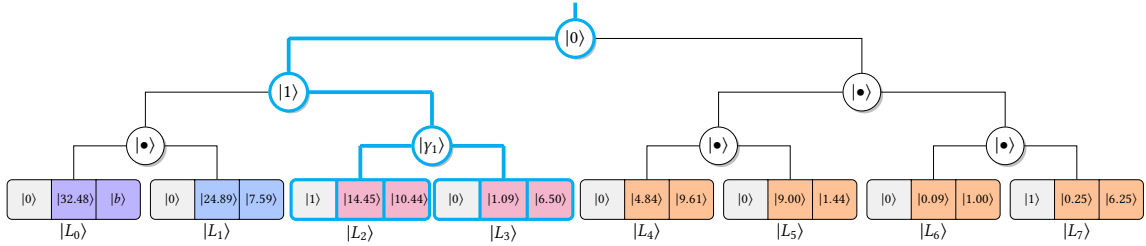
\begin{figure*}[t]
  \centering
  \tikzset{
    every label/.append style={font=\huge},
    node/.style={
      draw, fill=white, circle, inner sep=0pt, font=\huge,
      drop shadow, text width=3em, align=center
    },
    levelzeroq/.style={rectangle split part fill={neutral-left,PastelIBMPurple,PastelIBMPurple}},
    leveloneq/.style={rectangle split part fill={neutral-left,PastelIBMBlue,PastelIBMBlue}},
    leveltwoq/.style={rectangle split part fill={neutral-left,PastelIBMPink,PastelIBMPink}},
    levelthreeq/.style={rectangle split part fill={neutral-left,PastelIBMOrange,PastelIBMOrange}},
    leaf/.style={
      shape=rectangle split, rectangle split parts=3,
      rectangle split draw splits, rectangle split horizontal,
      rounded corners, draw, inner sep=4pt,
      minimum width=5em, minimum height=3.5em,
      font=\Large, text width=3em, align=center
    },
    hi/.style={draw=cyan,line width=3pt} 
  }
  \resizebox{\linewidth}{!}{%
    \begin{tikzpicture}
      \def\dx{8} \def\dy{1.5} \def\dxii{4} \def\leafsep{2}

      \node[node, draw=cyan, line width=2.5pt] (root) at (0,0) {$\ket{0}$};
      \coordinate (above) at ($(root)+(0,\dy/1.5)$);
      \draw[hi] (above) -- (root);      

      \node[node, draw=cyan, line width=2.5pt] (L)  at ($(root)+(-\dx,-\dy)$) {$\ket{1}$};
      \node[node] (R)  at ($(root)+(\dx,-\dy)$)  {$\ket{\bullet}$};

      \node[node] (LL) at ($(L)+(-\dxii,-\dy)$) {$\ket{\bullet}$};
      \node[node, draw=cyan, line width=2.5pt] (LR) at ($(L)+(\dxii,-\dy)$)  {$\ket{\gamma_1}$};
      \node[node] (RL) at ($(R)+(-\dxii,-\dy)$) {$\ket{\bullet}$};
      \node[node] (RR) at ($(R)+(\dxii,-\dy)$)  {$\ket{\bullet}$};

      \node[leaf, levelzeroq, label=below:{$\ket{L_0}$}] (LLL) at ($(LL)+(-\leafsep,-\dy)$)
        {$\ket{0}$\nodepart{two}$\ket{32.48}$\nodepart{three}$\ket{b}$};

      \node[leaf, leveloneq,  label=below:{$\ket{L_1}$}] (LLR) at ($(LL)+(\leafsep,-\dy)$)
        {$\ket{0}$\nodepart{two}$\ket{24.89}$\nodepart{three}$\ket{7.59}$};

      \node[leaf, draw=cyan, line width=2.5pt, leveltwoq,  label=below:{$\ket{L_2}$}] (LRL) at ($(LR)+(-\leafsep,-\dy)$)
        {$\ket{1}$\nodepart{two}$\ket{14.45}$\nodepart{three}$\ket{10.44}$};

      \node[leaf, draw=cyan, line width=2.5pt, leveltwoq,  label=below:{$\ket{L_3}$}] (LRR) at ($(LR)+(\leafsep,-\dy)$)
        {$\ket{0}$\nodepart{two}$\ket{1.09}$\nodepart{three}$\ket{6.50}$};

      \node[leaf, levelthreeq,label=below:{$\ket{L_4}$}] (RLL) at ($(RL)+(-\leafsep,-\dy)$)
        {$\ket{0}$\nodepart{two}$\ket{4.84}$\nodepart{three}$\ket{9.61}$};

      \node[leaf, levelthreeq,label=below:{$\ket{L_5}$}] (RLR) at ($(RL)+(\leafsep,-\dy)$)
        {$\ket{0}$\nodepart{two}$\ket{9.00}$\nodepart{three}$\ket{1.44}$};

      \node[leaf, levelthreeq,label=below:{$\ket{L_6}$}] (RRL) at ($(RR)+(-\leafsep,-\dy)$)
        {$\ket{0}$\nodepart{two}$\ket{0.09}$\nodepart{three}$\ket{1.00}$};

      \node[leaf, levelthreeq,label=below:{$\ket{L_7}$}] (RRR) at ($(RR)+(\leafsep,-\dy)$)
        {$\ket{1}$\nodepart{two}$\ket{0.25}$\nodepart{three}$\ket{6.25}$};

      \draw (root) -- ++(-\dx,0) coordinate(auxL) -- (L);
      \draw (root) -- ++(\dx,0)  coordinate(auxR) -- (R);
      \draw (L)  -- ++(-\dxii,0) coordinate(auxLL) -- (LL);
      \draw (L)  -- ++(\dxii,0)  coordinate(auxLR) -- (LR);
      \draw (R)  -- ++(-\dxii,0) coordinate(auxRL) -- (RL);
      \draw (R)  -- ++(\dxii,0)  coordinate(auxRR) -- (RR);
      \draw (LL) -- ++(-\leafsep,0) coordinate(auxLLL) -- (LLL);
      \draw (LL) -- ++(\leafsep,0)  coordinate(auxLLR) -- (LLR);
      \draw (LR) -- ++(-\leafsep,0) coordinate(auxLRL) -- (LRL);
      \draw (LR) -- ++(\leafsep,0)  coordinate(auxLRR) -- (LRR);
      \draw (RL) -- ++(-\leafsep,0) coordinate(auxRLL) -- (RLL);
      \draw (RL) -- ++(\leafsep,0)  coordinate(auxRLR) -- (RLR);
      \draw (RR) -- ++(-\leafsep,0) coordinate(auxRRL) -- (RRL);
      \draw (RR) -- ++(\leafsep,0)  coordinate(auxRRR) -- (RRR);

      \draw[hi] (root) -- (auxL);
      \draw[hi] (auxL) -- (L);

      \draw[hi] (L) -- (auxLR);
      \draw[hi] (auxLR) -- (LR);

      \draw[hi] (LR) -- (auxLRL);
      \draw[hi] (auxLRL) -- (LRL);
      \draw[hi] (LR) -- (auxLRR);
      \draw[hi] (auxLRR) -- (LRR);
    \end{tikzpicture}}
  \caption{Address register $\ket{0\,1\,\gamma_1 }_{\mathrm{a}}$, where $ \ket{\gamma_1} = \sqrt{24.89}\ket{0}_{\mathrm{a}}+ \sqrt{7.59}\ket{1}_{\mathrm{a}}$, accesses in superposition the memory cells $\ket{L_2}$ and $\ket{L_3}$ containing the two sibling pairs $\ket{14.45}\ket{10.44}$ and $\ket{1.09}\ket{6.50}$, respectively, that reside at height $h=2$ in $T$.}
  \label{fig:bbqram_path_01plus}\Description{}
\end{figure*}
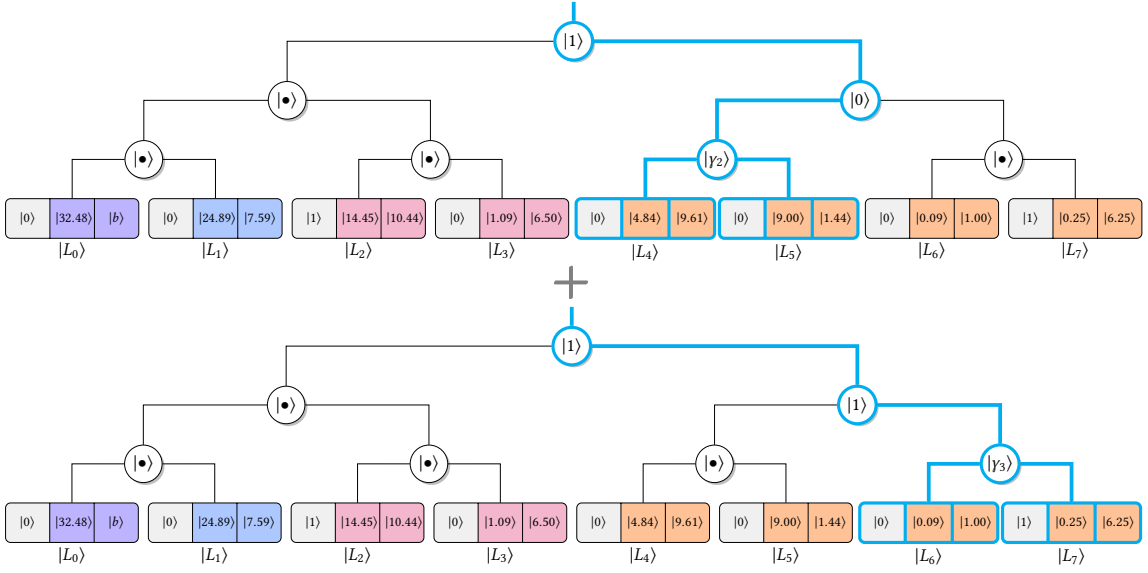
\begin{figure*}[t]
  \centering

  \tikzset{
    every label/.append style={font=\huge},
    node/.style={
      draw, fill=white, circle, inner sep=0pt, font=\huge,
      drop shadow, text width=3em, align=center
    },
    levelzeroq/.style={rectangle split part fill={neutral-left,PastelIBMPurple,PastelIBMPurple}},
    leveloneq/.style={rectangle split part fill={neutral-left,PastelIBMBlue,PastelIBMBlue}},
    leveltwoq/.style={rectangle split part fill={neutral-left,PastelIBMPink,PastelIBMPink}},
    levelthreeq/.style={rectangle split part fill={neutral-left,PastelIBMOrange,PastelIBMOrange}},
    leaf/.style={
      shape=rectangle split, rectangle split parts=3,
      rectangle split draw splits, rectangle split horizontal,
      rounded corners, draw, inner sep=4pt,
      minimum width=5em, minimum height=3.5em,
      font=\Large, text width=3em, align=center
    },
    hi/.style={draw=cyan,line width=3pt} 
  }

  \begin{subfigure}[t]{\textwidth}
    \centering
    \resizebox{\linewidth}{!}{%
      \begin{tikzpicture}[yscale=1.1]
        \def\dx{8} \def\dy{1.5} \def\dxii{4} \def\leafsep{2}

        \node[node, draw=cyan, line width=2.5pt] (root) at (0,0) {$\ket{1}$};
        \coordinate (above) at ($(root)+(0,\dy/1.5)$);
        \draw[hi] (above) -- (root);

        \node[node] (L)  at ($(root)+(-\dx,-\dy)$) {$\ket{\bullet}$};
        \node[node, draw=cyan, line width=2.5pt] (R)  at ($(root)+(\dx,-\dy)$)  {$\ket{0}$};

        \node[node] (LL) at ($(L)+(-\dxii,-\dy)$) {$\ket{\bullet}$};
        \node[node] (LR) at ($(L)+(\dxii,-\dy)$)  {$\ket{\bullet}$};
        \node[node, draw=cyan, line width=2.5pt] (RL) at ($(R)+(-\dxii,-\dy)$) {$\ket{\gamma_2}$};
        \node[node] (RR) at ($(R)+(\dxii,-\dy)$)  {$\ket{\bullet}$};

        \node[leaf, levelzeroq, label=below:{$\ket{L_0}$}] (LLL) at ($(LL)+(-\leafsep,-\dy)$)
          {$\ket{0}$\nodepart{two}$\ket{32.48}$\nodepart{three}$\ket{b}$};
        \node[leaf, leveloneq,  label=below:{$\ket{L_1}$}] (LLR) at ($(LL)+(\leafsep,-\dy)$)
          {$\ket{0}$\nodepart{two}$\ket{24.89}$\nodepart{three}$\ket{7.59}$};
        \node[leaf, leveltwoq,  label=below:{$\ket{L_2}$}] (LRL) at ($(LR)+(-\leafsep,-\dy)$)
          {$\ket{1}$\nodepart{two}$\ket{14.45}$\nodepart{three}$\ket{10.44}$};
        \node[leaf, leveltwoq,  label=below:{$\ket{L_3}$}] (LRR) at ($(LR)+(\leafsep,-\dy)$)
          {$\ket{0}$\nodepart{two}$\ket{1.09}$\nodepart{three}$\ket{6.50}$};
        \node[leaf, draw=cyan, line width=2.5pt, levelthreeq,label=below:{$\ket{L_4}$}] (RLL) at ($(RL)+(-\leafsep,-\dy)$)
          {$\ket{0}$\nodepart{two}$\ket{4.84}$\nodepart{three}$\ket{9.61}$};
        \node[leaf, draw=cyan, line width=2.5pt, levelthreeq,label=below:{$\ket{L_5}$}] (RLR) at ($(RL)+(\leafsep,-\dy)$)
          {$\ket{0}$\nodepart{two}$\ket{9.00}$\nodepart{three}$\ket{1.44}$};
        \node[leaf, levelthreeq,label=below:{$\ket{L_6}$}] (RRL) at ($(RR)+(-\leafsep,-\dy)$)
          {$\ket{0}$\nodepart{two}$\ket{0.09}$\nodepart{three}$\ket{1.00}$};
        \node[leaf, levelthreeq, label=below:{$\ket{L_7}$}] (RRR) at ($(RR)+(\leafsep,-\dy)$)
          {$\ket{1}$\nodepart{two}$\ket{0.25}$\nodepart{three}$\ket{6.25}$};

        \draw (root) -- ++(-\dx,0) coordinate(auxL) -- (L);
        \draw (root) -- ++(\dx,0)  coordinate(auxR) -- (R);
        \draw (L)  -- ++(-\dxii,0) coordinate(auxLL) -- (LL);
        \draw (L)  -- ++(\dxii,0)  coordinate(auxLR) -- (LR);
        \draw (R)  -- ++(-\dxii,0) coordinate(auxRL) -- (RL);
        \draw (R)  -- ++(\dxii,0)  coordinate(auxRR) -- (RR);
        \draw (LL) -- ++(-\leafsep,0) coordinate(auxLLL) -- (LLL);
        \draw (LL) -- ++(\leafsep,0)  coordinate(auxLLR) -- (LLR);
        \draw (LR) -- ++(-\leafsep,0) coordinate(auxLRL) -- (LRL);
        \draw (LR) -- ++(\leafsep,0)  coordinate(auxLRR) -- (LRR);
        \draw (RL) -- ++(-\leafsep,0) coordinate(auxRLL) -- (RLL);
        \draw (RL) -- ++(\leafsep,0)  coordinate(auxRLR) -- (RLR);
        \draw (RR) -- ++(-\leafsep,0) coordinate(auxRRL) -- (RRL);
        \draw (RR) -- ++(\leafsep,0)  coordinate(auxRRR) -- (RRR);

        \draw[hi] (root)--(auxR)--(R)
                  (R)--(auxRL)--(RL)
                  (RL)--(auxRLL)--(RLL)
                  (RL)--(auxRLR)--(RLR);
      \end{tikzpicture}%
    }
  \end{subfigure}
  \hfill
  \begin{subfigure}[t]{\textwidth}
    \centering
    \makebox[\textwidth][l]{\hspace{7.23cm}{\fontsize{30}{0}\selectfont \textcolor{gray}{+}}}
  \end{subfigure}
  \hfill
  \begin{subfigure}[t]{\textwidth}
    \centering
    \resizebox{\linewidth}{!}{%
      \begin{tikzpicture}[yscale=1.1]
        \def\dx{8} \def\dy{1.5} \def\dxii{4} \def\leafsep{2}

        \node[node, draw=cyan, line width=2.5pt] (root) at (0,0) {$\ket{1}$};
        \coordinate (above) at ($(root)+(0,\dy/1.5)$);
        \draw[hi] (above) -- (root);

        \node[node] (L)  at ($(root)+(-\dx,-\dy)$) {$\ket{\bullet}$};
        \node[node, draw=cyan, line width=2.5pt] (R)  at ($(root)+(\dx,-\dy)$)  {$\ket{1}$};

        \node[node] (LL) at ($(L)+(-\dxii,-\dy)$) {$\ket{\bullet}$};
        \node[node] (LR) at ($(L)+(\dxii,-\dy)$)  {$\ket{\bullet}$};
        \node[node] (RL) at ($(R)+(-\dxii,-\dy)$) {$\ket{\bullet}$};
        \node[node, draw=cyan, line width=2.5pt] (RR) at ($(R)+(\dxii,-\dy)$)  {$\ket{\gamma_3}$};

        \node[leaf, levelzeroq, label=below:{$\ket{L_0}$}] (LLL) at ($(LL)+(-\leafsep,-\dy)$)
          {$\ket{0}$\nodepart{two}$\ket{32.48}$\nodepart{three}$\ket{b}$};
        \node[leaf, leveloneq,  label=below:{$\ket{L_1}$}] (LLR) at ($(LL)+(\leafsep,-\dy)$)
          {$\ket{0}$\nodepart{two}$\ket{24.89}$\nodepart{three}$\ket{7.59}$};
        \node[leaf, leveltwoq,  label=below:{$\ket{L_2}$}] (LRL) at ($(LR)+(-\leafsep,-\dy)$)
          {$\ket{1}$\nodepart{two}$\ket{14.45}$\nodepart{three}$\ket{10.44}$};
        \node[leaf, leveltwoq,  label=below:{$\ket{L_3}$}] (LRR) at ($(LR)+(\leafsep,-\dy)$)
          {$\ket{0}$\nodepart{two}$\ket{1.09}$\nodepart{three}$\ket{6.50}$};
        \node[leaf, levelthreeq,label=below:{$\ket{L_4}$}] (RLL) at ($(RL)+(-\leafsep,-\dy)$)
          {$\ket{0}$\nodepart{two}$\ket{4.84}$\nodepart{three}$\ket{9.61}$};
        \node[leaf, levelthreeq,label=below:{$\ket{L_5}$}] (RLR) at ($(RL)+(\leafsep,-\dy)$)
          {$\ket{0}$\nodepart{two}$\ket{9.00}$\nodepart{three}$\ket{1.44}$};
        \node[leaf, draw=cyan, line width=2.5pt, levelthreeq,label=below:{$\ket{L_6}$}] (RRL) at ($(RR)+(-\leafsep,-\dy)$)
          {$\ket{0}$\nodepart{two}$\ket{0.09}$\nodepart{three}$\ket{1.00}$};
        \node[leaf, draw=cyan, line width=2.5pt, levelthreeq,label=below:{$\ket{L_7}$}] (RRR) at ($(RR)+(\leafsep,-\dy)$)
          {$\ket{1}$\nodepart{two}$\ket{0.25}$\nodepart{three}$\ket{6.25}$};

        \draw (root) -- ++(-\dx,0) coordinate(auxL) -- (L);
        \draw (root) -- ++(\dx,0)  coordinate(auxR) -- (R);
        \draw (L)  -- ++(-\dxii,0) coordinate(auxLL) -- (LL);
        \draw (L)  -- ++(\dxii,0)  coordinate(auxLR) -- (LR);
        \draw (R)  -- ++(-\dxii,0) coordinate(auxRL) -- (RL);
        \draw (R)  -- ++(\dxii,0)  coordinate(auxRR) -- (RR);
        \draw (LL) -- ++(-\leafsep,0) coordinate(auxLLL) -- (LLL);
        \draw (LL) -- ++(\leafsep,0)  coordinate(auxLLR) -- (LLR);
        \draw (LR) -- ++(-\leafsep,0) coordinate(auxLRL) -- (LRL);
        \draw (LR) -- ++(\leafsep,0)  coordinate(auxLRR) -- (LRR);
        \draw (RL) -- ++(-\leafsep,0) coordinate(auxRLL) -- (RLL);
        \draw (RL) -- ++(\leafsep,0)  coordinate(auxRLR) -- (RLR);
        \draw (RR) -- ++(-\leafsep,0) coordinate(auxRRL) -- (RRL);
        \draw (RR) -- ++(\leafsep,0)  coordinate(auxRRR) -- (RRR);

        \draw[hi] (root)--(auxR)--(R)
                  (R)--(auxRR)--(RR)
                  (RR)--(auxRRL)--(RRL)
                  (RR)--(auxRRR)--(RRR);
      \end{tikzpicture}%
    }
  \end{subfigure}
  \caption{The figure illustrates the superposition of two distinct access paths when querying the memory cells with the address register $\ket{1}_{\mathrm{a}}\otimes\left(\ket{0\gamma_2}_{\mathrm{a}}+\ket{1\gamma_3}_{\mathrm{a}}\right)$, where $\ket{\gamma_2} = \sqrt{14.45}\ket{0} + \sqrt{10.44}\ket{1}$ and $\ket{\gamma_3} = \sqrt{1.09}\ket{0} + \sqrt{6.50}\ket{1}$. This query accesses, in superposition, the memory cells labeled as $\ket{L_4}$, $\ket{L_5}$, $\ket{L_6}$, and $\ket{L_7}$ at level $h = 3$ of the segment tree $T$. The $+$ between the two figures highlights that the quantum system of BBQRAM is in a superposition of two access paths.}
  \label{fig:bbqram_path_1plusplus}\Description{}
\end{figure*}
and we proceed by retrieving in superposition $\ket{14.45}\ket{10.44}$ and $\ket{1.09}\ket{6.50}$ (see Figure~\ref{fig:bbqram_path_01plus}):
\begin{align*}
\frac{1}{\sqrt{32.48}}\left(
\sqrt{24.89}\ket{0}_{\mathrm{s}}\ket{14.45}_{\mathrm{l}}\ket{10.44}_{\mathrm{r}} \ket{0}_{\mathrm{v}}\ket{010}_{\mathrm{a}}+ \sqrt{7.59}\ket{0}_{\mathrm{s}}\ket{1.09}_{\mathrm{l}}\ket{6.50}_{\mathrm{r}} \ket{0}_{\mathrm{v}}\ket{011}_{\mathrm{a}}
\right).
\end{align*}

Then, we apply $U_{2CR}$ to the superposition of registers $\mathrm{l}$, $\mathrm{r}$, and $\mathrm{v}$:
\begin{align*}
\frac{1}{\sqrt{32.48}}\Bigg(
&\sqrt{24.89}\ket{0}_{\mathrm{s}}\ket{14.45}_{\mathrm{l}}\ket{10.44}_{\mathrm{r}}
    \left(
        \sqrt{\frac{14.45}{24.89}}\ket{0}_{\mathrm{v}}
        + \sqrt{\frac{10.44}{24.89}}\ket{1}_{\mathrm{v}}
    \right)\ket{010}_{\mathrm{a}} \\
    &+ \sqrt{7.59}\ket{0}_{\mathrm{s}}\ket{1.09}_{\mathrm{l}}\ket{6.50}_{\mathrm{r}}
    \left(
        \sqrt{\frac{1.09}{7.59}}\ket{0}_{\mathrm{v}}
        + \sqrt{\frac{6.50}{7.59}}\ket{1}_{\mathrm{v}}
    \right)\ket{011}_{\mathrm{a}}
\Bigg),
\end{align*}
which we rewrite as
\begin{align*}
\frac{1}{\sqrt{32.48}}\Bigg(
&\ket{0}_{\mathrm{s}}\ket{14.45}_{\mathrm{l}}\ket{10.44}_{\mathrm{r}}
    \left(
        \sqrt{14.45}\ket{0}_{\mathrm{v}}
        + \sqrt{10.44}\ket{1}_{\mathrm{v}}
    \right)\ket{010}_{\mathrm{a}} \\
    &+ \ket{0}_{\mathrm{s}}\ket{1.09}_{\mathrm{l}}\ket{6.50}_{\mathrm{r}}
    \left(
        \sqrt{1.09}\ket{0}_{\mathrm{v}}
        + \sqrt{6.50}\ket{1}_{\mathrm{v}}
    \right)\ket{011}_{\mathrm{a}}
\Bigg),
\end{align*}
and perform a left circular shift over the quantum register $\mathrm{v}$ and $\mathrm{a}$. This operation sets the address register $\mathrm{a}$ so as to access, in superposition, the memory cells $\ket{L_4}$, $\ket{L_5}$, $\ket{L_6}$, and $\ket{L_7}$ (see Figure~\ref{fig:bbqram_path_1plusplus}):
\begin{align*}
\frac{1}{\sqrt{32.48}}\Bigg(
&\ket{0}_{\mathrm{s}}\ket{0}_{\mathrm{l}}\ket{0}_{\mathrm{r}}
    \ket{0}_{\mathrm{v}}\ket{10}_{\mathrm{a}}\left(
        \sqrt{14.45}\ket{0}_{\mathrm{a}}
        + \sqrt{10.44}\ket{1}_{\mathrm{a}}
    \right) +\\
    &+ \ket{0}_{\mathrm{s}}\ket{0}_{\mathrm{l}}\ket{0}_{\mathrm{r}} 
    \ket{0}_{\mathrm{v}}
   \ket{11}_{\mathrm{a}} \left(
        \sqrt{1.09}\ket{0}_{\mathrm{a}}
        + \sqrt{6.50}\ket{1}_{\mathrm{a}}
    \right)
\Bigg),
\end{align*}
and we rewrite as:
\begin{align*}
\frac{1}{\sqrt{32.48}}\Bigg( 
& \sqrt{14.45}\ket{0}_{\mathrm{s}}\ket{0}_{\mathrm{l}}\ket{0}_{\mathrm{r}}\ket{0}_{\mathrm{v}}\ket{100}_{\mathrm{a}}  + \sqrt{10.44}\ket{0}_{\mathrm{s}}\ket{0}_{\mathrm{l}}\ket{0}_{\mathrm{r}}\ket{0}_{\mathrm{v}}\ket{101}_{\mathrm{a}} + \\
& + \sqrt{1.09}\ket{0}_{\mathrm{s}}\ket{0}_{\mathrm{l}}\ket{0}_{\mathrm{r}}\ket{0}_{\mathrm{v}}\ket{110}_{\mathrm{a}}  + \sqrt{6.50}\ket{0}_{\mathrm{s}}\ket{0}_{\mathrm{l}}\ket{0}_{\mathrm{r}}\ket{0}_{\mathrm{v}}\ket{111}_{\mathrm{a}} \Bigg).
\end{align*}
\begin{figure*}[!htb]
 \centering

  \tikzset{
    every label/.append style={font=\huge},
    node/.style={
      draw, fill=white, circle, inner sep=0pt, font=\huge,
      drop shadow, text width=3em, align=center
    },
    levelzeroq/.style={rectangle split part fill={neutral-left,PastelIBMPurple,PastelIBMPurple}},
    leveloneq/.style={rectangle split part fill={neutral-left,PastelIBMBlue,PastelIBMBlue}},
    leveltwoq/.style={rectangle split part fill={neutral-left,PastelIBMPink,PastelIBMPink}},
    levelthreeq/.style={rectangle split part fill={neutral-left,PastelIBMOrange,PastelIBMOrange}},
    leaf/.style={
      shape=rectangle split, rectangle split parts=3,
      rectangle split draw splits, rectangle split horizontal,
      rounded corners, draw, inner sep=4pt,
      minimum width=5em, minimum height=3.5em,
      font=\Large, text width=3em, align=center
    },
    hi/.style={draw=cyan,line width=3pt} 
  }
  \begin{subfigure}[t]{\textwidth}
    \centering
    \resizebox{\linewidth}{!}{%
      \begin{tikzpicture}[yscale=1.1]
        \def\dx{8} \def\dy{1.5} \def\dxii{4} \def\leafsep{2}

        \node[node, draw=cyan, line width=2.5pt] (root) at (0,0) {$\ket{0}$};
        \coordinate (above) at ($(root)+(0,\dy/1.5)$);
        \draw[hi] (above) -- (root);

        \node[node, draw=cyan, line width=2.5pt] (L)  at ($(root)+(-\dx,-\dy)$) {$\ket{0}$};
        \node[node] (R)  at ($(root)+(\dx,-\dy)$)  {$\ket{\bullet}$};

        \node[node, draw=cyan, line width=2.5pt] (LL) at ($(L)+(-\dxii,-\dy)$) {$\ket{\gamma_0}$};
        \node[node] (LR) at ($(L)+(\dxii,-\dy)$)  {$\ket{\bullet}$};
        \node[node] (RL) at ($(R)+(-\dxii,-\dy)$) {$\ket{\bullet}$};
        \node[node] (RR) at ($(R)+(\dxii,-\dy)$)  {$\ket{\bullet}$};

        \node[leaf, draw=cyan, line width=2.5pt, levelzeroq, label=below:{$\ket{L_0}$}] (LLL) at ($(LL)+(-\leafsep,-\dy)$)
          {$\ket{0}$\nodepart{two}$\ket{32.48}$\nodepart{three}$\ket{b}$};
        \node[leaf, draw=cyan, line width=2.5pt, leveloneq,  label=below:{$\ket{L_1}$}] (LLR) at ($(LL)+(\leafsep,-\dy)$)
          {$\ket{0}$\nodepart{two}$\ket{24.89}$\nodepart{three}$\ket{7.59}$};
        \node[leaf, leveltwoq,  label=below:{$\ket{L_2}$}] (LRL) at ($(LR)+(-\leafsep,-\dy)$)
          {$\ket{1}$\nodepart{two}$\ket{14.45}$\nodepart{three}$\ket{10.44}$};
        \node[leaf, leveltwoq,  label=below:{$\ket{L_3}$}] (LRR) at ($(LR)+(\leafsep,-\dy)$)
          {$\ket{0}$\nodepart{two}$\ket{1.09}$\nodepart{three}$\ket{6.50}$};
        \node[leaf,  levelthreeq,label=below:{$\ket{L_4}$}] (RLL) at ($(RL)+(-\leafsep,-\dy)$)
          {$\ket{0}$\nodepart{two}$\ket{4.84}$\nodepart{three}$\ket{9.61}$};
        \node[leaf, levelthreeq,label=below:{$\ket{L_5}$}] (RLR) at ($(RL)+(\leafsep,-\dy)$)
          {$\ket{0}$\nodepart{two}$\ket{9.00}$\nodepart{three}$\ket{1.44}$};
        \node[leaf, levelthreeq,label=below:{$\ket{L_6}$}] (RRL) at ($(RR)+(-\leafsep,-\dy)$)
          {$\ket{0}$\nodepart{two}$\ket{0.09}$\nodepart{three}$\ket{1.00}$};
        \node[leaf, levelthreeq, label=below:{$\ket{L_7}$}] (RRR) at ($(RR)+(\leafsep,-\dy)$)
          {$\ket{1}$\nodepart{two}$\ket{0.25}$\nodepart{three}$\ket{6.25}$};

        \draw (root) -- ++(-\dx,0) coordinate(auxL) -- (L);
        \draw (root) -- ++(\dx,0)  coordinate(auxR) -- (R);
        \draw (L)  -- ++(-\dxii,0) coordinate(auxLL) -- (LL);
        \draw (L)  -- ++(\dxii,0)  coordinate(auxLR) -- (LR);
        \draw (R)  -- ++(-\dxii,0) coordinate(auxRL) -- (RL);
        \draw (R)  -- ++(\dxii,0)  coordinate(auxRR) -- (RR);
        \draw (LL) -- ++(-\leafsep,0) coordinate(auxLLL) -- (LLL);
        \draw (LL) -- ++(\leafsep,0)  coordinate(auxLLR) -- (LLR);
        \draw (LR) -- ++(-\leafsep,0) coordinate(auxLRL) -- (LRL);
        \draw (LR) -- ++(\leafsep,0)  coordinate(auxLRR) -- (LRR);
        \draw (RL) -- ++(-\leafsep,0) coordinate(auxRLL) -- (RLL);
        \draw (RL) -- ++(\leafsep,0)  coordinate(auxRLR) -- (RLR);
        \draw (RR) -- ++(-\leafsep,0) coordinate(auxRRL) -- (RRL);
        \draw (RR) -- ++(\leafsep,0)  coordinate(auxRRR) -- (RRR);

        \draw[hi] (root)--(auxL)--(L)
                  (L)--(auxLL)--(LL)  
                  (LL)--(auxLLL)--(LLL)
                  (LL)--(auxLLR)--(LLR); 
      \end{tikzpicture}%
    }
  \end{subfigure}
  \hfill
  \begin{subfigure}[t]{\textwidth}
    \centering
    \makebox[\textwidth][l]{\hspace{7.23cm}{\fontsize{30}{0}\selectfont \textcolor{gray}{+}}}
  \end{subfigure}
  \hfill
  \begin{subfigure}[t]{\textwidth}
    \centering
    \resizebox{\linewidth}{!}{%
      \begin{tikzpicture}[yscale=1.1]
        \def\dx{8} \def\dy{1.5} \def\dxii{4} \def\leafsep{2}

        \node[node, draw=cyan, line width=2.5pt] (root) at (0,0) {$\ket{1}$};
        \coordinate (above) at ($(root)+(0,\dy/1.5)$);
        \draw[hi] (above) -- (root);

        \node[node, draw=cyan, line width=2.5pt] (L)  at ($(root)+(-\dx,-\dy)$) {$\ket{1}$};
        \node[node] (R)  at ($(root)+(\dx,-\dy)$)  {$\ket{\bullet}$};

        \node[node] (LL) at ($(L)+(-\dxii,-\dy)$) {$\ket{\bullet}$};
        \node[node, draw=cyan, line width=2.5pt] (LR) at ($(L)+(\dxii,-\dy)$)  {$\ket{\gamma_1}$};
        \node[node] (RL) at ($(R)+(-\dxii,-\dy)$) {$\ket{\bullet}$};
        \node[node] (RR) at ($(R)+(\dxii,-\dy)$)  {$\ket{\bullet}$};

        \node[leaf, levelzeroq, label=below:{$\ket{L_0}$}] (LLL) at ($(LL)+(-\leafsep,-\dy)$)
          {$\ket{0}$\nodepart{two}$\ket{32.48}$\nodepart{three}$\ket{b}$};
        \node[leaf, leveloneq,  label=below:{$\ket{L_1}$}] (LLR) at ($(LL)+(\leafsep,-\dy)$)
          {$\ket{0}$\nodepart{two}$\ket{24.89}$\nodepart{three}$\ket{7.59}$};
        \node[leaf,  draw=cyan, line width=2.5pt, leveltwoq,  label=below:{$\ket{L_2}$}] (LRL) at ($(LR)+(-\leafsep,-\dy)$)
          {$\ket{1}$\nodepart{two}$\ket{14.45}$\nodepart{three}$\ket{10.44}$};
        \node[leaf,  draw=cyan, line width=2.5pt, leveltwoq,  label=below:{$\ket{L_3}$}] (LRR) at ($(LR)+(\leafsep,-\dy)$)
          {$\ket{0}$\nodepart{two}$\ket{1.09}$\nodepart{three}$\ket{6.50}$};
        \node[leaf, levelthreeq,label=below:{$\ket{L_4}$}] (RLL) at ($(RL)+(-\leafsep,-\dy)$)
          {$\ket{0}$\nodepart{two}$\ket{4.84}$\nodepart{three}$\ket{9.61}$};
        \node[leaf,levelthreeq,label=below:{$\ket{L_5}$}] (RLR) at ($(RL)+(\leafsep,-\dy)$)
          {$\ket{0}$\nodepart{two}$\ket{9.00}$\nodepart{three}$\ket{1.44}$};
        \node[leaf, levelthreeq,label=below:{$\ket{L_6}$}] (RRL) at ($(RR)+(-\leafsep,-\dy)$)
          {$\ket{0}$\nodepart{two}$\ket{0.09}$\nodepart{three}$\ket{1.00}$};
        \node[leaf, levelthreeq, label=below:{$\ket{L_7}$}] (RRR) at ($(RR)+(\leafsep,-\dy)$)
          {$\ket{1}$\nodepart{two}$\ket{0.25}$\nodepart{three}$\ket{6.25}$};

        \draw (root) -- ++(-\dx,0) coordinate(auxL) -- (L);
        \draw (root) -- ++(\dx,0)  coordinate(auxR) -- (R);
        \draw (L)  -- ++(-\dxii,0) coordinate(auxLL) -- (LL);
        \draw (L)  -- ++(\dxii,0)  coordinate(auxLR) -- (LR);
        \draw (R)  -- ++(-\dxii,0) coordinate(auxRL) -- (RL);
        \draw (R)  -- ++(\dxii,0)  coordinate(auxRR) -- (RR);
        \draw (LL) -- ++(-\leafsep,0) coordinate(auxLLL) -- (LLL);
        \draw (LL) -- ++(\leafsep,0)  coordinate(auxLLR) -- (LLR);
        \draw (LR) -- ++(-\leafsep,0) coordinate(auxLRL) -- (LRL);
        \draw (LR) -- ++(\leafsep,0)  coordinate(auxLRR) -- (LRR);
        \draw (RL) -- ++(-\leafsep,0) coordinate(auxRLL) -- (RLL);
        \draw (RL) -- ++(\leafsep,0)  coordinate(auxRLR) -- (RLR);
        \draw (RR) -- ++(-\leafsep,0) coordinate(auxRRL) -- (RRL);
        \draw (RR) -- ++(\leafsep,0)  coordinate(auxRRR) -- (RRR);

        \draw[hi] (root)--(auxL)--(L)
                  (L)--(auxLR)--(LR)  
                  (LR)--(auxLRL)--(LRL) 
                  (LR)--(auxLRR)--(LRR);      
      \end{tikzpicture}%
    }
  \end{subfigure}
  \hfill
  \begin{subfigure}[t]{\textwidth}
    \centering
    \makebox[\textwidth][l]{\hspace{7.23cm}{\fontsize{30}{0}\selectfont \textcolor{gray}{+}}}
  \end{subfigure}
  \hfill
  \begin{subfigure}[t]{\textwidth}
    \centering
    \resizebox{\linewidth}{!}{%
      \begin{tikzpicture}[yscale=1.1]
        \def\dx{8} \def\dy{1.5} \def\dxii{4} \def\leafsep{2}

        \node[node, draw=cyan, line width=2.5pt] (root) at (0,0) {$\ket{1}$};
        \coordinate (above) at ($(root)+(0,\dy/1.5)$);
        \draw[hi] (above) -- (root);

        \node[node] (L)  at ($(root)+(-\dx,-\dy)$) {$\ket{\bullet}$};
        \node[node, draw=cyan, line width=2.5pt] (R)  at ($(root)+(\dx,-\dy)$)  {$\ket{0}$};

        \node[node] (LL) at ($(L)+(-\dxii,-\dy)$) {$\ket{\bullet}$};
        \node[node] (LR) at ($(L)+(\dxii,-\dy)$)  {$\ket{\bullet}$};
        \node[node, draw=cyan, line width=2.5pt] (RL) at ($(R)+(-\dxii,-\dy)$) {$\ket{\gamma_2}$};
        \node[node] (RR) at ($(R)+(\dxii,-\dy)$)  {$\ket{\bullet}$};

        \node[leaf, levelzeroq, label=below:{$\ket{L_0}$}] (LLL) at ($(LL)+(-\leafsep,-\dy)$)
          {$\ket{0}$\nodepart{two}$\ket{32.48}$\nodepart{three}$\ket{b}$};
        \node[leaf, leveloneq,  label=below:{$\ket{L_1}$}] (LLR) at ($(LL)+(\leafsep,-\dy)$)
          {$\ket{0}$\nodepart{two}$\ket{24.89}$\nodepart{three}$\ket{7.59}$};
        \node[leaf, leveltwoq,  label=below:{$\ket{L_2}$}] (LRL) at ($(LR)+(-\leafsep,-\dy)$)
          {$\ket{1}$\nodepart{two}$\ket{14.45}$\nodepart{three}$\ket{10.44}$};
        \node[leaf, leveltwoq,  label=below:{$\ket{L_3}$}] (LRR) at ($(LR)+(\leafsep,-\dy)$)
          {$\ket{0}$\nodepart{two}$\ket{1.09}$\nodepart{three}$\ket{6.50}$};
        \node[leaf, draw=cyan, line width=2.5pt, levelthreeq,label=below:{$\ket{L_4}$}] (RLL) at ($(RL)+(-\leafsep,-\dy)$)
          {$\ket{0}$\nodepart{two}$\ket{4.84}$\nodepart{three}$\ket{9.61}$};
        \node[leaf, draw=cyan, line width=2.5pt, levelthreeq,label=below:{$\ket{L_5}$}] (RLR) at ($(RL)+(\leafsep,-\dy)$)
          {$\ket{0}$\nodepart{two}$\ket{9.00}$\nodepart{three}$\ket{1.44}$};
        \node[leaf, levelthreeq,label=below:{$\ket{L_6}$}] (RRL) at ($(RR)+(-\leafsep,-\dy)$)
          {$\ket{0}$\nodepart{two}$\ket{0.09}$\nodepart{three}$\ket{1.00}$};
        \node[leaf, levelthreeq, label=below:{$\ket{L_7}$}] (RRR) at ($(RR)+(\leafsep,-\dy)$)
          {$\ket{1}$\nodepart{two}$\ket{0.25}$\nodepart{three}$\ket{6.25}$};

        \draw (root) -- ++(-\dx,0) coordinate(auxL) -- (L);
        \draw (root) -- ++(\dx,0)  coordinate(auxR) -- (R);
        \draw (L)  -- ++(-\dxii,0) coordinate(auxLL) -- (LL);
        \draw (L)  -- ++(\dxii,0)  coordinate(auxLR) -- (LR);
        \draw (R)  -- ++(-\dxii,0) coordinate(auxRL) -- (RL);
        \draw (R)  -- ++(\dxii,0)  coordinate(auxRR) -- (RR);
        \draw (LL) -- ++(-\leafsep,0) coordinate(auxLLL) -- (LLL);
        \draw (LL) -- ++(\leafsep,0)  coordinate(auxLLR) -- (LLR);
        \draw (LR) -- ++(-\leafsep,0) coordinate(auxLRL) -- (LRL);
        \draw (LR) -- ++(\leafsep,0)  coordinate(auxLRR) -- (LRR);
        \draw (RL) -- ++(-\leafsep,0) coordinate(auxRLL) -- (RLL);
        \draw (RL) -- ++(\leafsep,0)  coordinate(auxRLR) -- (RLR);
        \draw (RR) -- ++(-\leafsep,0) coordinate(auxRRL) -- (RRL);
        \draw (RR) -- ++(\leafsep,0)  coordinate(auxRRR) -- (RRR);

        \draw[hi] (root)--(auxR)--(R)
                  (R)--(auxRL)--(RL)
                  (RL)--(auxRLL)--(RLL)
                  (RL)--(auxRLR)--(RLR);
      \end{tikzpicture}%
    }
  \end{subfigure}
  \hfill
  \begin{subfigure}[t]{\textwidth}
    \centering
    \makebox[\textwidth][l]{\hspace{7.23cm}{\fontsize{30}{0}\selectfont \textcolor{gray}{+}}}
  \end{subfigure}
  \hfill
  \begin{subfigure}[t]{\textwidth}
    \centering
    \resizebox{\linewidth}{!}{%
      \begin{tikzpicture}[yscale=1.1]
        \def\dx{8} \def\dy{1.5} \def\dxii{4} \def\leafsep{2}

        \node[node, draw=cyan, line width=2.5pt] (root) at (0,0) {$\ket{1}$};
        \coordinate (above) at ($(root)+(0,\dy/1.5)$);
        \draw[hi] (above) -- (root);

        \node[node] (L)  at ($(root)+(-\dx,-\dy)$) {$\ket{\bullet}$};
        \node[node, draw=cyan, line width=2.5pt] (R)  at ($(root)+(\dx,-\dy)$)  {$\ket{1}$};

        \node[node] (LL) at ($(L)+(-\dxii,-\dy)$) {$\ket{\bullet}$};
        \node[node] (LR) at ($(L)+(\dxii,-\dy)$)  {$\ket{\bullet}$};
        \node[node] (RL) at ($(R)+(-\dxii,-\dy)$) {$\ket{\bullet}$};
        \node[node, draw=cyan, line width=2.5pt] (RR) at ($(R)+(\dxii,-\dy)$)  {$\ket{\gamma_3}$};

        \node[leaf, levelzeroq, label=below:{$\ket{L_0}$}] (LLL) at ($(LL)+(-\leafsep,-\dy)$)
          {$\ket{0}$\nodepart{two}$\ket{32.48}$\nodepart{three}$\ket{b}$};
        \node[leaf, leveloneq,  label=below:{$\ket{L_1}$}] (LLR) at ($(LL)+(\leafsep,-\dy)$)
          {$\ket{0}$\nodepart{two}$\ket{24.89}$\nodepart{three}$\ket{7.59}$};
        \node[leaf, leveltwoq,  label=below:{$\ket{L_2}$}] (LRL) at ($(LR)+(-\leafsep,-\dy)$)
          {$\ket{1}$\nodepart{two}$\ket{14.45}$\nodepart{three}$\ket{10.44}$};
        \node[leaf, leveltwoq,  label=below:{$\ket{L_3}$}] (LRR) at ($(LR)+(\leafsep,-\dy)$)
          {$\ket{0}$\nodepart{two}$\ket{1.09}$\nodepart{three}$\ket{6.50}$};
        \node[leaf, levelthreeq,label=below:{$\ket{L_4}$}] (RLL) at ($(RL)+(-\leafsep,-\dy)$)
          {$\ket{0}$\nodepart{two}$\ket{4.84}$\nodepart{three}$\ket{9.61}$};
        \node[leaf, levelthreeq,label=below:{$\ket{L_5}$}] (RLR) at ($(RL)+(\leafsep,-\dy)$)
          {$\ket{0}$\nodepart{two}$\ket{9.00}$\nodepart{three}$\ket{1.44}$};
        \node[leaf, draw=cyan, line width=2.5pt, levelthreeq,label=below:{$\ket{L_6}$}] (RRL) at ($(RR)+(-\leafsep,-\dy)$)
          {$\ket{0}$\nodepart{two}$\ket{0.09}$\nodepart{three}$\ket{1.00}$};
        \node[leaf, draw=cyan, line width=2.5pt, levelthreeq,label=below:{$\ket{L_7}$}] (RRR) at ($(RR)+(\leafsep,-\dy)$)
          {$\ket{1}$\nodepart{two}$\ket{0.25}$\nodepart{three}$\ket{6.25}$};

        \draw (root) -- ++(-\dx,0) coordinate(auxL) -- (L);
        \draw (root) -- ++(\dx,0)  coordinate(auxR) -- (R);
        \draw (L)  -- ++(-\dxii,0) coordinate(auxLL) -- (LL);
        \draw (L)  -- ++(\dxii,0)  coordinate(auxLR) -- (LR);
        \draw (R)  -- ++(-\dxii,0) coordinate(auxRL) -- (RL);
        \draw (R)  -- ++(\dxii,0)  coordinate(auxRR) -- (RR);
        \draw (LL) -- ++(-\leafsep,0) coordinate(auxLLL) -- (LLL);
        \draw (LL) -- ++(\leafsep,0)  coordinate(auxLLR) -- (LLR);
        \draw (LR) -- ++(-\leafsep,0) coordinate(auxLRL) -- (LRL);
        \draw (LR) -- ++(\leafsep,0)  coordinate(auxLRR) -- (LRR);
        \draw (RL) -- ++(-\leafsep,0) coordinate(auxRLL) -- (RLL);
        \draw (RL) -- ++(\leafsep,0)  coordinate(auxRLR) -- (RLR);
        \draw (RR) -- ++(-\leafsep,0) coordinate(auxRRL) -- (RRL);
        \draw (RR) -- ++(\leafsep,0)  coordinate(auxRRR) -- (RRR);

        \draw[hi] (root)--(auxR)--(R)
                  (R)--(auxRR)--(RR)
                  (RR)--(auxRRL)--(RRL)
                  (RR)--(auxRRR)--(RRR);
      \end{tikzpicture}%
    }
  \end{subfigure}
  \caption{The figure illustrates the superposition of four distinct access paths when querying the memory cells for retrieving sign qubits with the address register 
$\ket{00}_{\mathrm{a}}\ket{\gamma_0}_{\mathrm{a}} + \ket{01}_{\mathrm{a}}\ket{\gamma_1}_{\mathrm{a}}+\ket{10}_{\mathrm{a}}\ket{\gamma_2}_{\mathrm{a}} + \ket{11}_{\mathrm{a}}\ket{\gamma_3}_{\mathrm{a}}$. 
Here, $\ket{\gamma_0} = \sqrt{4.84}\ket{0} + \sqrt{9.61}\ket{1}$, 
$\ket{\gamma_1} = \sqrt{9.00}\ket{0} + \sqrt{1.44}\ket{1}$, 
$\ket{\gamma_2} = \sqrt{0.09}\ket{0} + \sqrt{1.00}\ket{1}$, 
and $\ket{\gamma_3} = \sqrt{0.25}\ket{0} + \sqrt{6.25}\ket{1}$. 
This query accesses, in superposition, the memory cells labeled 
$\ket{L_0}, \ket{L_1}, \ket{L_2}, \ket{L_3}, \ket{L_4}, \ket{L_5}, \ket{L_6},$ and $\ket{L_7}$ 
at level $h = 3$ of the segment tree $T$. 
The + between the figures highlights that the quantum system of BBQRAM is in a superposition of four access paths.}
  \label{fig:bbqram_path_plusplusplus}\Description{}
\end{figure*}
Now we retrieve the last set of sibling nodes in superposition:
\begin{align*}
\frac{1}{\sqrt{32.48}}\Bigg( 
& \sqrt{14.45}\ket{0}_{\mathrm{s}}\ket{4.84}_{\mathrm{l}}\ket{9.61}_{\mathrm{r}}\ket{0}_{\mathrm{v}}\ket{100}_{\mathrm{a}}  + \sqrt{10.44}\ket{0}_{\mathrm{s}}\ket{9.00}_{\mathrm{l}}\ket{1.44}_{\mathrm{r}}\ket{0}_{\mathrm{v}}\ket{101}_{\mathrm{a}} + \\
& + \sqrt{1.09}\ket{0}_{\mathrm{s}}\ket{0.09}_{\mathrm{l}}\ket{1.00}_{\mathrm{r}}\ket{0}_{\mathrm{v}}\ket{110}_{\mathrm{a}}  + \sqrt{6.50}\ket{0}_{\mathrm{s}}\ket{0.25}_{\mathrm{l}}\ket{6.25}_{\mathrm{r}}\ket{0}_{\mathrm{v}}\ket{111}_{\mathrm{a}} \Bigg),
\end{align*}
and apply $U_{2CR}$ on the registers $\mathrm{l}$, $\mathrm{r}$, and $\mathrm{v}$:
\begin{align*}
\frac{1}{\sqrt{32.48}}\Bigg( 
& \sqrt{14.45}\ket{0}_{\mathrm{s}}\ket{4.84}_{\mathrm{l}}\ket{9.61}_{\mathrm{r}}  
    \left(
        \sqrt{\frac{4.84}{14.45}}\ket{0}_{\mathrm{v}}
        + \sqrt{\frac{9.61}{14.45}}\ket{1}_{\mathrm{v}}
    \right)  
    \ket{100}_{\mathrm{a}} + \\
& + \sqrt{10.44}\ket{0}_{\mathrm{s}} 
    \ket{9.00}_{\mathrm{l}}  \ket{1.44}_{\mathrm{r}}    \left(
        \sqrt{\frac{9.00}{10.44}}\ket{0}_{\mathrm{v}}
        + \sqrt{\frac{1.44}{10.44}}\ket{1}_{\mathrm{v}}
    \right)  
    \ket{101}_{\mathrm{a}} + \\
& + \sqrt{1.09}\ket{0}_{\mathrm{s}}\ket{0.09}_{\mathrm{l}}\ket{1.00}_{\mathrm{r}} \left(
        \sqrt{\frac{0.09}{1.09}}\ket{0}_{\mathrm{v}}
        + \sqrt{\frac{1.00}{1.09}}\ket{1}_{\mathrm{v}}
    \right)  \ket{110}_{\mathrm{a}} +  \\ & + \sqrt{6.50}\ket{0}_{\mathrm{s}}\ket{0.25}_{\mathrm{l}}\ket{6.25}_{\mathrm{r}}          
     \left(
        \sqrt{\frac{0.25}{6.50}}\ket{0}_{\mathrm{v}}
        + \sqrt{\frac{6.25}{6.50}}\ket{1}_{\mathrm{v}}
    \right)  
    \ket{111}_{\mathrm{a}} \Bigg),
\end{align*}
and we rewrite the state as
\begin{align*}
\frac{1}{\sqrt{32.48}}\Bigg( 
& \ket{0}_{\mathrm{s}}\ket{4.84}_{\mathrm{l}}\ket{9.61}_{\mathrm{r}}  
    \left(
        \sqrt{4.84}\ket{0}_{\mathrm{v}}
        + \sqrt{9.61}\ket{1}_{\mathrm{v}}
    \right)  
    \ket{100}_{\mathrm{a}} + \ket{0}_{\mathrm{s}}   \ket{9.00}_{\mathrm{l}}  \ket{1.44}_{\mathrm{r}}   
    \left(
        \sqrt{9.00}\ket{0}_{\mathrm{v}}
        + \sqrt{1.44}\ket{1}_{\mathrm{v}}
    \right)  
    \ket{101}_{\mathrm{a}} + \\
& + \ket{0}_{\mathrm{s}}\ket{0.09}_{\mathrm{l}}\ket{1.00}_{\mathrm{r}} 
     \left(
        \sqrt{0.09}\ket{0}_{\mathrm{v}}
        + \sqrt{1.00}\ket{1}_{\mathrm{v}}
    \right)  \ket{110}_{\mathrm{a}} + \ket{0}_{\mathrm{s}}\ket{0.25}_{\mathrm{l}}\ket{6.25}_{\mathrm{r}}   
     \left(
        \sqrt{0.25}\ket{0}_{\mathrm{v}}
        + \sqrt{6.25}\ket{1}_{\mathrm{v}}
    \right)  
    \ket{111}_{\mathrm{a}} \Bigg).
\end{align*}

Then, we uncompute the registers $\mathrm{l}$, $\mathrm{r}$:
\begin{align*}
\frac{1}{\sqrt{32.48}}\Bigg( 
& \ket{0}_{\mathrm{s}}\ket{0}_{\mathrm{l}}\ket{0}_{\mathrm{r}}  
    \left(
        \sqrt{4.84}\ket{0}_{\mathrm{v}}
        + \sqrt{9.61}\ket{1}_{\mathrm{v}}
    \right)  
    \ket{100}_{\mathrm{a}} + \ket{0}_{\mathrm{s}}   \ket{0}_{\mathrm{l}}  \ket{0}_{\mathrm{r}}   
    \left(
        \sqrt{9.00}\ket{0}_{\mathrm{v}}
        + \sqrt{1.44}\ket{1}_{\mathrm{v}}
    \right)  
    \ket{101}_{\mathrm{a}} + \\
& + \ket{0}_{\mathrm{s}}\ket{0}_{\mathrm{l}}\ket{0}_{\mathrm{r}} 
     \left(
        \sqrt{0.09}\ket{0}_{\mathrm{v}}
        + \sqrt{1.00}\ket{1}_{\mathrm{v}}
    \right)  \ket{110}_{\mathrm{a}} + \ket{0}_{\mathrm{s}}\ket{0}_{\mathrm{l}}\ket{0}_{\mathrm{r}}          
     \left(
        \sqrt{0.25}\ket{0}_{\mathrm{v}}
        + \sqrt{6.25}\ket{1}_{\mathrm{v}}
    \right)  
    \ket{111}_{\mathrm{a}} \Bigg).
\end{align*}

Finally, we perform a left circular shift over $\mathrm{v}$ and $\mathrm{a}$ to set the address register $\mathrm{a}$ for retrieving in superposition the sign bits stored in the BBQRAM memory cells $\{\ket{L_z}\}_{z=0}^{K-1}$ (see Figure~\ref{fig:bbqram_path_plusplusplus}):
\begin{align*}
\frac{1}{\sqrt{32.48}}\Bigg( 
& \ket{0}_{\mathrm{s}}\ket{0}_{\mathrm{l}}\ket{0}_{\mathrm{r}} \ket{1}_{\mathrm{v}}
    \ket{00}_{\mathrm{a}}    \left(
        \sqrt{4.84}\ket{0}_{\mathrm{a}}
        + \sqrt{9.61}\ket{1}_{\mathrm{a}}
    \right)   + \ket{0}_{\mathrm{s}}   \ket{0}_{\mathrm{l}}  \ket{0}_{\mathrm{r}} \ket{1}_{\mathrm{v}}  
    \ket{01}_{\mathrm{a}} \left(
        \sqrt{9.00}\ket{0}_{\mathrm{a}}
        + \sqrt{1.44}\ket{1}_{\mathrm{a}}
    \right)   + \\
& + \ket{0}_{\mathrm{s}}\ket{0}_{\mathrm{l}}\ket{0}_{\mathrm{r}} \ket{1} _{\mathrm{v}}
    \ket{10}_{\mathrm{a}}     \left(
        \sqrt{0.09}\ket{0}_{\mathrm{a}}
        + \sqrt{1.00}\ket{1}_{\mathrm{a}}
    \right) 
    + \ket{0}_{\mathrm{s}}\ket{0}_{\mathrm{l}}\ket{0}_{\mathrm{r}}  \ket{1}_{\mathrm{v}}        
    \ket{11}_{\mathrm{a}}
    \left(
        \sqrt{0.25}\ket{0}_{\mathrm{a}}
        + \sqrt{6.25}\ket{1}_{\mathrm{a}}
    \right)  
    \Bigg),
\end{align*}
which we can refactorize as
\begin{align*}
\frac{1}{\sqrt{32.48}}\Big( 
& \quad\sqrt{4.84}\ket{0}_{\mathrm{s}}\ket{0}_{\mathrm{l}}\ket{0}_{\mathrm{r}}   \ket{1}_{\mathrm{v}}\ket{000}_{\mathrm{a}} + \sqrt{9.61}\ket{0}_{\mathrm{s}}\ket{0}_{\mathrm{l}}\ket{0}_{\mathrm{r}}  \ket{1}_{\mathrm{v}}\ket{001}_{\mathrm{a}} + \\
& +\sqrt{9.00}\ket{0}_{\mathrm{s}}\ket{0}_{\mathrm{l}}\ket{0}_{\mathrm{r}}  \ket{1}_{\mathrm{v}}\ket{010}_{\mathrm{a}}  +\sqrt{1.44}\ket{0}_{\mathrm{s}}\ket{0}_{\mathrm{l}}\ket{0}_{\mathrm{r}}  \ket{1}_{\mathrm{v}}\ket{011}_{\mathrm{a}} + \\
& +\sqrt{0.09}\ket{0}_{\mathrm{s}}\ket{0}_{\mathrm{l}}\ket{0}_{\mathrm{r}}  \ket{1}_{\mathrm{v}}\ket{100}_{\mathrm{a}}  +\sqrt{1.00}\ket{0}_{\mathrm{s}}\ket{0}_{\mathrm{l}}\ket{0}_{\mathrm{r}}  \ket{1}_{\mathrm{v}}\ket{101}_{\mathrm{a}} + \\
& +\sqrt{0.25}\ket{0}_{\mathrm{s}}\ket{0}_{\mathrm{l}}\ket{0}_{\mathrm{r}}  \ket{1}_{\mathrm{v}}\ket{110}_{\mathrm{a}}  +\sqrt{6.25}\ket{0}_{\mathrm{s}}\ket{0}_{\mathrm{l}}\ket{0}_{\mathrm{r}}  \ket{1}_{\mathrm{v}}\ket{111}_{\mathrm{a}} \Big),
\end{align*}
and rewrite as 
\begin{align*}
\frac{1}{5.698}\Big( 
& 2.2\ket{0}_{\mathrm{s}}\ket{0}_{\mathrm{l}}\ket{0}_{\mathrm{r}}   \ket{1}_{\mathrm{v}}\ket{000}_{\mathrm{a}} +3.1\ket{0}_{\mathrm{s}}\ket{0}_{\mathrm{l}}\ket{0}_{\mathrm{r}}  \ket{1}_{\mathrm{v}}\ket{001}_{\mathrm{a}} +3.0\ket{0}_{\mathrm{s}}\ket{0}_{\mathrm{l}}\ket{0}_{\mathrm{r}}  \ket{1}_{\mathrm{v}}\ket{010}_{\mathrm{a}} +1.2\ket{0}_{\mathrm{s}}\ket{0}_{\mathrm{l}}\ket{0}_{\mathrm{r}}  \ket{1}_{\mathrm{v}}\ket{011}_{\mathrm{a}} + \\
& +0.3\ket{0}_{\mathrm{s}}\ket{0}_{\mathrm{l}}\ket{0}_{\mathrm{r}}  \ket{1}_{\mathrm{v}}\ket{100}_{\mathrm{a}} 
 +1.0\ket{0}_{\mathrm{s}}\ket{0}_{\mathrm{l}}\ket{0}_{\mathrm{r}}  \ket{1}_{\mathrm{v}}\ket{101}_{\mathrm{a}}  +0.5\ket{0}_{\mathrm{s}}\ket{0}_{\mathrm{l}}\ket{0}_{\mathrm{r}}  \ket{1}_{\mathrm{v}}\ket{110}_{\mathrm{a}}  +2.5\ket{0}_{\mathrm{s}}\ket{0}_{\mathrm{l}}\ket{0}_{\mathrm{r}}  \ket{1}_{\mathrm{v}}\ket{111}_{\mathrm{a}} \Big),
\end{align*}
where $\frac{1}{5.689}$ coincides with the Frobenius norm of $A$ and the amplitudes associated with each superposition of the address register $\mathrm{a}$ match the entries of the matrix $A$ up to a sign. Next, we retrieve the signs located in the leftmost qubit of each memory cell into the register $\mathrm{s}$:
\begin{align*}
\frac{1}{5.698}\Big( 
& 2.2\ket{0}_{\mathrm{s}}\ket{0}_{\mathrm{l}}\ket{0}_{\mathrm{r}}   \ket{1}_{\mathrm{v}}\ket{000}_{\mathrm{a}}  +3.1\ket{0}_{\mathrm{s}}\ket{0}_{\mathrm{l}}\ket{0}_{\mathrm{r}} \ket{1}_{\mathrm{v}}\ket{001}_{\mathrm{a}}  +3.0\ket{1}_{\mathrm{s}}\ket{0}_{\mathrm{l}}\ket{0}_{\mathrm{r}}  \ket{1}_{\mathrm{v}}\ket{010}_{\mathrm{a}} +1.2\ket{0}_{\mathrm{s}}\ket{0}_{\mathrm{l}}\ket{0}_{\mathrm{r}}  \ket{1}_{\mathrm{v}}\ket{011}_{\mathrm{a}} + \\
& +0.3\ket{0}_{\mathrm{s}}\ket{0}_{\mathrm{l}}\ket{0}_{\mathrm{r}}  \ket{1}_{\mathrm{v}}\ket{100}_{\mathrm{a}}  +1.0\ket{0}_{\mathrm{s}}\ket{0}_{\mathrm{l}}\ket{0}_{\mathrm{r}}  \ket{1}_{\mathrm{v}}\ket{101}_{\mathrm{a}}  +0.5\ket{0}_{\mathrm{s}}\ket{0}_{\mathrm{l}}\ket{0}_{\mathrm{r}}  \ket{1}_{\mathrm{v}}\ket{110}_{\mathrm{a}}  +2.5\ket{1}_{\mathrm{s}}\ket{0}_{\mathrm{l}}\ket{0}_{\mathrm{r}}  \ket{1}_{\mathrm{v}}\ket{111}_{\mathrm{a}} \Big).
\end{align*}

The memory layout ensures that each superposition indexing a negative entry $a_{i,j} \in A$ has the sign qubit $\mathrm{s}$ in the state $\ket{1}$. Hence, we apply a controlled-$Z$ gate ($CZ$) with $\mathrm{s}$ as the control qubit and $\mathrm{v}$ as the target qubit. This operation introduces a phase of $-1$ when the qubit $\mathrm{s}$ is in the state $\ket{1}$, thereby setting the correct sign to each amplitude:
\begin{align*}
\frac{1}{5.698}\Big( 
& 2.2 \ket{0}_{\mathrm{s}}\ket{0}_{\mathrm{l}}\ket{0}_{\mathrm{r}}  \ket{1}_{\mathrm{v}}\ket{000}_{\mathrm{a}} + 3.1\ket{0}_{\mathrm{s}}\ket{0}_{\mathrm{l}}\ket{0}_{\mathrm{r}}  \ket{1}_{\mathrm{v}}\ket{001}_{\mathrm{a}} - 3.0\ket{1}_{\mathrm{s}}\ket{0}_{\mathrm{l}}\ket{0}_{\mathrm{r}}  \ket{1}_{\mathrm{v}}\ket{010}_{\mathrm{a}} 
+ 1.2 \ket{0}_{\mathrm{s}}\ket{0}_{\mathrm{l}}\ket{0}_{\mathrm{r}}  \ket{1}_{\mathrm{v}}\ket{011}_{\mathrm{a}} + \\
& + 0.3 \ket{0}_{\mathrm{s}}\ket{0}_{\mathrm{l}}\ket{0}_{\mathrm{r}}  \ket{1}_{\mathrm{v}}\ket{100}_{\mathrm{a}}  + 1.0 \ket{0}_{\mathrm{s}}\ket{0}_{\mathrm{l}}\ket{0}_{\mathrm{r}}  \ket{1}_{\mathrm{v}}\ket{101}_{\mathrm{a}}  + 0.5 \ket{0}_{\mathrm{s}}\ket{0}_{\mathrm{l}}\ket{0}_{\mathrm{r}}  \ket{1}_{\mathrm{v}}\ket{110}_{\mathrm{a}}  - 2.5\ket{1}_{\mathrm{s}}\ket{0}_{\mathrm{l}}\ket{0}_{\mathrm{r}}  \ket{1}_{\mathrm{v}}\ket{111}_{\mathrm{a}} \Big).
\end{align*}
We uncompute registers $\mathrm{s}$ and  $\mathrm{v}$:
\begin{align*}
\frac{1}{5.698}\Big( 
& 2.2  \ket{0}_{\mathrm{s}}\ket{0}_{\mathrm{l}}\ket{0}_{\mathrm{r}}  \ket{0}_{\mathrm{v}}\ket{000}_{\mathrm{a}}  +3.1\ket{0}_{\mathrm{s}}\ket{0}_{\mathrm{l}}\ket{0}_{\mathrm{r}}  \ket{0}_{\mathrm{v}}\ket{001}_{\mathrm{a}}  -3.0 \ket{0}_{\mathrm{s}}\ket{0}_{\mathrm{l}}\ket{0}_{\mathrm{r}}  \ket{0}_{\mathrm{v}}\ket{010}_{\mathrm{a}}  +1.2 \ket{0}_{\mathrm{s}}\ket{0}_{\mathrm{l}}\ket{0}_{\mathrm{r}}  \ket{0}_{\mathrm{v}}\ket{011}_{\mathrm{a}} + \\
& +0.3 \ket{0}_{\mathrm{s}}\ket{0}_{\mathrm{l}}\ket{0}_{\mathrm{r}}  \ket{0}_{\mathrm{v}}\ket{100}_{\mathrm{a}}  +1.0 \ket{0}_{\mathrm{s}}\ket{0}_{\mathrm{l}}\ket{0}_{\mathrm{r}}  \ket{0}_{\mathrm{v}}\ket{101}_{\mathrm{a}}  +0.5 \ket{0}_{\mathrm{s}}\ket{0}_{\mathrm{l}}\ket{0}_{\mathrm{r}}  \ket{0}_{\mathrm{v}}\ket{110}_{\mathrm{a}}  -2.5 \ket{0}_{\mathrm{s}}\ket{0}_{\mathrm{l}}\ket{0}_{\mathrm{r}}  \ket{0}_{\mathrm{v}}\ket{111}_{\mathrm{a}} \Big),
\end{align*}
and we refactorize the state as
\begin{align*}
 \frac{1}{5.689}  \ket{0}_{\mathrm{s}}\ket{0}_{\mathrm{l}}\ket{0}_{\mathrm{r}} \ket{0}_{\mathrm{v}}\Big(2.2 \ket{000}_{\mathrm{a}}  +3.1\ket{001}_{\mathrm{a}} - 3.0\ket{010}_{\mathrm{a}} + 1.2 \ket{011}_{\mathrm{a}} + \\ 
 + 0.3\ket{100}_{\mathrm{a}} + 1.00 \ket{101}_{\mathrm{a}} + 0.5 \ket{110}_{\mathrm{a}} - 2.5 \ket{111}_{\mathrm{a}} \Big).
\end{align*}
Finally, we trace out registers $\mathrm{s}$, $\mathrm{l}$, $\mathrm{r}$, and $\mathrm{v}$, and return to the matrix coordinates $\mathrm{i}$ and $\mathrm{j}$:
\begin{align*}
\frac{1}{5.698}  \Big(2.2 \ket{0}_{\mathrm{i}}\ket{00}_{\mathrm{j}}  +3.1\ket{0}_{\mathrm{i}}\ket{01}_{\mathrm{j}} - 3.0\ket{0}_{\mathrm{i}}\ket{10}_{\mathrm{j}} + 1.2 \ket{0}_{\mathrm{i}}\ket{11}_{\mathrm{j}} + \\
+0.3\ket{1}_{\mathrm{i}}\ket{00}_{\mathrm{j}} + 1.00 \ket{1}_{\mathrm{i}}\ket{01}_{\mathrm{j}} + 0.5 \ket{1}_{\mathrm{i}}\ket{10}_{\mathrm{j}} - 2.5 \ket{1}_{\mathrm{i}}\ket{11}_{\mathrm{j}} \Big),
\end{align*}
obtaining the encoding of $A \in \mathbb{R}^{2\times 4}$ in a quantum state as described in Theorem~\ref{theo:efficient_state_prep}.

\section{Conclusion and Future Work}\label{sec:conclusion}
This work showed that quantum algorithms can operate under the assumption of negligible data loading costs when employing the hardware QRAM model of BBQRAM.
In particular, we explored the challenge of providing an efficient state preparation procedure that is aware of the underlying QRAM architecture. We demonstrated that a matrix $A \in \mathbb{R}^{M\times N}$ can be encoded in $\mathcal{O}(\log_2^2(MN))$ time and using $\Theta(\log_2(MN))$ qubits, {with constant working resources and $\mathcal{O}(MN)$ memory cells within BBQRAM}. Specifically, we designed a memory layout that maps segment tree nodes to BBQRAM memory cells while preserving the tree's hierarchical structure and enabling data retrieval in logarithmic time. Then, we defined a set of quantum primitives for retrieving precomputed amplitudes and sign bits in superposition. These primitives enable the algorithm to orchestrate data access patterns that exploit both the hierarchical organization of the segment tree and the parallelism inherent in quantum memory; which is essential for an efficient amplitude encoding algorithm.
This framework sets forth a new paradigm for designing quantum algorithms that are \emph{architecture-aware} from the ground up. Our explicit $\mathcal{O}(\log^2(MN))$ time bound sharpens previous results and provides implementers with concrete guidance for resource planning.

Several areas for improvement are possible future research directions. First, the framework can be extended to handle complex matrices by incorporating the phase information in the segment tree and adapting the state preparation procedure accordingly. Second, specialized memory layouts for sparse matrices could be explored to store only non-zero elements, potentially reducing memory requirements. {Third, exploring alternatives that do not require the $U_{2CR}$ unitary is a valuable direction. Indeed, despite its theoretical $\tilde{O}(1)$ cost under fixed precision, the underlying fast-arithmetic circuits can lead to non-negligible overhead in practical implementations.} Finally, adaptive precision schemes that dynamically adjust bit width based on algorithmic requirements, rather than using fixed precision for all values, could further optimize performance. 

This research direction aims to bridge the gap between theoretical quantum algorithms and practical implementations by addressing scalability challenges, and designing more efficient classical-to-quantum encoding algorithms that are aware of the quantum memory. {As quantum hardware matures and quantum algorithms grow in complexity, architecture-aware approaches to state preparation --- such as the one presented in this work --- offer a complementary perspective to circuit-based methods, and can inform the co-design of quantum software and memory architectures.}

\section*{Acknowledgements}
We really want to thank Alessandro Luongo and Alexander Hue for the in-depth discussions that helped shape the research questions explored in this manuscript.

This work is supported by the National Centre on HPC, Big Data and Quantum Computing - SPOKE 10 (Quantum Computing) and received funding from the European Union Next-GenerationEU - National Recovery and Resilience Plan (NRRP) – MISSION \emph{4} COMPONENT \emph{2}, INVESTMENT N. \emph{1.4} – CUP N. \emph{I53C22000690001}. A. Berti was supported by INdAM - GNCS Project N. \emph{E53C24001950001}. F. Ghisoni was supported by the `National Quantum Science Technology Institute' (NQSTI, PE4) within the PNRR  project \emph{PE0000023}.

\bibliographystyle{ACM-Reference-Format}
\bibliography{main}


\end{document}